\documentclass{aa}  

\usepackage{graphicx}
\usepackage{natbib}
\usepackage{txfonts}
\bibpunct{(}{)}{;}{a}{}{,} % to follow the A&A style
\usepackage{longtable,lscape}
\usepackage[figuresright]{rotating}
\usepackage{breqn}
\usepackage{multirow}
\usepackage{rotating}
\usepackage{lscape}

\def\hi{\relax \ifmmode {\mbox H\,{\scshape i}}\else H\,{\scshape i}\fi}

\def\hii{\relax \ifmmode {\mbox H\,{\scshape ii}}\else H\,{\scshape ii}\fi}

\def\nii{\relax \ifmmode {\mbox N\,{\scshape ii}}\else N\,{\scshape ii}\fi}

\def\oi{\relax \ifmmode {\mbox O\,{\scshape i}}\else O\,{\scshape i}\fi}

\def\oii{\relax \ifmmode {\mbox O\,{\scshape ii}}\else O\,{\scshape ii}\fi}

\def\oiii{\relax \ifmmode {\mbox O\,{\scshape iii}}\else O\,{\scshape iii}\fi}

\def\cii{\relax \ifmmode {\mbox C\,{\scshape ii}}\else C\,{\scshape ii}\fi}

\def\ciii{\relax \ifmmode {\mbox C\,{\scshape iii}}\else C\,{\scshape iii}\fi}

\def\civ{\relax \ifmmode {\mbox C\,{\scshape iv}}\else C\,{\scshape iv}\fi}

\def\hei{\relax \ifmmode {\mbox He\,{\scshape i}}\else He\,{\scshape i}\fi}

\def\heii{\relax \ifmmode {\mbox He\,{\scshape ii}}\else He\,{\scshape ii}\fi}

\def\mgii{\relax \ifmmode {\mbox Mg\,{\scshape ii}}\else Mg\,{\scshape ii}\fi}

\def\sii{\relax \ifmmode {\mbox S\,{\scshape ii}}\else S\,{\scshape ii}\fi}

\def\neiii{\relax \ifmmode {\mbox Ne\,{\scshape iii}}\else Ne\,{\scshape iii}\fi}

\def\ariv{\relax \ifmmode {\mbox Ar\,{\scshape iv}}\else Ar\,{\scshape iv}\fi}

\def\ni{\relax \ifmmode {\mbox N\,{\scshape i}}\else N\,{\scshape i}\fi}

\def\ariii{\relax \ifmmode {\mbox Ar\,{\scshape iii}}\else Ar\,{\scshape iii}\fi}

\def\caii{\relax \ifmmode {\mbox Ca\,{\scshape ii}}\else Ca\,{\scshape ii}\fi}

\begin{document} 

    \title{Recovering star formation histories:  Integrated-light analyses vs stellar colour-magnitude diagrams \thanks{Based on observations obtained at the 3.6m ESO telescope on La Silla (Chile) and with the Hubble Space Telescope, operated by NASA.}}
    \titlerunning{Recovering star formation histories}

    \author{T. Ruiz-Lara \inst{1, 2}, I. P\'erez \inst{1, 2}, C. Gallart \inst{3, 4}, D. Alloin \inst{5}, M. Monelli \inst{3, 4}, M. Koleva \inst{6}, E. Pompei \inst{7}, M. Beasley \inst{3, 4}, P. S\'anchez-Bl\'azquez \inst{8}, E. Florido \inst{1, 2}, A. Aparicio \inst{3, 4}, E. Fleurence \inst{9}, E. Hardy \inst{10, 11}, S. Hidalgo \inst{3, 4}, \and D. Raimann \inst{12}}
    %\author{T. Ruiz-Lara \inst{1, 2}, I. P\'erez \inst{1, 2}, C. Gallart \inst{3, 4}, D. Alloin \inst{5}, M. Monelli \inst{3, 4}, M. Koleva \inst{6}, E. Pompei \inst{7}, E. Fleurence \inst{7}, M. Beasley \inst{3, 4}, P. S\'anchez-Bl\'azquez \inst{8}, E. Florido \inst{1, 2}, A. Aparicio \inst{3, 4}, S. Hidalgo \inst{3, 4}, D. Raimann \inst{9}, P. Demarque \inst{10}, E. Hardy \inst{11, 12}, U. Fritze \inst{13}, \and R. Zinn \inst{10} }

    \authorrunning{T. Ruiz-Lara et al.}

    \institute{\inst{1} Departamento de F\'isica Te\'orica y del Cosmos, Universidad de Granada, Campus de Fuentenueva, E-18071 Granada, Spain \\
       \email{ruizlara@ugr.es} \\
    \inst{2} Instituto Carlos I de F\'isica Te\'orica y computacional, Universidad de Granada, E-18071 Granada, Spain \\ 
    \inst{3} Instituto de Astrof\'isica de Canarias, Calle V\'ia L\'actea s/n, E-38205 La Laguna, Tenerife, Spain \\
    \inst{4} Departamento de Astrof\'isica, Universidad de La Laguna, E-38200 La Laguna, Tenerife, Spain \\
    \inst{5} Observatoire de Haute Provence, Institut Pytheas CNRS-Universit\'e Aix-Marseille, 04870 Saint Michel l'Observatoire, France \\
    \inst{6} Sterrenkundig Observatorium, Krijgslaan 281, B-9000 Gent, Ghent University, Belgium \\
    \inst{7} European Southern Observatory, Alonso de Cordova 3107, Casilla 19001, Vitacura, Santiago 19, Chile \\
    \inst{8} Departamento de F\'isica Te\'orica, Universidad Aut\'onoma de Madrid, E-28049 Cantoblanco, Spain \\ 
    \inst{9} Cours Charlemagne Nancy, 11 avenue Saint-Michel 54220 Malzeville, France \\
    \inst{10} National Radio Astronomy Observatory Avenida Nueva Costanera 4091, Vitacura, Santiago, Chile \\
    \inst{11} Departamento de Astronom\'ia, Universidad de Chile Casilla 36-D, Santiago, Chile \\
    \inst{12} Universidade do Estado de Santa Catarina, Rua Beloni Trombeta Zanin 680E, Bairro Santo Ant\^onio, Chapec\'o 89815-630, SC, Brazil \\}
    %\inst{10} Department of Astronomy, Yale University P.O. Box 208101, New Haven, CT 06520-8101 \\
    %\inst{13} Centre for Astrophysics Research, University of Hertfordshire, College Lane, Hatfield AL10 9AB \\}

    \date{Received ---; accepted ---}

% \abstract{}{}{}{}{} 
% 5 {} token are mandatory
 
    \abstract
     {Accurate star formation histories (SFHs) of galaxies are fundamental for understanding the build-up of their stellar
content. However, the most accurate SFHs - those obtained from colour-magnitude
diagrams (CMDs) of resolved stars reaching the oldest main sequence turnoffs (oMSTO) - are presently limited to a few systems in the Local Group. It is therefore crucial to determine the reliability and range of applicability of SFHs derived from integrated light spectroscopy, as this affects our understanding of unresolved galaxies from low to high redshift.}
     {To evaluate the reliability of current full spectral fitting techniques in deriving SFHs from integrated light spectroscopy by comparing SFHs from integrated spectra to those obtained from deep CMDs of resolved stars.} %The LMC bar characteristics allow us to obtain high quality integrated spectra and Colour-Magnitude Diagrams (CMD).}}
     {We have obtained a high signal--to--noise (S/N $\sim$ 36.3 per \AA) integrated spectrum of a field in the bar of the Large Magellanic Cloud (LMC) using EFOSC2 at the 3.6 meter telescope at La Silla Observatory. For this same field, resolved stellar data reaching the oMSTO are available.  We have compared the star formation rate (SFR) as a function of time and the age-metallicity relation (AMR) obtained from the integrated spectrum using {\tt STECKMAP}, and the CMD using the IAC-star/MinnIAC/IAC-pop set of routines. For the sake of completeness we also use and discuss other synthesis codes ({\tt STARLIGHT} and {\tt ULySS}) to derive the SFR and AMR from the integrated LMC spectrum.}
     {We find very good agreement (average differences $\sim$ 4.1 $\%$) between the SFR(t) and the AMR obtained using {\tt STECKMAP} on the integrated light spectrum, and the CMD analysis. {\tt STECKMAP} minimizes the impact of the age-metallicity degeneracy and has the advantage of preferring smooth solutions to recover complex SFHs by means of a penalized $\chi^2$. We find that the use of single stellar populations (SSPs) to recover the stellar content, using for instance {\tt STARLIGHT} or {\tt ULySS} codes, hampers the reconstruction of the SFR(t) and AMR shapes, yielding larger discrepancies with respect to the CMD results. These discrepancies can be reduced if spectral templates based on known and complex SFHs instead of just single SSPs are employed.}
     {}


    \maketitle
%
%________________________________________________________________

\section{Introduction}
\label{intro}

The study of the star formation histories (SFH) of galaxies is a key element in understanding their past evolution. How the baryonic component of a galaxy has formed and evolved should be reflected in its stellar populations. 
%Thus, the accurate determination of the SFH of galaxies is one of the main breakthroughs in understanding their evolution. 
Due to observational constraints, different methodologies are applied to the study of the stellar content in galaxies \citep[see][and references therein]{2013seg..book..353P}. Intrinsically, these methodologies are very different and are affected by different sources of errors. 

Deep Colour-Magnitude Diagrams (CMD) reaching the oldest main sequence turnoff (oMSTO) are generally regarded as the most direct and reliable observables in order to obtain a detailed SFH of a galaxy \citep{Gallart05ARAA}. This is because at magnitudes brighter\footnote {Below the oMSTO, there is basically no age information, but the luminosity function of the low main sequence can be used to obtain information on the initial mass function (IMF) of low mass stars \citep[e.g.,][]{2013ApJ...763..110K}} than the oMSTO, stars along the main sequence are distributed in a sequence of age: short lived, young, massive stars are the bluest and brightest, and less massive and therefore longer lived stars are progressively fainter and redder. While there remains some age-metallicity degeneracy in the positions of the stars on the main sequence, it is significantly less than that seen in other stellar evolutionary phases such as the red giant branch (RGB), or the horizontal-branch. In such features, the position of the stars are mainly determined by metallicity and there is very little age sensitivity. By combining the information of the position of the stars in the main sequence of a CMD reaching the oMSTO, with the number counts across it, it is possible to minimize the remaining age-metallicity degeneracy and obtain accurate, detailed and reliable SFHs, including at early times in the galaxy's history \citep[e.g.,][]{Gallart99sfh, Dolphin2002, Cole2007, Noel2009, 2010ApJ...722.1864M, monelli10b, hidalgo11, 2012A&A...544A..73D, meschin14, Skillman2014}. However, this kind of analysis is so far limited to a few dozens of nearby systems, those within a distance of $\simeq$ 1-2 Mpc \citep[see][for an updated census of the Local Group, and distance of its members to the Milky Way]{2005MNRAS.356..979M}. The wide variety of morphologies and characteristics found in galaxies forces us to study systems at larger distances, where the analysis of individual stars is unfeasible. In these systems, because of the limited spatial resolution, we need to derive the stellar content using colours or spectral information coming from integrated stellar populations \citep[e.g.,][among many others]{1996A&A...313..377D, 2004ApJS..152..175M, 2011A&A...529A..64P, 2011MNRAS.415..709S}.

In order to characterise the composite stellar populations of unresolved galaxies, and in particular, of galaxies at high redshift, significant effort has been put since the early 80's to interpret  integrated stellar populations \citep[e.g.,][]{1985ApJ...296..340P, 1988A&A...195...76B}. Broad band colours were first used to separate between old/young, metal poor/rich stellar populations \citep[][]{1973ApJ...179..427S, 1976ApJ...203...52T, 1996AJ....111.2238P}. However, this type of analysis is affected by the known age-metallicity degeneracy and the effect of dust \citep[e.g.,][]{1994ApJS...95..107W, 2009MNRAS.395.1669G}. Spectroscopic analysis can help to minimize that degeneracy. The very first spectroscopic approach was developed by \citet[][]{1959PASP...71...83D}, later followed by \citet[][]{1962ApJ...135..715S} and \citet[][]{1971A&A....10..401A}. The use of line equivalent widths or line--strength indices, taking into account their dependence on metallicity and age, became popular in the 80's and 90's \citep[e.g.,][]{1984AJ.....89.1238R, 1985ApJS...57..711F, 1986A&A...162...21B, 1986A&AS...66..171B, 1988A&A...195...76B, 1993ApJS...86..153G, 1994ApJS...95..107W, 1994A&A...283..805B, 1996ApJS..106..307V, 1997ApJS..111..377W, 2003MNRAS.341...33K}. Indices have been used to obtain single stellar population (SSP) equivalent values (age and metallicity) or even to derive the whole SFH shape \citep[e.g.,][]{2000MNRAS.311...37P}. The use of indices has been relatively successful in the characterization of stellar populations in ``simple'' systems such as globular clusters or elliptical galaxies \citep[e.g.,][]{2007MNRAS.379..445P, 2010MNRAS.408...97K}. Yet, these studies i) are still hampered by the age-metallicity degeneracy hindering the derivation of a reliable SFH, ii) are biased towards the youngest stellar populations which contribute much to the light while involving a small mass fraction, iii) make use of a limited part of the observed spectra, and iv) are quite limited when replicating the whole shape of the SFH. Combinations of different indices allow us to better constrain the stellar information, but this approach is still affected by the above limitations. Other approaches must be used to study the actual shape of the SFH in galaxies with complex stellar compositions. 

With the goal of enlarging the spectral coverage (and thereby maximizing the information used from the observed spectra) and of being able to reproduce the SFH of complex systems, spectral energy distribution (SED) fitting codes have been developed such as {\tt MOPED} \citep[][]{2000MNRAS.317..965H, 2001MNRAS.327..849R}, {\tt VESPA} \citep[][]{2007MNRAS.381.1252T} or {\tt STARLIGHT} \citep[][]{2005MNRAS.358..363C} along with other approaches. In this approach,  the overall shape of the observed spectrum is fitted through a combination of stellar population models. These codes are sensitive to problems in the data such as flux calibration or extinction errors, as they also take into account the continuum in the fit. At the same time, the first so-called ``full spectrum fitting codes'' \citep[e.g.,][STECKMAP and ULySS respectively]{2006MNRAS.365...46O, 2006MNRAS.365...74O, 2009A&A...501.1269K} became available. These techniques avoid problems in the flux calibration and extinction in the spectra by fitting a polynomial to the shape of the continuum. Both SED and full spectrum fitting codes are better at reducing the impact of the age-metallicity degeneracy than line--strength indices analysis as they maximize the information used from the observed spectrum \citep[][]{2008MNRAS.385.1998K, 2011MNRAS.415..709S, 2015arXiv150406128B}. Along with this evolution in the fitting codes, there has been a huge progress in the modeling of stellar populations \citep[e.g.,][]{2003MNRAS.344.1000B, 2005ApJS..160..176L, 2007ApJS..171..146S, 2009ApJ...699..486C, 2010MNRAS.404.1639V} partly based on the improvement of stellar libraries \citep[e.g.,][]{2001A&A...369.1048P, 2003A&A...402..433L, 2004ApJS..152..251V, 2006MNRAS.371..703S, 2007astro.ph..3658P}, isochrones and evolutionary tracks \citep[e.g.,][]{2000A&AS..141..371G, 2004ApJ...612..168P, 2012MNRAS.427..127B, 2013A&A...558A..46P}, and studies on the IMF \citep[][]{1955ApJ...121..161S, 1996ApJS..106..307V, 2001MNRAS.322..231K, 2013MNRAS.436.3309W, 2013MNRAS.435.2274W, 2014ApJ...784..162P}.

Various workers have tested the various  SED and full spectrum fitting codes using artificial spectra \citep[e.g.,][]{2005MNRAS.358..363C, 2006MNRAS.365...74O, 2011MNRAS.415..709S, 2011EAS....48...87K}. In particular, \citet[][]{2011EAS....48...87K} test {\tt STARLIGHT}, {\tt STECKMAP}, and {\tt ULySS} with 48 mock spectra with different known SFHs. A general result of these studies is that the final success in the recovery of the stellar population content lies in the quality of the spectrum: the better the S/N and the resolution of the observed spectra, the more reliable the stellar population determination will be. 

To further test the reliability of full spectrum fitting techniques, it is of crucial importance to thoroughly compare the (in principle) more reliable CMD results to those obtained using these codes with actual data.  This exercise should reveal where they fail and where they succeed, and thus inform about the spectral ranges and features which are more likely to improve these techniques. Such analysis is not straightforward. Not only do CMDs containing stars as faint as those on the oMSTO need to be studied, but also high resolution and high $S/N$ integrated spectra must be obtained. The latter is hampered by the often low surface brightness of nearby resolved systems. Although previous studies have tried to do something similar in star clusters \citep[e.g.,][]{1999AJ....118.1268G, 2002MNRAS.336..168B, 2006MNRAS.366..295D, 2006A&A...448.1023S, 2010MNRAS.403..797G} and in dwarf galaxies \citep[e.g.,][]{2010MNRAS.406.1152M, 2012MNRAS.423..406G}, it is necessary to improve those tests by applying them to more complex systems (i.e. systems with complex SFH) with available data of individual stars down to the oMSTO in the CMD.
 
We have performed such a test using one of the few fully resolved bright galaxies where this test can be carried out, the Large Magellanic Cloud (hereafter LMC). In particular, we study a region of the LMC bar as a ``guinea pig'' for which we can compare complex SFH's derived through integrated-light synthesis techniques and through the CMD of the resolved stellar population. We emphasize the fact that, in this paper, the SFH in the LMC bar region has been derived from the CMD without knowing the results from the integrated spectrum analysis and vice versa, making this a ``blind test'' in order to avoid bias toward a common solution by adjusting the fitting parameters.

Throughout this paper we will consider the SFH as composed by three main functions: one is the star formation rate as a function of time, SFR(t); the second is the chemical enrichment history (age metallicity relation, AMR); and the third is the initial mass function (IMF).

In this paper, we show the LMC bar SFH derived from full spectrum fitting techniques applied to its integrated spectrum and we compare it to the CMD results from {\it Hubble Space Telescope} (HST) data. We have focused our work on the  {\tt STECKMAP} results. This code gives more probability to smooth solutions \citep[by means of a penalised $\chi^2$ fitting algorithm, see][for further information]{2006MNRAS.365...46O, 2006MNRAS.365...74O}. The smoothness of the recovered SFH makes the comparison with the CMD results easier and more sensible. However, we will also compare the results obtained using other spectral fitting codes such as {\tt ULySS} and {\tt STARLIGHT}. In Sect. \ref{obs} we explain our target choice as well as describe the observations and data reduction procedure to obtain the composite CMD and the integrated spectrum. In Sects. \ref{SFH} and \ref{comp} we derive the SFH following both approaches and compare the similarities and differences found, with special emphasis on how the final solutions are effected by using different {\tt STECKMAP} input parameters. Results coming from other codes ({\tt ULySS} and {\tt STARLIGHT}) are shown and analyzed in Sect. \ref{others}. Some caveats, parallel analysis, implications, future work, and main conclusions are discussed in Sects. \ref{disc} and \ref{conclusions}. 

%__________________________________________________________________

\section{Observations and data reduction}
\label{obs}

\begin{figure*}
\centering
\includegraphics[scale=0.2]{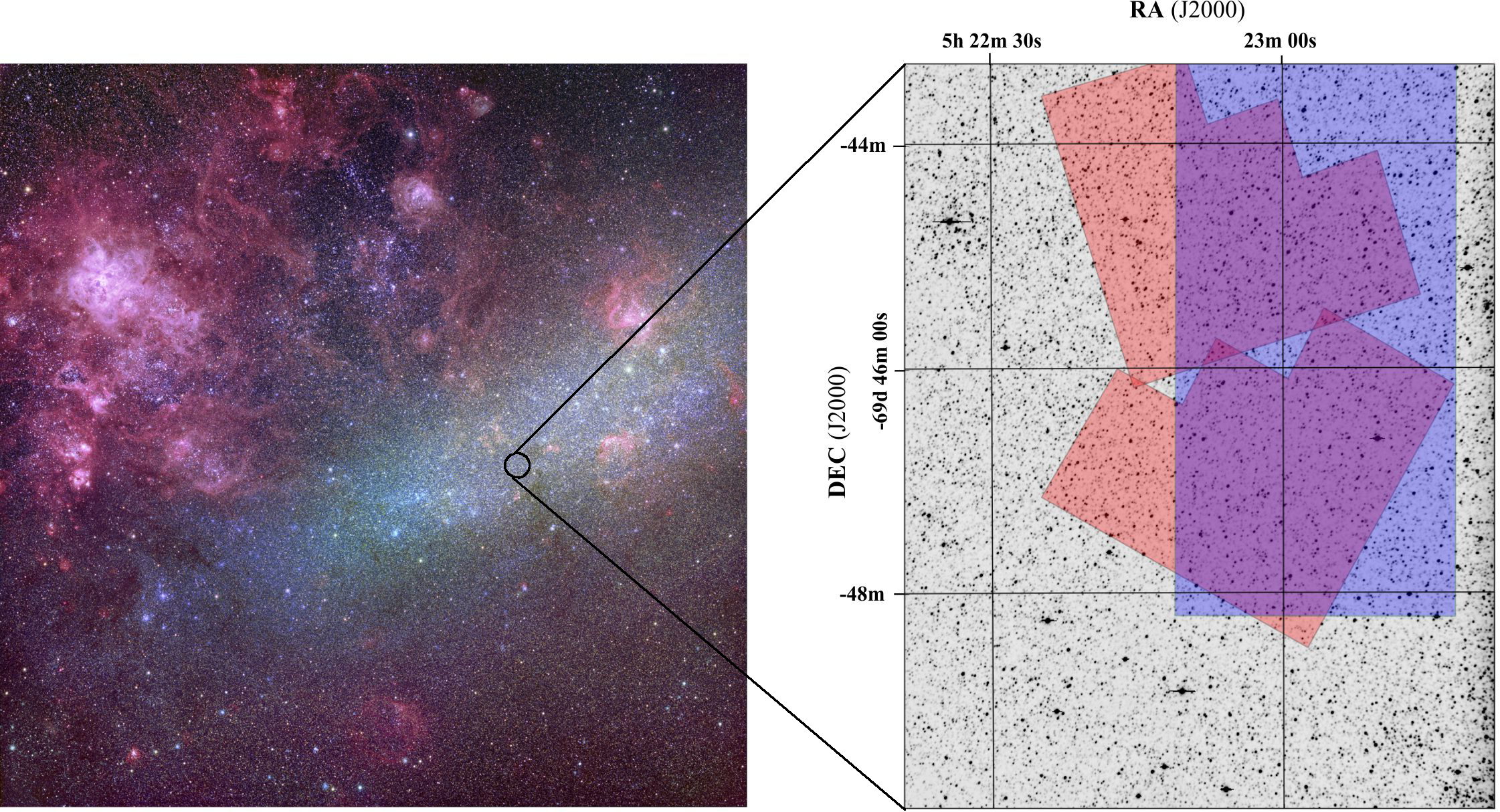}
\caption{Left panel: Image of the LMC bar and its surroundings (Credit: John Gleason). The location in the center of the bar of the field studied in this paper is indicated. Right panel: Positions of the two WFPC2 fields (red shaded areas), and the $2.5\arcmin\times5\arcmin$ area covered by sweeping the slit in the East-West direction (blue shaded area), superimposed on a VIMOS B-band image.}
\label{LMC_image}
\end{figure*}

To proceed with the comparison between integrated and resolved stellar population approaches in the case of a complex stellar population, we selected a field in the LMC bar (see Fig.\ref{LMC_image}). The centre of the LMC bar is bright enough as to obtain a high quality integrated-light spectrum with high S/N, which we did from observations at the 3.6\,m ESO telescope on La Silla using EFOSC2 \citep{2002Ap&SS.281..109A}. It is also sufficiently close and resolved for a CMD reaching the oMSTO to be secured with the HST. Such a CMD has been already published by  \citet{Smecker-Hane2002} and various SFHs have been derived by different groups, and published in \citet[][]{2002ASPC..274..535S}. We present a newly derived SFH with more sophisticated analysis techniques here.

When the spectroscopic observations were obtained (15 years ago) the only HST field available with sufficient surface brightness to carry out this project was the field observed with the WFPC2.  Unfortunately, the poor dynamic range of the camera limits our observations of the brightest stars. However, we can overcome this limitation with the methodology used to analyse the data (see  Sect.~\ref{SFH_CMD}).

\subsection{Resolved stellar populations}
\label{obs_CMD}

The photometry and artificial star tests necessary to reconstruct the SFH of the LMC bar region were obtained from the "HST Local Group Stellar Photometry Archive" \citep[][]{2006ApJS..166..534H}, maintained by J. Holtzmann \footnote{\url{http://astronomy.nmsu.edu/holtz/archival/html/lg.html}}. In particular, we downloaded the data of two WFPC2 pointings, namely the u4b112 and u4b115 fields, located at the center of the LMC bar (see Fig.\ref{LMC_image}) and originally observed within the GO program 7382 (P.I. T. Smecker-Hane). Table  \ref{tab:tab1} details the observing log, including the position of the two fields, the integration time in both the $F555W$ and $F814W$ filters, and the date of the observations.

\begin{figure}
\centering
\includegraphics[scale=0.45]{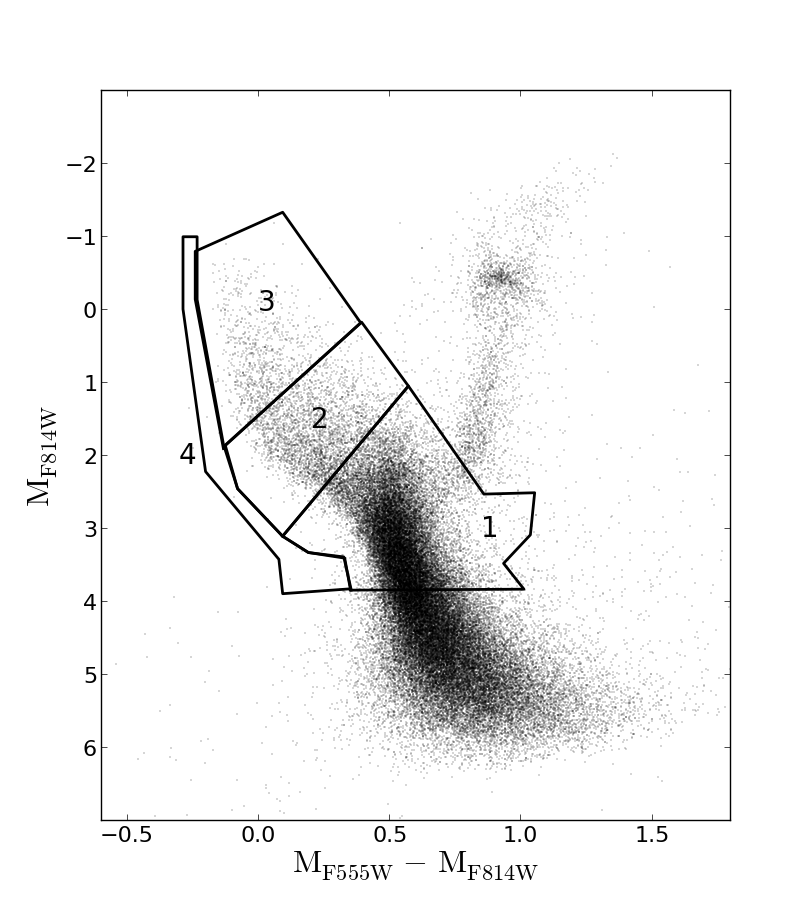}
\caption{$M_{F814W}$ vs $M_{F555W} - M_{F814W}$ CMD based on WFPC2 data. The four polygons show the regions used to derive the SFH using the IAC-star/MinnIAC/IAC-pop suite of routines (see text for details).}
\label{fig:cmd}
\end{figure}

The ($M_{F814W}$, $M_{F555W}-M_{F814W}$) CMD is presented in Fig.~\ref{fig:cmd}. The photometry reaches down to $M_{F814W} \sim $ 6, well below the oMSTO. A prominent bright main sequence is visible up to $M_{F814W} \sim $ -1, indicating that star formation continued until very recent epochs in this region. The red giant branch (RGB), comprising stars older than $\sim$ 1 Gyr, is also well populated up to $M_{F814W} \sim $~-2. Finally, we also highlight the presence of a prominent red clump of centrally He-burning stars ($M_{F814W} \sim $ -0.5), while the old horizontal branch is barely populated. We have marked four regions in the figure that will be discussed in Sect.~\ref{SFH_CMD}.

The lack of stars brighter than $M_{F814W}\sim$-1 and $M_{F814W}\sim$-2, along the main sequence and the RGB respectively, is likely due to saturation. Indeed, visual inspection of the brightest sources in the field confirms that a number of  bright stars ($\sim$ 200) are saturated in both filters. The region of the CMD affected by saturation will not be used in the SFH derivation.  At the metallicity of the youngest LMC stars, our saturation magnitude of $F814W \sim$ -1 on the main sequence corresponds to the turnoff of a $\sim 0.3$ Gyr stellar population. Because the bright massive stars younger than this age do not appear in the CMD,  this implies that we are not using all the possible information to derive the SFH for these ages. However, because lower mass stars of the same ages are present at fainter magnitudes, the SFH can be still obtained from them, under the assumption of a given IMF.

\begin{table*}[t]
\begin{center}
\begin{tabular}{l|cccccc}
\hline \hline
\noalign{\vspace{0.1 truecm}}
Field   & R.A.       &  Dec                        & Exp. Time -- $F555W$   & Exp. Time -- $F814W$       &   Date          \\
        & $h~m~ss.s$ &  $~~\deg~~\arcmin~~\arcsec$ & $s$                    & $s$                        &                 \\
\noalign{\vspace{0.1 truecm}}
\hline
\noalign{\vspace{0.1 truecm}}
u4b112  & 05:22:57  & --69:46:53                   & 4$\times$500           &  2$\times$300+2$\times$700 &  Nov 27, 1997   \\
u4b115  & 05:22:55  & --69:42:51                   & 4$\times$500           &  2$\times$300+2$\times$700 &  Jan 05, 1999   \\
\noalign{\vspace{0.1 truecm}}                               
\hline
\end{tabular}
\caption{Observing log of the photometric data.}
\label{tab:tab1}
\end{center}
\end{table*}

\subsection{Integrated spectrum}
\label{obs_int}

\begin{figure*}
\centering
\includegraphics[scale=0.45]{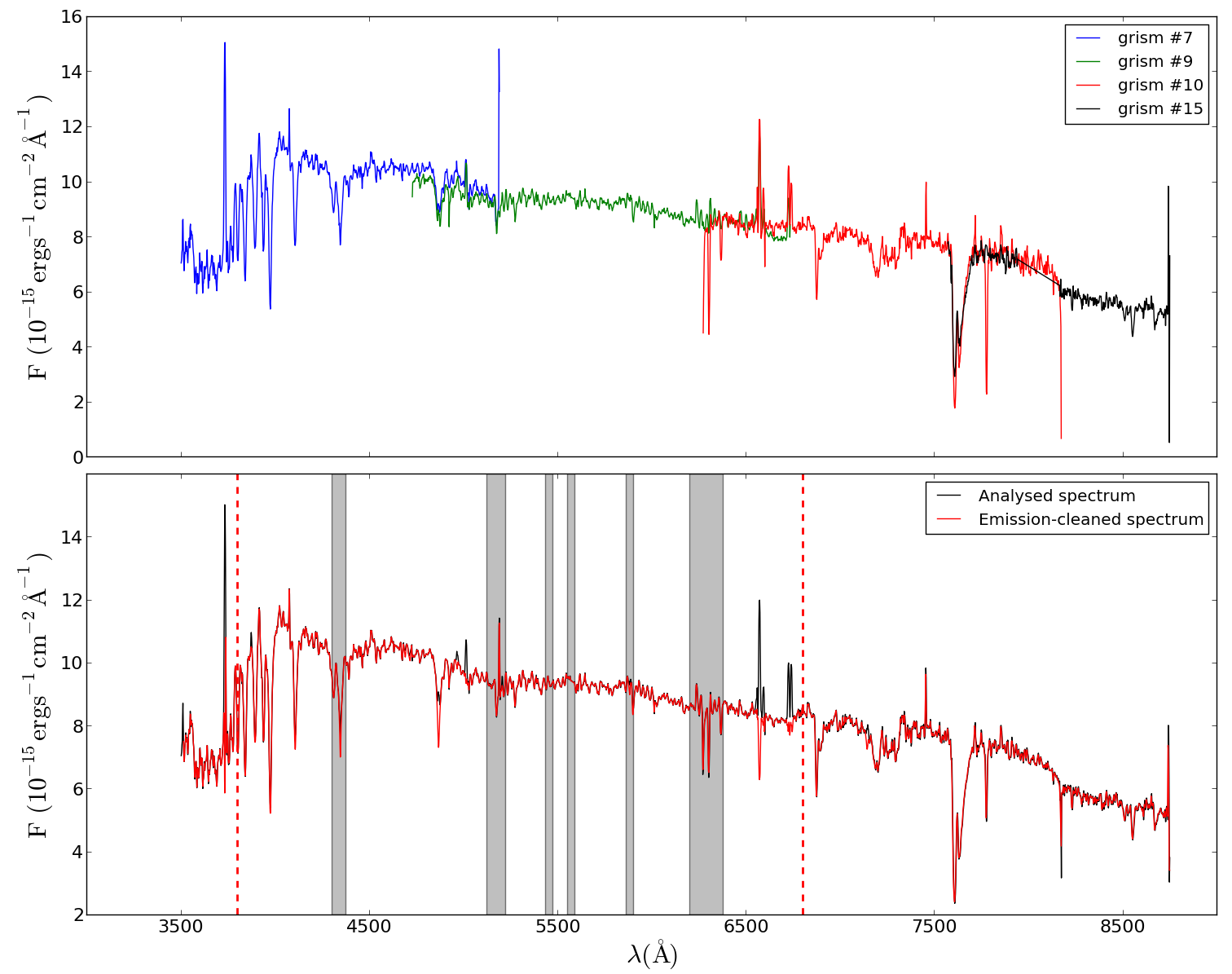}
\caption{The composite spectrum of the LMC bar field used in the integrated stellar population analysis. Top row: The spectrum is divided according to the four different grisms used (different colours). Blue for the grism \#\,7 ($3500$ $-$ $5185$ \AA), green for grism \#\,9 ($4725$ $-$ $6730$ \AA), red for grism \#\,10 ($6275$ $-$ $8160$ \AA), and black for grism \#\,15 ($7565$ $-$ $8745$). For each grism the corresponding sky spectrum, i.e. the spectrum taken in the sky field, has been subtracted. Only the LMC bar contribution is shown here. Bottom row: Fully-reduced composite spectrum of the LMC bar (black) and fully-reduced, emission-cleaned composite spectrum (red). For the stellar content analysis we use the wavelength range from 3800 to 6800 \AA $ $ (see text for details, dashed vertical red lines). The shaded regions of the spectrum are the masked regions (not considered in the fit).}
\label{LMC_spec}
\end{figure*}

Integrated-light spectra of the LMC bar and a sky field were collected at the 3.6\,m ESO telescope on La Silla, using EFOSC \citep[][]{1984Msngr..38....9B}, on 18-20 December 2000. The J2000 coordinates of the observed target field are: $\alpha = 05:23:17$ and $\delta = -69:45:42$. We used a North-South 5$'$ long and 1.5$"$ wide slit, which was swept along the East-West direction: this allowed us to cover a full area of 2.5$'$ by 5$'$, which approximately matches the WFPC2 pointings (see Figure 1). Four different grisms were used, spanning in total the wavelength range 3500 to 8745\AA $ $ (see Fig. \ref{LMC_spec}). We smoothed the spectra from the four grisms to a common final dispersion of 1.9$\pm$0.1 $\AA$/pix (see Fig. \ref{LMC_spec}). A mean exposure time of 5400\,s, divided in three separate exposures, was adopted for each grism, both for the target-field and also for the sky-field. The sky-field spectrum was taken 6$^{\circ}$ North of the target-field position. The sky position was chosen as a compromise between the sky-field being relatively close and thus representative of the foreground Milky Way (MW) contamination and sky light present in the LMC bar field, and to mimimize the contribution from LMC light. We note, however, that with a galactocentric radius of at least  16$^{\circ}$ \citep{2009IAUS..256...51M, 2010AJ....140.1719S}, even at 6$^{\circ}$ from its center, the LMC still presents a well populated CMD, with many intermediate-age stars \citep{2008ApJ...682L..89G, 2010AJ....140.1719S}. 

The 2D-spectra corresponding to the 4 grisms for both the target and the sky fields  were reduced through standard techniques, using {\tt MIDAS} and {\tt IRAF} \footnote{{\tt IRAF} is distributed by the National Optical Astronomy Observatory, which is operated by the Association of Universities for Research in Astronomy (AURA) under cooperative agreement with the National Science Foundation.} packages. A full two-dimensional wavelength calibration was built up to correct for geometrical distortions. The spectra were flux calibrated using the spectrophotometric standard EG 21 and LTT 4816; the error in flux calibration is estimated to be around 10$\%$. After this step, the 1D-spectra of the 4 grisms were matched together and the quality of the match controlled through the overlapping wavelength regions.

To check whether small number statistics in the sampling of minority stellar populations in the target field (such as the insertion of a few young stars dominating the final spectrum but with little mass contribution) could lead us to important fluctuations in the SFH, we performed the following test. From the integrated-light, spatially resolved 2D-spectra, we extracted two series of integrated-light 1D-spectra, one for the target-field and one for the sky-field, with extraction windows of 5$'$ and of 2.5$'$.   We ended up with four 1D-spectra over the 3500--8745\AA $ $ range (FWHM $\sim$ 10 \AA), two of them corresponding to the integrated light in the LMC bar field over spatial areas of respectively $2.5'\times5'$ (field 1) and $2.5'\times2.5'$ (field 2, included in field 1) and the other two corresponding to the integrated light in the sky-field with the same extraction windows. At face value, the sky-subtracted spectra of the target-field through the two different extraction windows exhibit very similar characteristics. In terms of the shape of their spectra, we scaled the field 2 spectrum to match the higher flux level of field 1 spectrum. The average difference of both spectra relative to field 1 spectrum, over the whole spectral range, is 1.6\% (i.e. both scaled spectra are almost identical).  We then derived the SFH of both spectra (field 1 spectrum and the scaled, field 2 spectrum) using ({\tt STECKMAP}, see Sect. \ref{SFH_int}). The residuals between both recovered SFR(t) are within our error values (see Sect. \ref{SFH_int} for further information). This analysis indicates that we are properly sampling the stellar content in the analyzed fields, in spite of the very small fraction of the galaxy's light present in them. Therefore, we decided to analyze in detail the composite stellar population corresponding to the spectrum covering the $2.5'\times5'$ area (40\,pc$\times$80\,pc), which has a better S/N ratio as a consequence of the larger spatial coverage of an area with intrinsically the same stellar content. The corresponding final, sky-subtracted spectrum, is shown in Fig. \ref{LMC_spec}. The final signal-to-noise ratio of this spectrum is $\sim$ 36.3 (per \AA).

\section{Determination of the star formation history}
\label{SFH}

We have obtained the SFH using the data described in Sects.~\ref{obs_CMD} and \ref{obs_int}, following well developed and tested methodologies to study the stellar content via CMD \citep{iacstar,monelli10b} and spectroscopic analysis \citep[][Ruiz-Lara in prep.]{2011MNRAS.415..709S, 2014A&A...570A...6S, 2015MNRAS.446.2837S}. This study is meant to be a blind test between both approaches. Thus, all the results obtained and described in this section have been analyzed and studied independently (without knowing the results from the other approach) avoiding any possible bias to a common solution.

\subsection{CMD analysis}
\label{SFH_CMD}

The SFH from the CMD was obtained using the IAC-star/MinnIAC/IAC-pop \citep{iacstar,iacpop,hidalgo11,monelli10b} suite of routines. This SFH derivation is based on the comparison, through a $\chi^2$ minimization, of the distribution of stars in the observed and in a model CMD. The model CMD was obtained from a synthetic CMD computed with IAC-star, after simulation of the observational errors. For the calculation of the synthetic CMD, which contains 9$\times$10$^6$ stars, we assumed a constant star formation rate at all ages within 0 and 13.5 Gyr, and flat metallicity distribution between Z = 0.0002 and 0.02. For consistency with the analysis of the integrated spectrum, we used the Padova stellar evolution library \citep{girardi00, marigo08}. Finally, we assumed a Kroupa initial mass function \citep{2001MNRAS.322..231K}, and a binary fraction of 40\% \citep{monelli10b}. The observational errors were simulated using the information from the artificial stars tests. These allow for the inclusion of not only the completeness of the photometry, but also, and more importantly, how the photometric errors displace a star from its original position in the CMD. The model CMD was divided into simple stellar populations by adopting a basic set of age and metallicity bins, whose limits are:  age\,(Gyr)\,=\,[0, 0.25,  0.5,  1.0,  1.5 to 13.5 in steps of 1 Gyr] and  Z\,=\,[0.0002, 0.0006, 0.001 to 0.01 in steps of 0.001, 0.015, 0.02]. We adopted a distance modulus (m-M)$_0$ = 18.5 mag and reddening E(B-V)\,=\,0.1 mag to shift the observed CMD to the absolute magnitude plane.

Following previous investigations of other dwarf galaxies \citep{hidalgo09, monelli10b, monelli10c, hidalgo11, meschin14}, the SFH was derived using only the main sequence (MS) and sub-giant branch (SGB) regions, avoiding the regions where the completeness is less than $\sim$50\%. Figure \ref{fig:cmd} presents the CMD with the regions adopted for SFH derivation (``{\it bundles\/}'') highlighted. The star counts are performed in each bundle by dividing it into boxes of different sizes. The finest sampling was used in bundle 1 (0.01 mag and 0.2 mag for the colour and magnitude, respectively) and bundle 2 (0.02, 0.1). Bundle 3 was sampled in (0.2,0.5) boxes and bundle 4 was used as a single box. This strategy is an optimal one taking into account that i) the physics of main sequence stars is best understood, and ii) the largest number of stars in the CMD are located in the lower main sequence. Of the 69\,209 observed stars, 27\,556 are counted in the four bundles.

To take into account possible errors in the distance, reddening, and photometric or model calibration, a number of different solutions were derived introducing small shifts in colour and magnitude, in a grid of 25 positions within $\pm$0.06 mag and $\pm$0.15 mag in colour and magnitude, respectively. In order to minimize the effect of the sampling choice, in each position of the grid 24 solutions were calculated varying the assumed age and metallicity bins of the simple stellar populations, and moving the boxes inside the bundles. In each position of the grid, the 24 solutions are averaged and the corresponding $\overline{\chi^2}$ is calculated as the average of the $\chi^2$ of the solutions. The minimum value of $\overline{\chi^2}$ ($\overline{\chi^2_{min}}$) indicates the position in the grid where the best solution is obtained, which turned out to be $\Delta[(M_{F555W}-M_{F814W}), \Delta(M_{F814W})]=[0.03,0]$.  

Figure \ref{SFH_steck_3} shows the SFH derived from this analysis. This figure (blue colours) displays the SFR\,(t) and Z\,(t) projections of the SFH, as the average of the 24 individual solutions calculated at the position of the grid where $\overline{\chi^2_{min}}$ was obtained \citep[][]{hidalgo11}. We should note here that the averaging of the 24 solutions has the effect of smoothing the final SFH. Error intervals are calculated from the dispersion of these 24 solutions, {\it plus} any solution in the grid that differs in  $\chi^{2}$ by less than 1$\sigma$. This provides errors equal or in excess to the so-called {\it several solutions} criterion which was shown by \citet{iacpop} to produce reliable estimates of total internal errors. Figure~\ref{SFH_steck_3} shows that the SFR\,(t) is relatively smooth over the whole time interval, with a slight change at around 4.5~Gyr ago an increased SFR\,(t) at later times. This is reminiscent of the period of low star formation activity followed by a later increase, found by other studies in LMC regions located at larger galactocentric distances \citep[e.g.,][ and references therein]{meschin14}. The intermediate-age SFR\,(t) in the LMC bar, however, seems to differ substantially from that of the disc at different galactocentric distances, in which the two main periods of star formation, separated by an epoch of lower star formation activity, are barely seen in the bar and replaced by a much flatter SFR\,(t). The SFR\,(t) between 3.5 and 0.25\,Gyr obtained here is rather flat followed by a decrease for stars younger than 0.25\,Gyr. The AMR is similar to the one reported in previous studies \citep[e.g.,][]{1991AJ....101..515O, 2000A&A...360..133D, 2006AJ....132.1630G}, showing an exponential decline from $Z$ $\sim$ $0.0175$ at young ages to $Z$ $\sim$ $0.0004$ at old ages. The metallicity dispersion in the AMR is found to increase toward younger ages.

\subsection{Integrated spectrum analysis}
\label{SFH_int}

We used the integrated spectrum of the LMC bar (see Sect. \ref{obs_int}) to obtain its SFR\,(t) and AMR employing state-of-the-art techniques taking advantage of the large spectral coverage of our data along with some of the most detailed stellar libraries and models.

\subsubsection{Emission line cleaning ({\tt GANDALF})}
\label{emi_lin_cle}

When studying integrated stellar populations it is important to take into account any contributions from ionised gas (see Fig. \ref{LMC_spec}). To obtain a reliable SFH we must remove such a contribution, in order to be able to use also those regions affected by emission ($H\beta$, [\oiii] $\lambda$ $5007$, $H\alpha$, etc.) in the stellar population analysis. To this end we used {\tt GANDALF} \citep[Gas AND Absorption Line Fitting, ][]{2006MNRAS.366.1151S} for the emission line removal. In this step we made use of the \citet{2010MNRAS.404.1639V} models based on the {\tt MILES} library \footnote{The models are publicly available at \url{http://miles.iac.es}} \citep[][]{2006MNRAS.371..703S, 2007MNRAS.374..664C} and computed using the scaled-solar isochrones of \citet[][]{girardi00}. The models are generated following a Kroupa Universal IMF \citep[][]{2001MNRAS.322..231K}. {\tt GANDALF} simultaneously fits the absorption and emission features present in the studied spectrum. It previously runs {\tt pPXF} \citep[Penalized Pixel Fitting,][]{2004PASP..116..138C} to obtain the stellar kinematics and a combination of spectral templates matching the observed spectrum. The stellar kinematics obtained here will be used later on (see Sect.~\ref{cons_tests}). Once the absorption spectrum is taken into account, additional gaussians are included in the fit to obtain the kinematics, shapes, and fluxes of the different emission lines. 

\subsubsection{Star formation history ({\tt STECKMAP})}
\label{steckmap_section}

The emission-cleaned integrated stellar spectrum was then analysed using full spectrum fitting techniques by means of {\tt STECKMAP} \footnote{{\tt STECKMAP} can be downloaded at \url{http://astro.u-strasbg.fr/~ocvirk/}} \citep[STEllar Content and Kinematics via MAximum a Posteriori,][]{2006MNRAS.365...46O, 2006MNRAS.365...74O}. The spectrum is fitted using a Bayesian minimisation method (by means of a penalised $\chi^2$) to obtain the stellar combination that matches the observed spectrum via a maximum a {\it posteriori} algorithm. The equation to minimise is:

\begin{equation}
  Q_{\mu} = {\chi^2(s(x,Z,g)) + P_{\mu}(x,Z,g)},
\label{Q_def}
\end{equation}

where {\it s} is the modelled spectrum which depends on the age distribution ({\it x}), the AMR ({\it Z}), and the broadening function accounting for the kinematics ({\it g}). {\tt STECKMAP} has the advantage of being a non-parametric code, i.e. it does not assume an {\it a priori} shape for the solution (such as stellar age distribution, AMR, and the line-of-sight velocity distribution), but gives more probability to smoother solutions. This smoothness is accomplished through the penalisation function (function P$_\mu$ in equation \ref{Q_def}) and the function P (a function that gives high values to high oscillating solutions while small values to smooth solutions of {\it x}, {\it Z}, or {\it g}). The penalisation function P$_\mu$ is defined as:

\begin{equation}
  P_{\mu}(x,Z,g) = {\mu_{x} P(x) + \mu_{Z} P(Z) + \mu_{v} P(g)}.
\label{P_def}
\end{equation}

The parameters $\mu_x$, $\mu_Z$, and $\mu_v$ are the smoothing parameters for the Stellar Age Distribution (SAD), the AMR, and the line-of-sight velocity distribution (LOSVD), respectively. The different values that these parameters can adopt allow us to change the smoothness of the accepted solutions. The function P also can adopt different shapes \citep[for further information see][]{2006MNRAS.365...46O, 2006MNRAS.365...74O}. This code accounts for the continuum shape of the spectrum using a polynomial fitting (avoiding sources of error such as flux calibration or extinction errors). 

{\tt STECKMAP} outputs allowed us to reconstruct the SFR(t) and the AMR that best fit the integrated spectrum (see Fig.~\ref{SFH_steck_3}). Errors were computed by means of a series of 25 Monte Carlo simulations. Once {\tt STECKMAP} has determined the best combination of stellar populations to fit the observed spectrum ({\it ``best model''}), we added noise to the {\it ``best model''} spectrum and ran {\tt STECKMAP} again. This procedure was repeated 25 times. We compute the error of the stellar mass fraction at each age and metallicity bin as the standard deviation of the 25 resulting values.    

\subsubsection{Robustness of the {\tt STECKMAP} results}
\label{cons_tests}   

There are several input parameters to consider while running {\tt STECKMAP} that might affect the SFH reconstruction (e.g. spectral templates, S/N, the smoothing parameters, etc.). In this section we check the effect of the various input parameters as well as probe the robustness of the {\tt STECKMAP} results. 

\begin{table*}
\centering
\begin{tabular}{lcccllcc}
\hline\hline
Test & Models & Age range (yr) & Kinematics & $\mu_x$ & $\mu_Z$ & $\mu_v$ & $rms$ \\  \hline
1    & \citet[][]{2010MNRAS.404.1639V}             & [0.063$\times$10${^9}$, 17.8$\times$10$^{9}$] & FIX & 10$^{-2}$       & 10$^{2}$        & - (**)   &  0.1215   \\
2   & \citet[][]{2010MNRAS.404.1639V} +   & [0.001$\times$10${^9}$, 17.8$\times$10$^{9}$] & FIX & 10$^{-2}$       & 10$^{2}$        & -    &  0.1298  \\
   & \citet{2005MNRAS.357..945G}   &  &  &        &         &         \\
3   & \citet[][]{2010MNRAS.404.1639V}             & [0.063$\times$10${^9}$, 13.5$\times$10$^{9}$] & FIX & 10$^{-2}$       & 10$^{2}$        & -    &  0.1216  \\
4   & \citet[][]{2010MNRAS.404.1639V} +   & [0.001$\times$10${^9}$, 13.5$\times$10$^{9}$] & FIX & 10$^{-2}$       & 10$^{2}$        & -    &  0.1298  \\
   &  \citet{2005MNRAS.357..945G}   &  &  &       &       &        \\
5   & \citet[][]{2010MNRAS.404.1639V}             & [0.063$\times$10${^9}$, 17.8$\times$10$^{9}$] & FIT & 10$^{-2}$       & 10$^{2}$        & 10$^{-2}$    &  0.1670  \\
6   & \citet[][]{2010MNRAS.404.1639V} +   & [0.001$\times$10${^9}$, 17.8$\times$10$^{9}$] & FIT & 10$^{-2}$       & 10$^{2}$        & 10$^{-2}$   &   0.1764  \\
   & \citet{2005MNRAS.357..945G}   &  & &       &        &        \\
7   & \citet[][]{2010MNRAS.404.1639V}             & [0.063$\times$10${^9}$, 13.5$\times$10$^{9}$] & FIT & 10$^{-2}$       & 10$^{2}$        & 10$^{-2}$   &   0.1768  \\
8   & \citet[][]{2010MNRAS.404.1639V} +   & [0.001$\times$10${^9}$, 13.5$\times$10$^{9}$] & FIT & 10$^{-2}$       & 10$^{2}$        & 10$^{-2}$   &  0.1766   \\
   &  \citet{2005MNRAS.357..945G}   &  &  &        &        &        \\
9   & \citet[][]{2010MNRAS.404.1639V}             & [0.063$\times$10${^9}$, 17.8$\times$10$^{9}$] & FIX & 10$^{2}$        & 10$^{2}$        & -           &    0.1218   \\
10  & \citet[][]{2010MNRAS.404.1639V}             & [0.063$\times$10${^9}$, 17.8$\times$10$^{9}$] & FIX & 10$^{-2}$       & 10$^{-2}$       & -          &    0.1217   \\
11  & \citet[][]{2010MNRAS.404.1639V}             & [0.063$\times$10${^9}$, 17.8$\times$10$^{9}$] & FIX & 10$^{-6}$   & 10$^5$     & -           &    0.1224   \\
12  & \citet[][]{2010MNRAS.404.1639V}             & [0.063$\times$10${^9}$, 17.8$\times$10$^{9}$] & FIX & 10$^{-8}$ & 10$^{-8}$ & -    &    0.1217   \\
13  & \citet[][]{2010MNRAS.404.1639V}             & [0.063$\times$10${^9}$, 17.8$\times$10$^{9}$] & FIX & 10$^4$      & 10$^5$     & -       &    0.1223 \\
14  & \citet[][]{2010MNRAS.404.1639V}             & [0.063$\times$10${^9}$, 17.8$\times$10$^{9}$] & FIX & 10$^{15}$       & 10$^{14}$       & -          &    0.1224   \\
15  & \citet[][]{2010MNRAS.404.1639V}             & [0.063$\times$10${^9}$, 17.8$\times$10$^{9}$] & FIX & 10$^{-15}$      & 10$^{-14}$      & -         &    0.1217   \\
16  & \citet[][]{2010MNRAS.404.1639V}             & [0.063$\times$10${^9}$, 17.8$\times$10$^{9}$] & FIX & 10$^{-2}$       & 10$^{6}$    & -          &    0.1227   \\ 
17  & \citet[][]{2010MNRAS.404.1639V}             & [0.063$\times$10${^9}$, 13.5$\times$10$^{9}$] & FIX & 10$^{2}$        & 10$^{2}$        & -           &    0.1218   \\
18  & \citet[][]{2010MNRAS.404.1639V}             & [0.063$\times$10${^9}$, 13.5$\times$10$^{9}$] & FIX & 10$^{-2}$       & 10$^{-2}$       & -          &    0.1218   \\
19  & \citet[][]{2010MNRAS.404.1639V}             & [0.063$\times$10${^9}$, 13.5$\times$10$^{9}$] & FIX & 10$^{-6}$   & 10$^5$     & -           &    0.1226   \\
20  & \citet[][]{2010MNRAS.404.1639V}             & [0.063$\times$10${^9}$, 13.5$\times$10$^{9}$] & FIX & 10$^{-8}$ & 10$^{-8}$ & -    &    0.1218   \\
21  & \citet[][]{2010MNRAS.404.1639V}             & [0.063$\times$10${^9}$, 13.5$\times$10$^{9}$] & FIX & 10$^4$      & 10$^5$     & -       &    0.1224   \\
22  & \citet[][]{2010MNRAS.404.1639V}             & [0.063$\times$10${^9}$, 13.5$\times$10$^{9}$] & FIX & 10$^{15}$       & 10$^{14}$       & -          &    0.1249   \\
23  & \citet[][]{2010MNRAS.404.1639V}             & [0.063$\times$10${^9}$, 13.5$\times$10$^{9}$] & FIX & 10$^{-15}$      & 10$^{-14}$      & -         &    0.1218  \\
24 (*) & \citet[][]{2010MNRAS.404.1639V}             & [0.063$\times$10${^9}$, 13.5$\times$10$^{9}$] & FIX & 10$^{-2}$       & 10$^{6}$    & -          &    0.1225   \\    \hline
\end{tabular}
\caption{Set of characteristics and parameters we have used in the different tests to eximane the robustness of the {\tt STECKMAP} solutions. First column is the number of the test. Second column shows the models used in each test, i.e. \citet[][]{2010MNRAS.404.1639V} or \citet[][]{2010MNRAS.404.1639V}+\citet{2005MNRAS.357..945G}. Third column represents the age range. In the fourth column we highlight the fact of fitting or fixing the kinematics. $\mu_x$ stands for 'smoothing parameter for the stellar age distribution', $\mu_Z$ stands for 'smoothing parameter for the age-Z relation', and $\mu_v$ stands for 'smoothing parameter for the line-of-sight velocity distribution'. The last column represents the quality of each of the tests by means of its residuals $rms$ computed as the mean values of the absolute differences between the data and the fit. (*) Test that we have chosen to compare with the CMD results (see text for details). (**) As we are fixing the stellar kinematics, and thus, not fitting it, this parameter is not applicable.} 
\label{steck_tests}
\end{table*}

We have explored the entire input parameter space in 24 different tests to choose the best combination of parameters for the {\tt STECKMAP} run (see Sect.~\ref{SFH_int}). All these tests can be interpreted as a way of testing the robustness and consistency of the {\tt STECKMAP} results. The key parameters that can affect the final results are: i) the set of spectral templates (models and age range); ii) the smoothing parameters; iii) whether the stellar kinematics are fixed or fitted: although we prefer to fix the stellar kinematics in order to stabilise the solution and minimise well reported degeneracies \citep[][]{2011MNRAS.415..709S}, we will check how the processes of fitting simultaneously the stellar content and kinematics affects the final  results.

Table~\ref{steck_tests} summarizes the main input parameters for each test and the corresponding rms. In the tests we explored the whole parameter space divided into three main blocks: i) we checked the effect of using different stellar models, the age range used during the fit, and the simultaneous recovery of the stellar kinematics (tests 1 to 8); ii) based on the previous block and our expertise \citep[][Ruiz-Lara in prep.]{2011MNRAS.415..709S, 2014A&A...570A...6S, 2015MNRAS.446.2837S} we used the \citet[][]{2010MNRAS.404.1639V} models (age-range : 0.063$\times$10${^9}$ to 17.8$\times$10$^{9}$ yr) and fixed the stellar kinematics (tests 9 to 16) while exploring the smoothing parameters ($\mu_x$ and $\mu_Z$) with values ranging from 10$^{-15}$ to 10$^{15}$; and finally iii), in the third block (tests 17 to 24) we used the same ingredients as in the second block but limited the age range to 13.5$\times$10$^{9}$ yr (to match the CMD analysis age range). In all the tests we use a square Laplacian smoothing kernel for the shape of the penalisation function, P. 

A visual inspection of the fits, and the quantitative value of their $rms$ ($\sim$ 0.12 with little dispersion, see Table~\ref{steck_tests}) for all the tests, led us to conclude that there is no easy way to choose the combination of input parameters that best suits our data. In terms of the reconstructed SFR(t) and AMR shapes, all of the tests show very similar results (compatible within errors), except for tests 5, 6, 7, and 8 (see appendix~\ref{plots_tests}) that display the largest discrepancies (also in terms of their $rms$). These tests share the property that we fit simultaneously the stellar kinematics, which considerably hampers the correct SFH reconstruction. This behaviour highlights the overall stability of the {\tt STECKMAP} solutions, since as the general shape (if not the precise details) of the solutions is quite similar among tests. 

As there is no clear set of input parameters to favour over the others, we have decided to stick to the following reasonable set of input parameters: i) We fixed the stellar kinematics to the values found with {\tt pPXF} following \citet[][]{2011MNRAS.415..709S}. ii) We matched the age range used in the CMD analysis using SSP model templates with ages from 0.063$\times$10${^9}$ to 13.5$\times$10$^{9}$ yr. iii) We focused on the results using {\tt MILES}. In Figure~\ref{comp_models} we show a comparison between the commonly used {\tt MILES} models \citep[][]{2010MNRAS.404.1639V}, {\tt MILES} with the extension towards younger stellar populations from \citet[][]{2005MNRAS.357..945G}, and the \citet[][]{2003MNRAS.344.1000B} (BC03) models. We find small differences between models regarding the SFR(t). Larger differences are found in the recovered AMR, with discrepancies of the order of the errors. In particular, the inclusion of the \citet[][]{2005MNRAS.357..945G} models (based on theoretical stellar libraries) with ages younger than 63 Myr might affect the SFH recovery depending on the real amount of young stars present in the observed field. If we do not include those models, in principle, the contribution of such stars would be included in older bins modifying the recovered SFH shape. Further investigation is  needed to understand whether the impacts on the solutions do (or do not) make sense.

\begin{figure}
\centering
\includegraphics[scale=0.37]{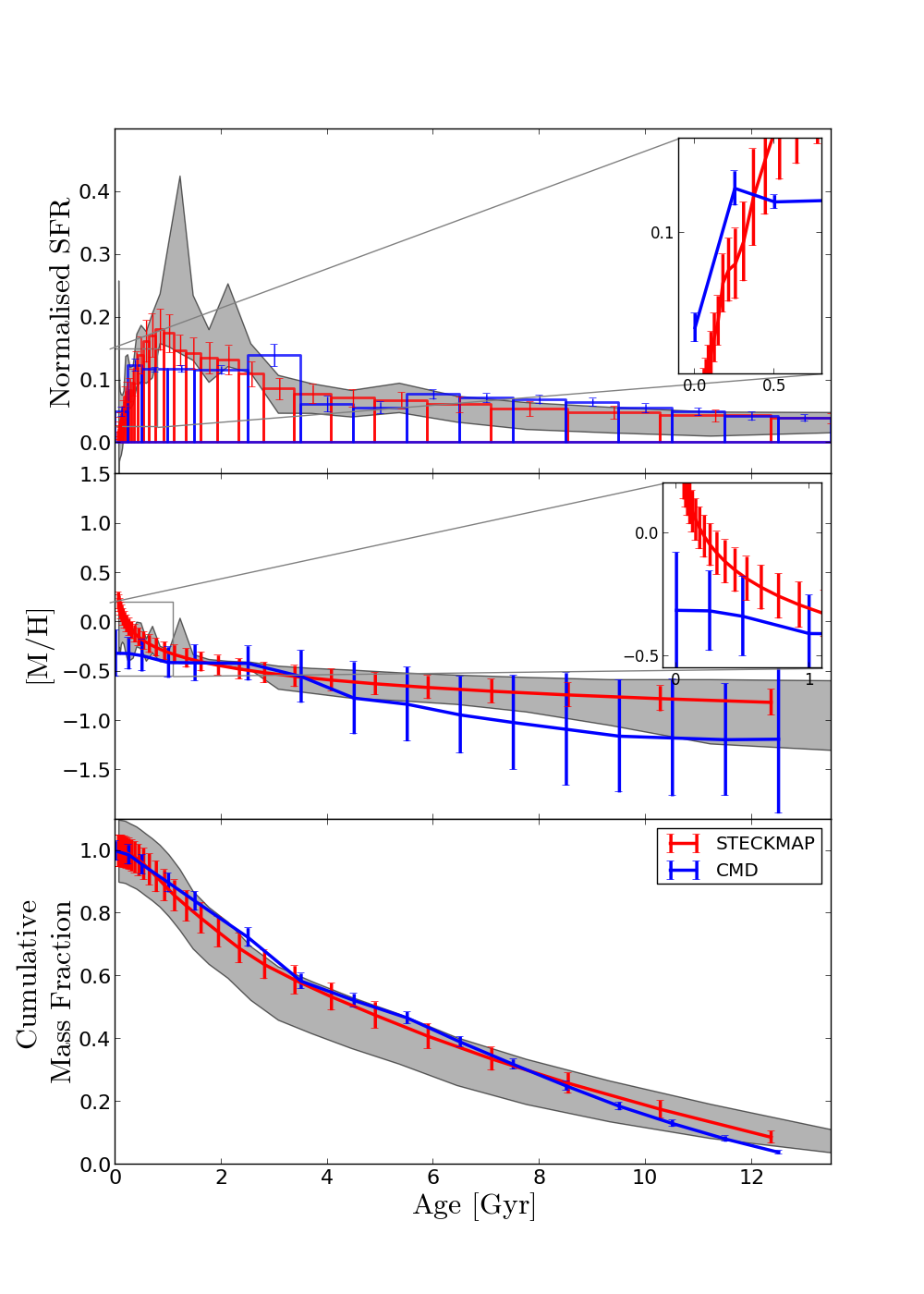} \\
\caption{Comparison between the LMC bar SFH from the CMD and the integrated spectrum using {\tt STECKMAP} (test 24). The three panels show, from top to bottom, the normalised SFR(t), the age–metallicity relation, and the cumulative mass fraction; with a zoom at young ages for the normalised SFR(t) and the AMR. We plot the envelope of the histograms in the main plots in the insets. In order to make a fair comparison a normalisation is needed. We normalise the SFR\,(t) in such a way that the sum of the areas of the different rectangles (SFR $[M_{\odot}/yr]$ $\times $ $\Delta t$) is 1 (upper panel). The AMR plot represents the average metallicity at every age bin. Error bars are 1$\sigma$ of the resulting distribution of solutions from a series of 25 Monte Carlo simulations in the case of STECKMAP and 600 different solutions by applying small shifts in the CMD for the CMD analysis (see text for details). The shaded regions correspond to the mean values and standard deviations of all the solutions using {\tt MILES} and ages ranging from 0.063 to 13.5 Gyr (tests 3, 7, 17, 18, 19, 20, 21, 22, 23, and 24).}
\label{SFH_steck_3}
\end{figure}

\begin{figure}
\centering
\includegraphics[scale=0.38]{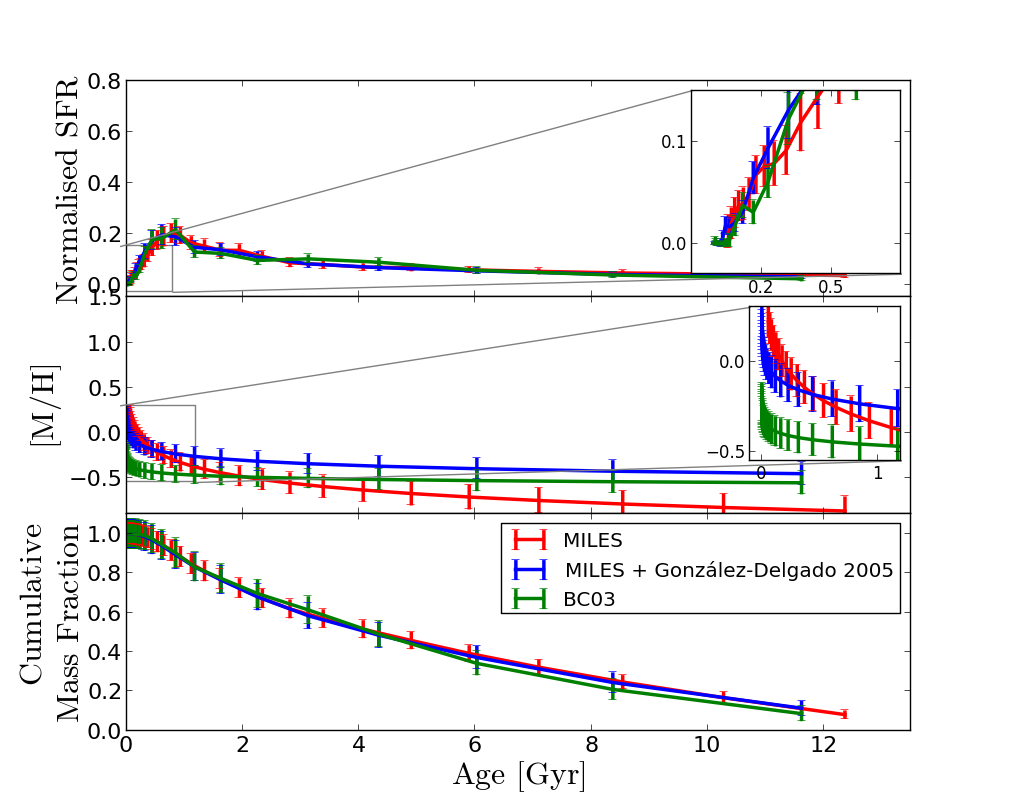} 
\caption{Comparison of the different SFHs from {\tt STECKMAP}, paying special attention to the choice of models (\citet[][]{2010MNRAS.404.1639V}, \citet[][]{2010MNRAS.404.1639V} + \citet{2005MNRAS.357..945G}, or \citet[][]{2003MNRAS.344.1000B}). A zoom at the younger ages for the SFR\,(t) and the AMR is applied. The rest of {\tt STECKMAP} input parameters as in test 24 (see table~\ref{steck_tests}). See caption of Fig. \ref{SFH_steck_3} for a complete explanation.}
\label{comp_models}
\end{figure}

Bearing this in mind, we show test~24 as an example of the recovered SFH. We have used this test just for illustration purposes, but most of the considered tests could have been used without modifying main conclusions. Inspecting Fig.~\ref{SFH_steck_3} (red lines) we can see that the LMC bar displays an almost constant SFR\,(t) since its formation until $\sim$ 4~Gyr ago according to the analysis of the integrated spectrum. Afterwards, a progressive increase in the SFR\,(t) is found with a peak at $\sim$ 1~Gyr ago followed by a drop in the SFR\,(t) to the present day. The AMR exhibits an exponential increase with old stellar populations showing the lowest metallicity ($[M/H]$ $\sim -0.6$) and young stars the highest metallicities ($[M/H]$ $\sim$ 0.2). We must highlight that this exponential behaviour in the AMR shape is mainly caused by our choice of smoothing parameters ($\mu_Z$ is 10$^{6}$ for test 24). The only difference between tests 17 to 24 is the choice of values of the smoothing parameters. As expected (see appendix~\ref{plots_tests}), high values of $\mu_x$ give smooth SFR(t) shapes (tests 17, 21, and 22) and high values of $\mu_Z$ gives smooth AMR shapes, (tests 17, 19, 21, 22, and 24) while low values give highly oscillating results. In order to easily compare with the rest of solutions we have plotted a shaded region corresponding to the average of all the solutions using {\tt MILES} models and ages ranging from 0.063 to 13.5 Gyr. As can be seen, our chosen test is within that shaded region showing again the robustness of the {\tt STECKMAP} results.

\section{Comparison between CMD and integrated spectrum analysis}
\label{comp}

In Sect.~\ref{SFH} we derived the SFHs in a region of the LMC bar using two different techniques.
In one technique we used the resolved stars in a deep CMD and compared them to theoretical CMDs based on theoretical 
isochrones, in the other we applied full spectral fitting techniques using stellar population synthesis models.
These analyses were performed independently in a blind test. The comparison between the SFH from both 
approaches will then allow us to test how well modern spectral fitting techniques recover the characteristics of a 
complex stellar population. 

\subsection{Star formation rate, SFR\,(t)}
\label{comp_SFH}

The upper panel of Fig.~\ref{SFH_steck_3} shows the SFR\,(t) in the LMC bar region obtained using CMD and {\tt STECKMAP} typical analyses, as discussed in Sect.~\ref{SFH}. The overall shape of SFR\,(t) is qualitatively similar in both derivations, with an almost continuous star formation since the earliest epochs till 4 Gyr ago, when small differences between the SFH obtained from both approaches appear. The CMD analysis reveals a rather flat SFR\,(t) between $\sim$\,0.25 and $\sim$\,3.5 Gyr with a clear drop in the star formation during the last 0.25 Gyr. The {\tt STECKMAP} analysis shows a SFR\,(t) that can be described as a gaussian skewed towards older ages with the peak around 1 Gyr. Although the youngest populations are better sampled in the case of the integrated analysis than in the case of the resolved analysis, the decline of SFR\,(t) at ages younger than 0.25 Gyr is consistent in both cases. The lower panel of Fig.~\ref{SFH_steck_3} shows the cumulative mass fraction, as a function of time, obtained following both approaches. Small differences can be noted between the curves representing the mass build-up of this region of the galaxy. They are, however, consistent within the error bars and hence not significant.

\subsection{Age-metallicity relation}
\label{comp_AMR}

In Fig. \ref{SFH_steck_3} (middle panel) we show the AMRs from the CMD and the {\tt STECKMAP} study with a zoom at young ages. The overall shape and the metallicity range are consistent. Even though larger metallicities are systematically found in the STECKMAP analysis, they are consistent within the error bars for most of the time interval, except the last $\simeq$0.25 Gyr. In the spectrum solution, there is an upturn from $\sim$\,1.5 Gyr to now, which is not found in the CMD analysis, where the metallicity remains constant and always below the {\tt STECKMAP} values. We find the main similarities in the age range between $\sim$\,1.5 and $\sim$\,3.5 Gyr. At ages older than $\sim$\,3.5 Gyr the AMR from the integrated spectrum displays a very shallow negative gradient while the AMR from the analysis of the CMD shows a steeper negative gradient and lower metallicities. A plausible explanation might be found in the choice of the smoothing parameters. The best test uses a smoothness parameter in the AMR of 10$^6$, thus, very smooth solutions for the AMR shape, as the one found, are preferred. If we compare this solution (Fig.~\ref{SFH_steck_3}) with other solutions using different smoothing parameters (see Sect.~\ref{cons_tests} and appendix~\ref{plots_tests}), we can see that this young, metal-rich component disappears for those tests with a low value of the AMR smoothness parameter. Thus, the main differences regarding the AMR shape (above outlined) might be an artifact from the high smoothing parameter imposed on the AMR. In fact, tests 17, 18, 20, and 23 (tests with a lower $\mu_Z$) all show quite similar AMR shapes when compared to the CMD results. This is also true of the SFR(t) shapes.

\section{Results from other available codes}
\label{others}

For the sake of completeness, we have also used other available full spectrum and SED fitting codes (see Sect. \ref{intro}). However, as this is not meant to be an exhaustive comparison between all the available codes for analyzing stellar content from integrated spectra, here we analyze the results from those codes we (the authors) are more familiar with. The input parameters for these tests are as similar to the {\tt STECKMAP} input parameters as possible.

\subsection{\tt ULySS}
\label{ulyss}

{\tt ULySS}\footnote{{\tt ULySS} can be downloaded at \url{http://ulyss.univ-lyon1.fr}} \citep{2009A&A...501.1269K} is a full spectrum fitting code which uses Levenberg-Marquardt minimization \citep[from][]{2009ASPC..411..251M} to fit a linear combination of non-linear parameters. It parametrizes the inverse problem as

\begin{dmath}
  F_{Obs}(\lambda) = P_{n}(\lambda) \times \bigg(LOSVD(v_{sys}, \sigma, h3, h4) \otimes \sum_{i=0}^{i=m} W_i \, Cmp_i(a_1, a_2, ...,\lambda) \bigg),
\label{eqn:ulyss}
\end{dmath}

where F$_{Obs}$($\lambda$) is the observed flux at every value of the wavelength ($\lambda$); P$_{n}$ is the multiplicative polynomial and $n$ its degree; LOSVD is the stellar line-of-sight-velocity-distribution that depends on kinematics parameters such as the systemic velocity (v$_{sys}$), velocity dispersion ($\sigma$), and higher momenta (h3, h4); $Cmp_i$ are the different components or SSPs dependent on age ($a_1$) and metallicity ($a_2$) with different weights ($W_i$) computed during the fit.

\begin{figure}
\centering
\includegraphics[scale=0.38]{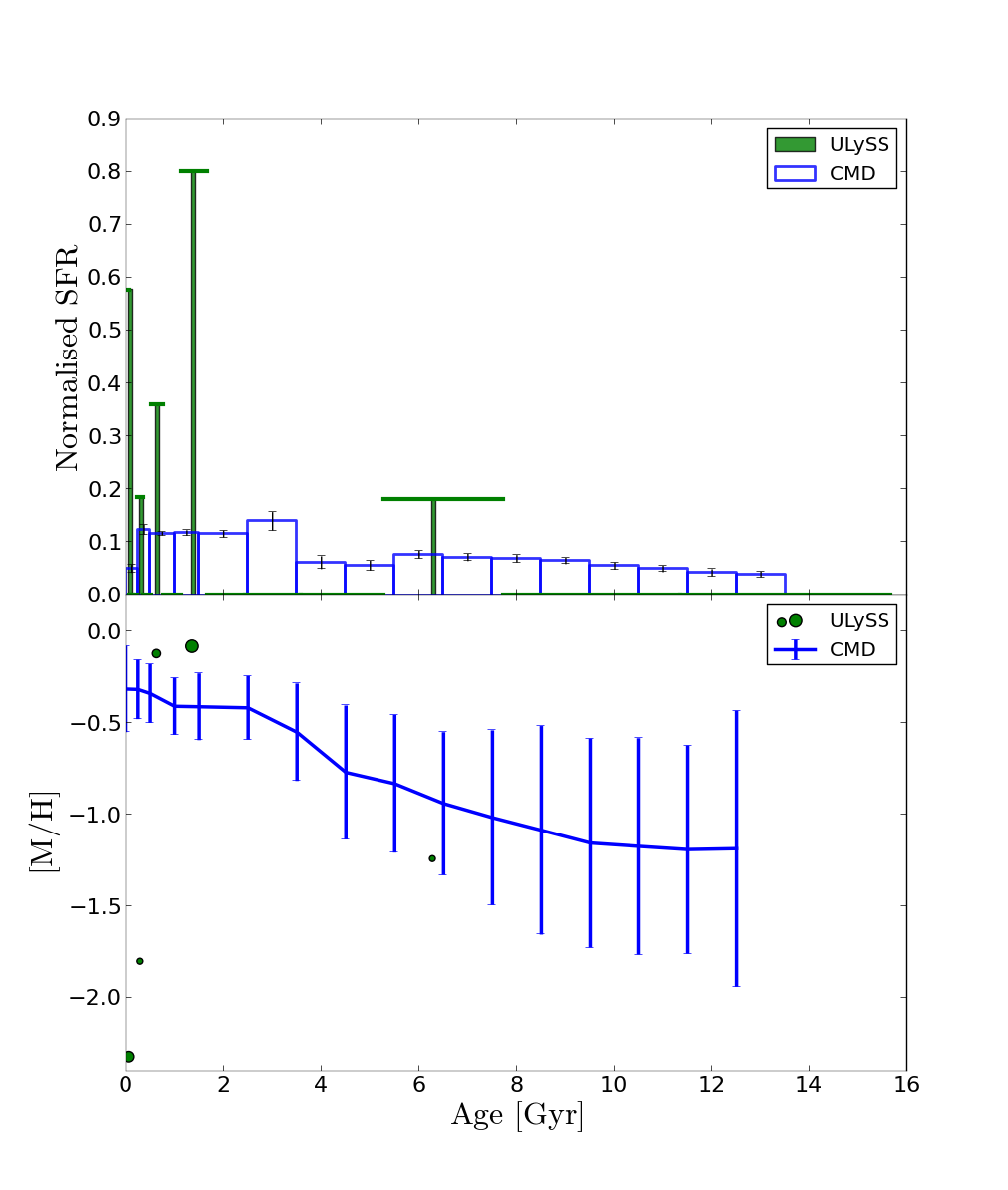} \\
\caption{Top panel: SFH using {\tt ULySS} with \citet[][]{2010MNRAS.404.1639V} models younger than 13.5 Gyr (``SSP'' test). Bottom panel: Age-metallicity relation from the {\tt ULySS} results. SFH: Note that the representation of the SFH is different that in the case of {\tt STECKMAP} (see Fig.~\ref{SFH_steck_3}). In this case we are using SSPs, with no smoothing applied, and thus, vertical lines are chosen to show the mass contribution of the different SSPs instead of a bar plot. However, for a fair comparison, a similar normalization has been applied. Horizontal lines represent the $\Delta t$ used in this case (computed based on stellar population models used). For further information see Fig.~\ref{SFH_steck_3}. AMR: Green points represent the age and metallicity of the different single stellar populations with a non-zero weight in the fit. The point size is proportional to such weight.}
\label{ulyss_blind}
\end{figure}

We have carried out different tests with different input parameters using {\tt ULySS}. In a first approach, our model was computed as a sub-set of $i=15$ SSPs broadened by a LOSVD from a sub-set of the \citet[][]{2010MNRAS.404.1639V} models (as for the {\tt STECKMAP} analysis). While the ages of the SSPs were fixed between 63 Myr and 13.5 Gyr (equally log-spaced) the metallicities were left free and they could vary between the limits of the models (-2.3 and 0.2 dex, the sub-set of models comprises 105 different SSPs). We used the emission-cleaned spectrum obtained in Sect.\,\ref{SFH_int} and fixed the stellar kinematics. We will refer to this test as ``SSP'' as it is based on SSPs as spectral templates. The best fit SFR\,(t) and AMR are shown in Fig. \ref{ulyss_blind}. The overall shape of the SFR\,(t) is qualitatively similar to the {\tt STECKMAP} and CMD reconstructions (see Fig. \ref{ulyss_blind}), although showing sporadic bursts of star formation, which is expected as we are using a combination of SSPs. {\tt ULySS} fails at replicating the AMR at ages younger than $\sim$\,0.5 Gyr; the metallicities recovered are lower than those inferred from the CMD and the {\tt STECKMAP} approaches. However, at intermediate and old ages (older than 0.25 Gyr) {\tt ULySS} AMR results are in fair agreement with the CMD and STECKMAP ones.

\begin{figure}
\centering
\includegraphics[scale=0.38]{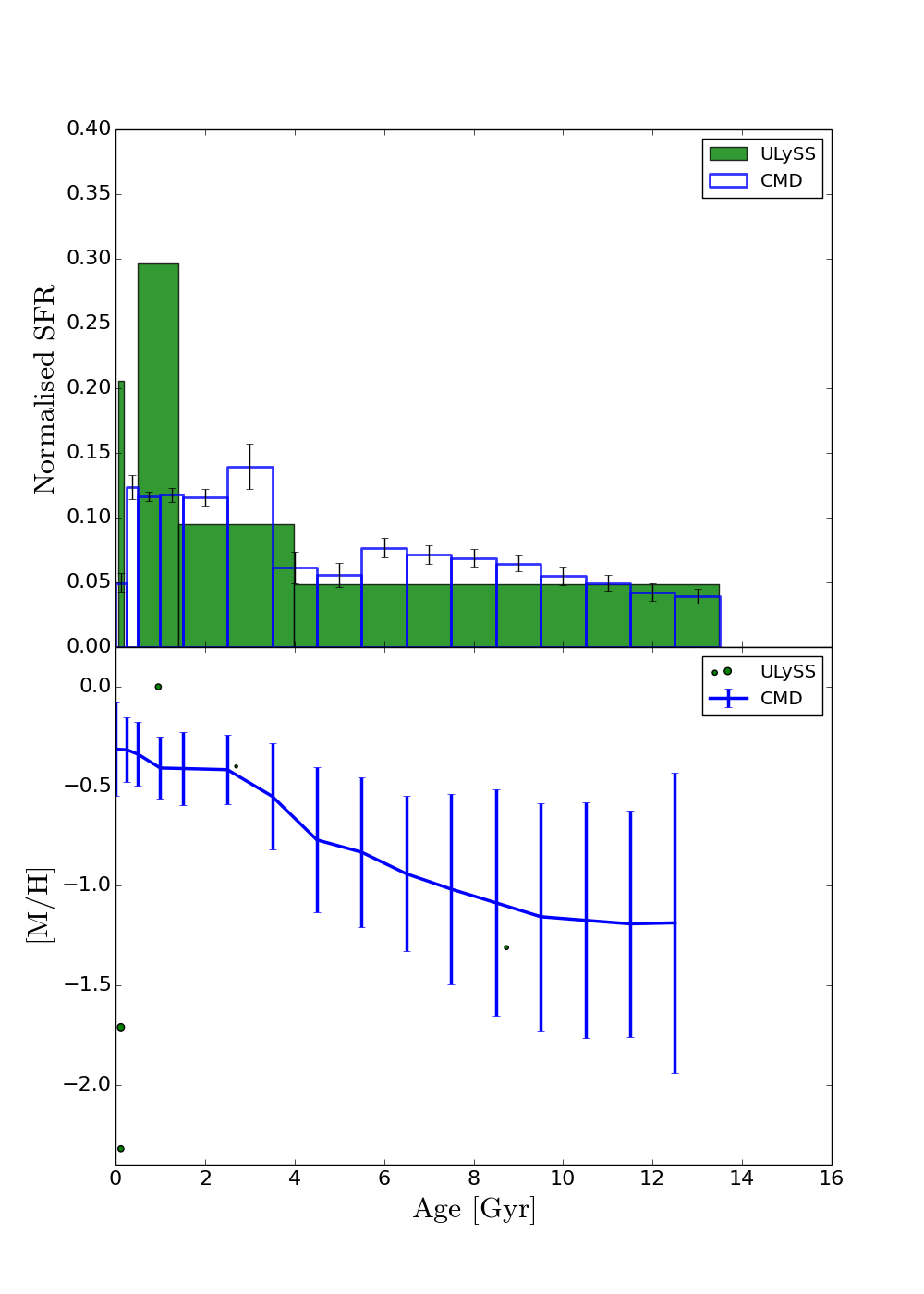} \\
\caption{Same as Fig.~\ref{ulyss_blind}, but using as spectral templates a set of 35 spectra generated from a constant SFR\,(t) and single metallicity (see text for details). However, as in this case we are not using SSPs as spectral templates, we use again a similar representation to the one used with {\tt STECKMAP} (see Fig.~\ref{SFH_steck_3}).}
\label{ulyss_best}
\end{figure}

Although the SSP approach is widely used when recovering the stellar content from integrated spectra, we have also employed a more complex set of spectral templates, more similar to the ones used in the CMD analysis. We made use of the IAC {\tt MILES} webtools \footnote{\url{http://miles.iac.es/pages/webtools/get-spectra-for-a-sfh.php}} to create spectral templates via user defined SFHs. We decided to use 35 spectra of populations with constant SFR\,(t) between $0.063-0.178$, $0.178-0.501$, $0.501-1.413$, $1.413-3.981$, and $3.981-13.49$\,Gyr (equally log-spaced in intervals of 0.45 dex) and fixed metallicities ranging from $-2.32$ to 0.22 dex [M/H]. These spectral templates may be  a better representation of the continuous mode of star formation expected in real galaxies than discrete SSPs.  We will name these new spectral templates ``complex SPs'' (complex stellar populations). Thus, we use these $i=35$ ``complex SPs'' ($Cmp_i$), following the same approach outlined above (input spectrum, wavelength range, fixed stellar kinematics, etc). We will refer to this test as ``constant SFR''. The results are shown in Fig. \ref{ulyss_best}. The use of complex SPs considerably improves the recovery of the shape of the SFR\,(t). The main discrepancies are found at ages younger than 0.5\,Gyr, with an absence of populations with ages between $0.178-0.501$\,Gyr and an excess contribution in the youngest bin ($0.063-0.178$\,Gyr). The AMR is similar to the one we obtain using SSPs, with the youngest ages showing too low metallicities. 

\subsection{\tt STARLIGHT}
\label{starlight}

We have also used {\tt STARLIGHT} \citep[][]{2005MNRAS.358..363C}~\footnote{{\tt STARLIGHT} can be downloaded at \url{http://astro.ufsc.br/starlight/}} to obtain the SFH from the observed integrated spectrum. {\tt STARLIGHT} is a SED fitting code that tries to match the observed spectrum (continuum and spectral features) by means of a combination of stellar models. It also fits the reddening and stellar kinematics simultaneously. 

The details of how {\tt STARLIGHT} works are given in \citet[]{2004MNRAS.355..273C} and \citet[][]{2005MNRAS.358..363C}. {\tt STARLIGHT} mainly requires as input an observed spectrum, a configuration file, a mask file, a set of $N_\star$ base spectra (templates), and a reddening law. Essentially, the code then tries to obtain the SFH, reddening, and stellar kinematics by the minimisation of a $\chi^2$:

\begin{equation}
  \chi^2 = \sum_\lambda \Bigl [   (O_\lambda - M_\lambda)w_\lambda   \Bigr ] ^2,
\label{chi_def}
\end{equation}

where O$_\lambda$ is the observed flux, M$_\lambda$ is the modelled flux, and w$_\lambda$ is the weight (0 for masked regions) at the wavelength $\lambda$. {\tt STARLIGHT} normalises the observed spectrum and the spectral templates at a given wavelength, in our case we use the window between 5590 and 5680 \AA $ $ for the observed spectrum, and the flux at 5635 \AA $ $ for the template normalization. 
The expression that this code uses for the modelled spectrum is as follows:

\begin{equation}
  M_\lambda = M_{\lambda 0} \Bigl [  \sum_{j=1}^{N_\star} x_j b_{j,\lambda} r_\lambda      \Bigr ] \otimes G(v_\star, \sigma_\star),
\label{model_def}
\end{equation}

where M$_{\lambda 0}$ is the synthetic flux at the normalization wavelength; M$_\lambda$ is the modelled flux at $\lambda$; $x_j$ is the weight of the j$^{th}$ element of the set of base spectra; $b_{j,\lambda} r_\lambda$ is the normalised reddened-spectrum for such j$^{th}$ component where $r_\lambda = 10^{-0.4(A_\lambda - A_{\lambda 0})}$ is the extinction term. This first part of the expression accounting for the stellar content and the reddening is convolved ($\otimes$) by the LOSVD of the stellar component ($G(v_\star, \sigma_\star)$). We made use of the LMC reddening law provided by the {\tt STARLIGHT} package and presented in \citet[][]{2003ApJ...594..279G}.

\begin{figure}
\centering
\includegraphics[scale=0.38]{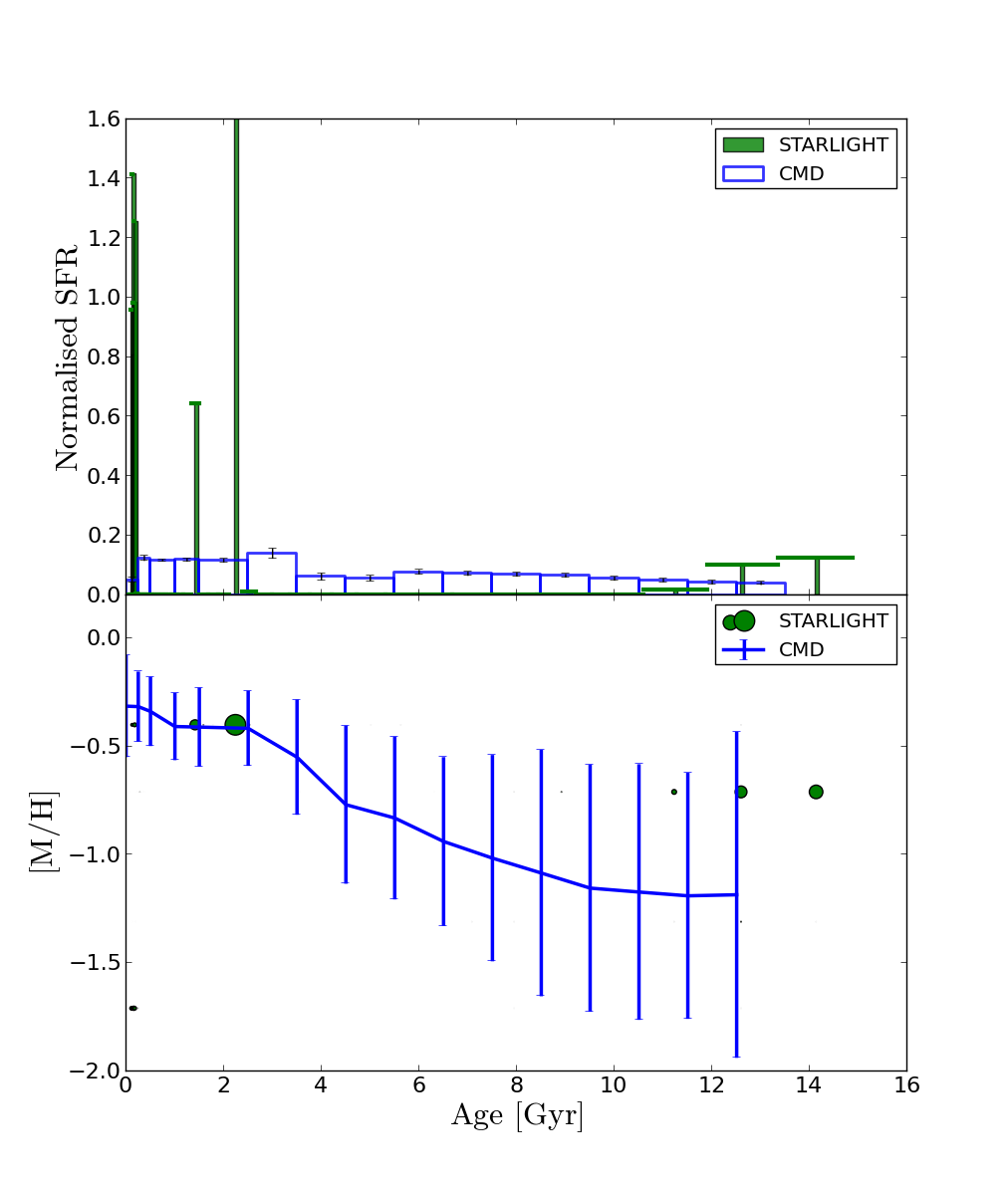} \\
\caption{Top panel: SFH from {\tt STARLIGHT} with \citet[][]{2010MNRAS.404.1639V} models younger than 13.5 Gyr. Bottom panel: Age-metallicity relation from the {\tt STARLIGHT} results. SFH: Note that the representation of the SFH is different that in the case of {\tt STECKMAP} (see Fig.~\ref{SFH_steck_3}). In this case we are using SSPs, with no smoothing applied, and thus, vertical lines are chosen to show the mass contribution of the different SSPs instead of a bar plot. However, for a fair comparison, a similar normalization has been applied. Horizontal lines represent the $\Delta t$ used in this case (computed based on the base of stellar models used). For further information see Fig.~\ref{SFH_steck_3}. AMR: Green points represent the Age and Metallicity of the different single stellar populations with a non-zero weight in the fit. The point size is proportional to their weights.}
\label{star_blind}
\end{figure}

We have analysed the LMC bar integrated spectrum following a recipe similar to the one discussed in Sects.~\ref{SFH_int} and \ref{ulyss}; this test will be called ``SSP'' as it will be based on SSPs as spectral templates. We used the same set of models  \footnote{In the case of {\tt STARLIGHT}, as well as for {\tt STECKMAP}, we use the entire set of models, [ages (Gyr)]x[M/H] = [0.0631, 0.0708, 0.0794, 0.0891, 0.1000, 0.1122, 0.1259, 0.1413,  0.1585, 0.1778, 0.1995, 0.2239, 0.2512, 0.2818, 0.3162,  0.3548, 0.3981, 0.4467, 0.5012, 0.5623, 0.6310, 0.7079, 0.7943, 0.8913, 1.0000,  1.1220, 1.2589,  1.4125, 1.5849, 1.7783, 1.9953, 2.2387, 2.5119, 2.8184, 3.1623, 3.5481, 3.9811, 4.4668, 5.0119, 5.6234, 6.3096, 7.0795, 7.9433, 8.9125, 10.0000, 11.2202, 12.5893, 14.12]x[-2.32, -1.71, -1.31, -0.71, -0.4, 0.0, 0.22]}  \citep[][up to 13.5 Gyr]{2010MNRAS.404.1639V} as in previous cases, removed the emission line contribution using {\tt GANDALF}, fixed the stellar kinematics to the {\tt pPXF} values, and run {\tt STARLIGHT} to match every spectral feature in the wavelength range from 3800 to 6800 \AA. We masked transition regions between grisms and sky features, but not the emission lines as they have meaningful information after applying {\tt GANDALF}. We find that the results from {\tt STARLIGHT}  show some important differences when compared to the CMD and {\tt STECKMAP} results (see Fig. \ref{star_blind}). The recovered SFR\,(t) shows some episodic bumps of star formation younger than 2 Gyr and a predominant old stellar population (older than 10 Gyr), while an intermediate population between 2 and 10 Gyr is not found. The main discrepancies regarding the AMR are found at ages younger than 0.5 Gyr. As in the case of {\tt ULySS}, the discontinuity in the recovered SFR\,(t) is a direct consequence of the use of SSPs as spectral templates.

These discrepancies encouraged us to carry out a set of 36 tests modifying different input parameters and procedures to check the reliability of the results. i) We tested the effect of changing the set of model templates and the age range. ii) To reduce the degrees of freedom in the fit we also inspected the effect of imposing an {\it a priori} AMR from \citet[][]{2008AJ....135..836C}, using a carefully selected set of spectral templates from the different models. iii) Considering that {\tt STARLIGHT} also fits the continuum shape of the spectrum, we have also tried to obtain stellar content with this code avoiding the emission line removal step with {\tt GANDALF}, masking the emission lines instead in order to test if the discrepancies are caused by a bad emission line subtraction. iv) We also allowed {\tt STARLIGHT} to fix or fit the stellar kinematics. 

We obtained a wide variety of solutions from these different tests. A clear bi-modality is found in the SFR\,(t) shape using the \citet[][]{2010MNRAS.404.1639V} and \citet[][]{2010MNRAS.404.1639V} plus \citet[][]{2005MNRAS.357..945G} models, with contributions from young and old populations (younger than 2 Gyr and older than 8 Gyr) and a lack of an intermediate component. However, the use of BC03 models results in a more spread-out, or smoother SFR\,(t) with contributions at all ages. Imposing an observed AMR strongly restricts the number of templates used, and gives poor fits. Fixing the kinematics gives slightly better results than fitting for the kinematics. Using {\tt GANDALF} or masking the emission lines plays a minor role in the recovery of the stellar content.

\begin{figure}
\centering
\includegraphics[scale=0.38]{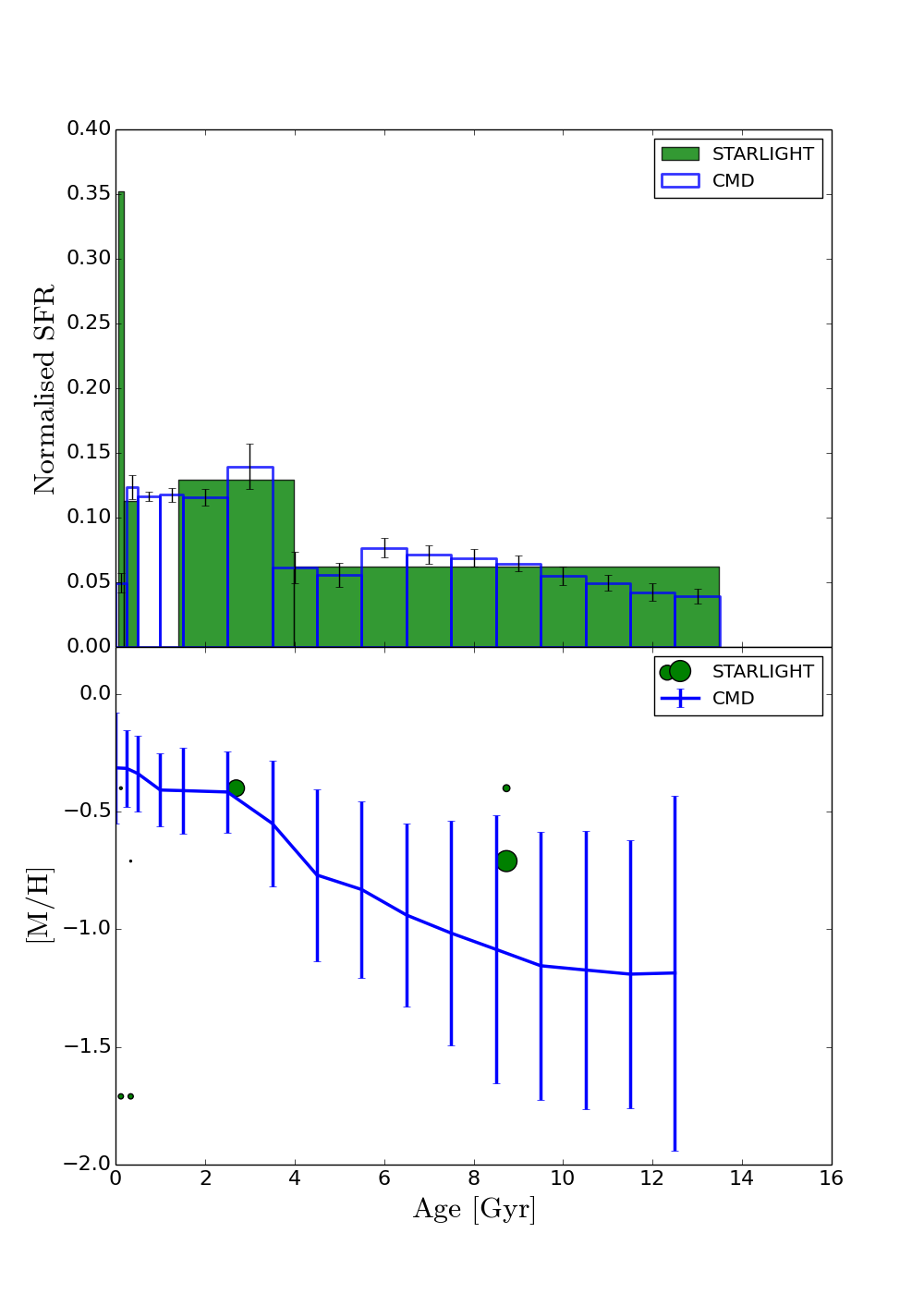} \\
\caption{Same as Fig. \ref{star_blind}, but using as spectral templates a set of 35 spectra generated from a constant SFR\,(t) and single metallicity (see text for details). However, as in this case we are not using SSPs as spectral templates, we use again a similar representation to the one used with {\tt STECKMAP} (see Fig.~\ref{SFH_steck_3}).}
\label{star_best}
\end{figure}

In addition, as we did with {\tt ULySS} (see Sect. \ref{ulyss}), we performed a further test where we used complex SPs, rather than SSPs, as spectral templates (see Sect.~\ref{ulyss}). We name this test ``constant SFR'' as previously done with {\tt ULySS}. The results of this test are shown in Fig. \ref{star_best}. Using complex SPs we have been able to obtain a closer approximation to the CMD results regarding the SFR\,(t) shape. However, we find a lack of stars with ages ranging from $0.501-1.413$\,Gyr and an excess of stars in the youngest age bin ($0.063-0.178$\,Gyr). The obtained AMR is very similar to the one recovered using SSPs as spectral templates (again, similar to the {\tt ULySS} results), displaying the same issues at the youngest ages.

\vspace{5mm}

\section{Discussion}
\label{disc}

The LMC bar is an appropriate astronomical object to perform a comparison between the SFH obtained for a composite stellar population by means of the information from its resolved stars and from its integrated light. The reason is twofold: i) we can obtain a CMD reaching the oMSTO using HST data, and ii) its high surface brightness allows us to observe a high quality integrated spectrum. By analyzing the CMD and integrated spectrum of the LMC bar we derived and compared its SFHs. In this work we mainly focused on the comparison between CMD analysis and full spectrum fitting analysis using {\tt STECKMAP} \citep[][]{2006MNRAS.365...46O, 2006MNRAS.365...74O}. In principle, the stellar populations of the LMC bar represent a challenging case for inversion codes analyzing integrated stellar spectra because of its complex nature and rich stellar diversity. However, the agreement in the SFR(t) and the AMR between the CMD and the integrated spectrum analysis strongly supports the use of state-of-the-art full spectrum fitting codes in deriving the SFH of complex stellar systems. The results of our ``blind tests'' are that the results of both approaches are consistent when trying to recover the SFH of the same object. However, during our analysis we identified several issues that should be further studied. 

The most striking difference between the STECKMAP and the CMD results appears in the derived AMRs. Both AMRs show a metallicity monotonically increasing with time, but the AMR derived from the CMD starts at lower metallicities ($\simeq$ -1.2 dex) and does not reach solar metallicity at young ages. The AMR obtained with STECKMAP starts at [M/H] $\simeq$ -0.75 dex at old ages, remains quite flat for most of the time range and increases steeply in the last $\simeq$ 2 Gyr, reaching metallicity slightly over solar at the present time. The overall SFR\,(t) shape also displays some differences for populations younger than 4.0\,Gyr. The SFR\,(t) is almost flat between $\sim$\,0.25 and $\sim$\,3.5 Gyr from the CMD analysis, whereas from the {\tt STECKMAP} analysis we see a skewed gaussian towards older ages with the peak at $\sim$ 1 Gyr. There are several possible explanations for these differences in SFR\,(t) obtained from the two methods: 

a) The CMD analysis produces precise results when a deep CMD reaching to below the oMSTO is used (as in this case). The methodology described in Sect. \ref{SFH_CMD} has been extensively tested by recovering SFHs using mock CMDs with known stellar content \citep[e.g.][]{iacpop, 2010ApJ...722.1864M, monelli10b, hidalgo11}. However, these analyses are not error-free. Differences in the solutions are found due to slight changes in the analysis method (e.g. way of dealing with errors, bundle definition, minimisation algorithm). Also, tests with mock stellar populations reveal that an age-dependent smoothing of the SFH features is produced, with age resolution worsening at old ages, due to the limitations in the age resolution intrinsic to the method \citep[e.g. see figure 8 in][]{hidalgo11}. Having a large number of stars in the CMD is important for a reliable solution, and we cannot exclude the possibility that the shape of the SFR\,(t) at young ages is not well recovered due to small number statistics. This may be aggravated by the fact that the brightest, youngest stars in the main sequence are saturated, and therefore do not appear in the CMD. Even though the SFR\,(t) at the corresponding ages may still be recovered from lower mass stars, which are fainter on the main sequence, some information is obviously lost. Thus, further investigation using a CMD taken over a larger area, and covering the whole magnitude range would be of interest.

b) There could be a lack of young stars in the field where the integrated spectrum was observed. This might produce a deficit of flux coming from the youngest populations, and therefore might result in the failure of {\tt STECKMAP} to reproduce the AMR at such young ages. However, we note that {\tt STECKMAP}, {\tt STARLIGHT} and {\tt ULySS} all find some contribution at very young ages (see Fig. \ref{SFH_steck_3}). In addition, a visual inspection of the observed spectrum shows some helium absorption lines (\hei$\lambda$3819 and \hei$\lambda$4922), which are signatures of the presence of young, hot stars. Therefore, it does not seem that a deficit of signal from young stars is the cause of the differences in the metallicity inferred for the very young population.

\begin{figure*}
\centering
\includegraphics[scale=0.4]{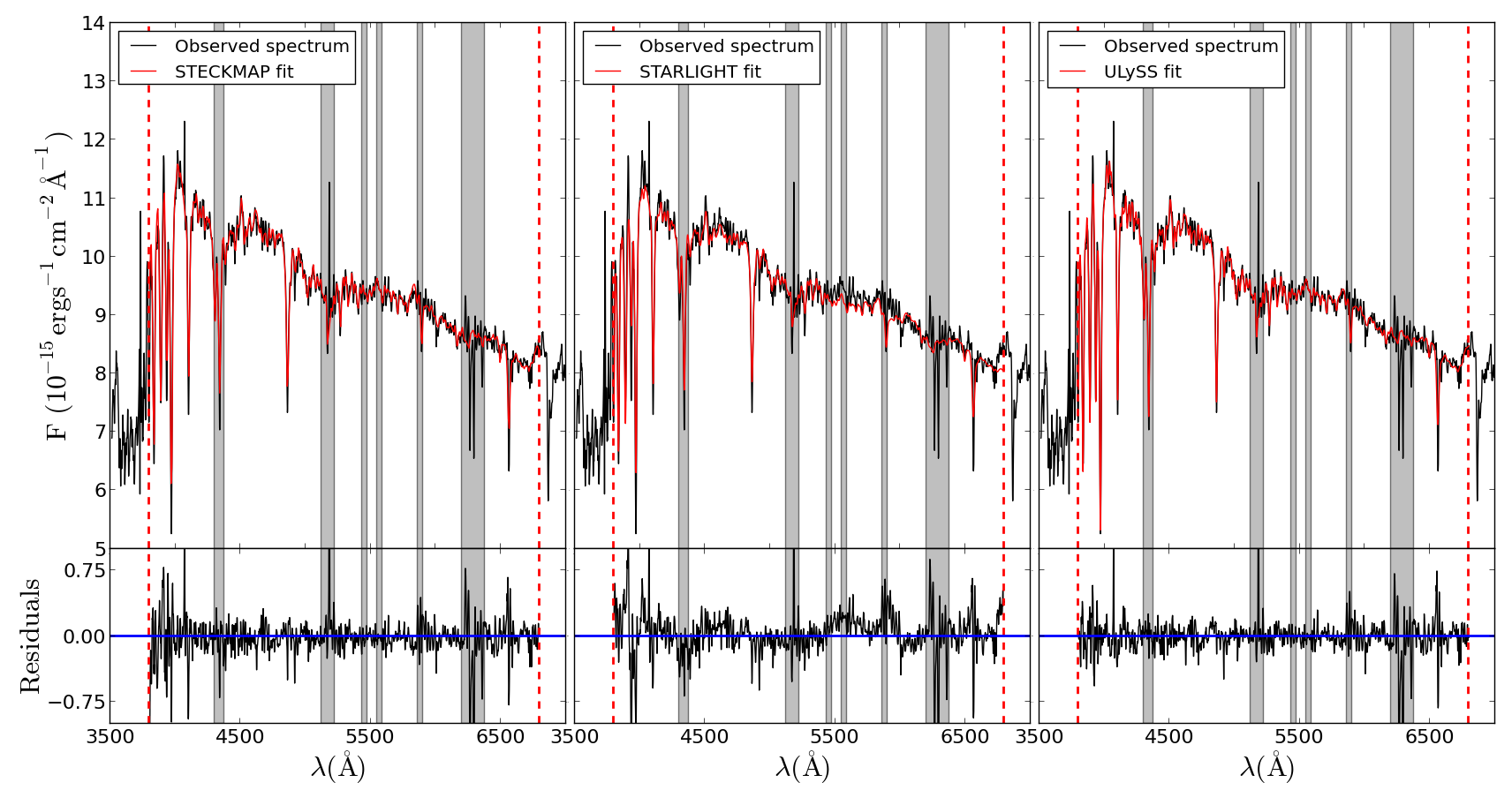}
\caption{Top panels: Comparison between the observed spectrum and the fit from {\tt STECKMAP} (left), {\tt STARLIGHT} (middle), and {\tt ULySS} (right). For {\tt STARLIGHT} and {\tt ULySS} we are using the ``constant SFR'' approach. Black: Fully-reduced, emission-cleaned composite spectrum of the LMC bar. Red: Fit from the different codes. Bottom panels: Residuals of those fits computed as observed - best model. For the stellar content analysis we use the wavelength range from 3800 to 6800 \AA $ $ (see text for details, dashed vertical red lines). The shaded regions of the spectrum are the masked regions (not considered in the fits). The residuals are in units of 10$^{-15}$ erg$^{-1}$ cm$^{-2}$ \AA$^{-1}$.}
\label{LMC_fits}
\end{figure*}

%c) At very young ages (e.g., younger than $\sim200$ Myr), it is hard to discriminate metallicity using integrated spectra since metal-lines are very weak \citep[see figs. 6 and 19 in ][]{2010MNRAS.404.1639V}. This might suggest that we are able to overcome the age-metallicity degeneracy in most of the studied age range, except in the very young age regime.

c) The AMR shape is especially affected by the smoothing penalty function in the AMR ($\mu_z$). In particular, the shape of our recovered AMR using {\tt STECKMAP} (see Fig.~\ref{SFH_steck_3}) is very smoothed as a consequence of imposing a large smoothing parameter ($\mu_z$ = 10$^6$). The discrepancies between this AMR and that from the CMD analysis in the younger and older edges of our age range could be due to the effect of this parameter. In fact, tests with $\mu_z$ above 10$^2$ currently present supersolar stars and primordial metallicities greater than $[M/H]$ = -1.0. Other tests with different smoothing parameters for the AMR (see appendix~\ref{plots_tests}) are better able to reproduce the AMR shape derived in the CMD analysis while still providing a good fit for the SFR(t).

d) As shown in Sect.~\ref{cons_tests}, different input parameters give similar SFH results and very accurate fits (see Fig.~\ref{LMC_fits} and table~\ref{steck_tests}) when {\tt STECKMAP} is applied to a high quality spectrum. The overall shape of the recovered SFH is reasonably consistent for the different tests although slight differences do arise between them. Therefore, we conclude that the shape of the recovered SFH from high quality spectra is well reproduced regardless of the input parameters.

Despite these small discrepancies, the similarities between the CMD and the {\tt STECKMAP} results are very reassuring. The manner in which {\tt STECKMAP} deals with the intrinsically ill-posed inversion problem (regularization through a penalized $\chi^2$) has proven to be very powerful. The smoothed solutions from {\tt STECKMAP} are perhaps physically ``sensible''; star formation in galaxies is a complex mechanism, and a consequence of continuous galaxy evolution. In {\tt STARLIGHT} and {\tt ULySS} no such smoothing is implemented, and thus, the use of SSPs as spectral templates naturally leads to a SFR\,(t) characterized by discrete bursts of star formation. Although the use of SSPs as a base for reconstructing the SFH of galaxies is a good first approximation, these SSPs are idealized realizations of bursts of star formation that are not necessarily expected in nature. The inclusion of some kind of smoothing techniques in these codes might eliminate these discontinuities.

Since the exact determination of the shape of the SFR\,(t) is difficult in {\tt STARLIGHT} and {\tt ULySS} if SSPs are employed, we used a set of spectral templates computed assuming populations with a range of ages (complex SPs). Although the shape of the AMR is very similar irrespective of whether we use SSPs or complex SPs, the shape of the SFR\,(t) obtained using this second approach is closer to the CMD and {\tt STECKMAP} SFR\,(t) shapes. The use of complex SPs seems to in some sense mimic the effect of imposing a smoothing on the solution (see Figs. \ref{ulyss_best} and \ref{star_best}) and it is preferred to the use of SSPs as it allows us to obtain some intermediate contribution that otherwise is impossible to detect. This suggests that these kind of spectral templates may be better in order to study the stellar content of external complex systems (e.g. galaxies, not star clusters).

We have also evaluated whether, in spite of not always producing consistent SFR(t) shapes, the different methods are able to reproduce the relative contribution in mass of young, intermediate, and old components (the age intervals for this test are defined to match the analysis {\it \`a la Bica}, see Appendix \ref{bica}).  Table \ref{percentages} shows the contributions to the total stellar mass for different age ranges, as inferred using all the analysed approaches. The outcomes from the CMD, {\tt STECKMAP}, {\tt STARLIGHT} and {\tt ULySS} (``SSP'' and ``complex SSP'' tests) display roughly similar percentages of stars in each age range, although slightly larger discrepancies are found in the {\tt STARLIGHT} analysis.

A common issue in the {\tt ULySS} and {\tt STARLIGHT} results is the excess of young stars (see Figs. \ref{ulyss_blind}, \ref{ulyss_best}, \ref{star_blind}, and \ref{star_best}). As discussed in Sect.\,\ref{obs_CMD}, some bright stars (F814W $\sim$ -1) are saturated, and thus, not taken into account in the SFH recovery from the CMD. This saturation magnitude roughly corresponds to the main sequence turnoff of a $\sim$ 0.3\,Gyr old population. However, we consider in our calculation other stars with the same age that are not saturated (lower mass and therefore, fainter stars), so we can still recover the information at the youngest ages. As this saturation affects the brightest (and youngest) stars which might be dominating in the light integrated spectrum, the excess of young stars found with {\tt ULySS} and {\tt STARLIGHT} may be a consequence of this limitation. We note that this issue is somehow observed in the {\tt STECKMAP} results with low smoothing parameters.  This should be further investigated with the comparison with other local group galaxy using similar analysis to this work.

\begin{table}
\centering
\begin{tabular}{llll}
\hline \hline
Approach &  Young & Intermediate & Old \\ 
 & ($\%$) & $(\%)$ & $\%$  \\ \hline
CMD & 4.3 & 49.0 & 46.7 \\
{\tt STECKMAP} & 4.2 & 49.4 & 46.4 \\
{\tt ULySS} ``SSP'' & 3.9 & 51.4 & 44.7 \\
{\tt ULySS} ``const. SFR'' & 2.4 & 51.3 & 46.3 \\
{\tt STARLIGHT} ``SSP'' & 9.0 & 54.9 & 36.1 \\
{\tt STARLIGHT} ``const. SFR'' & 7.7 & 33.1 & 59.2 \\
Bica (*) & 2.0 & 10.0 & 88.0 \\ \hline
\end{tabular}
\caption{Percentages of the total stellar mass in three different stellar sub-populations: young (younger than 0.5 Gyr), intermediate (older than 0.5 Gyr and younger than 5.0 Gyr), and old (older than 5.0 Gyr). In the cases of {\tt ULySS} and {\tt STARLIGHT} we show the results from the SSP and the mock spectral template approaches. We include the Bica (*) analyses for historical comparison. For information about the Bica analysis see Appendix \ref{bica}.}
\label{percentages}
\end{table}

\begin{figure}
\centering
\includegraphics[scale=0.35]{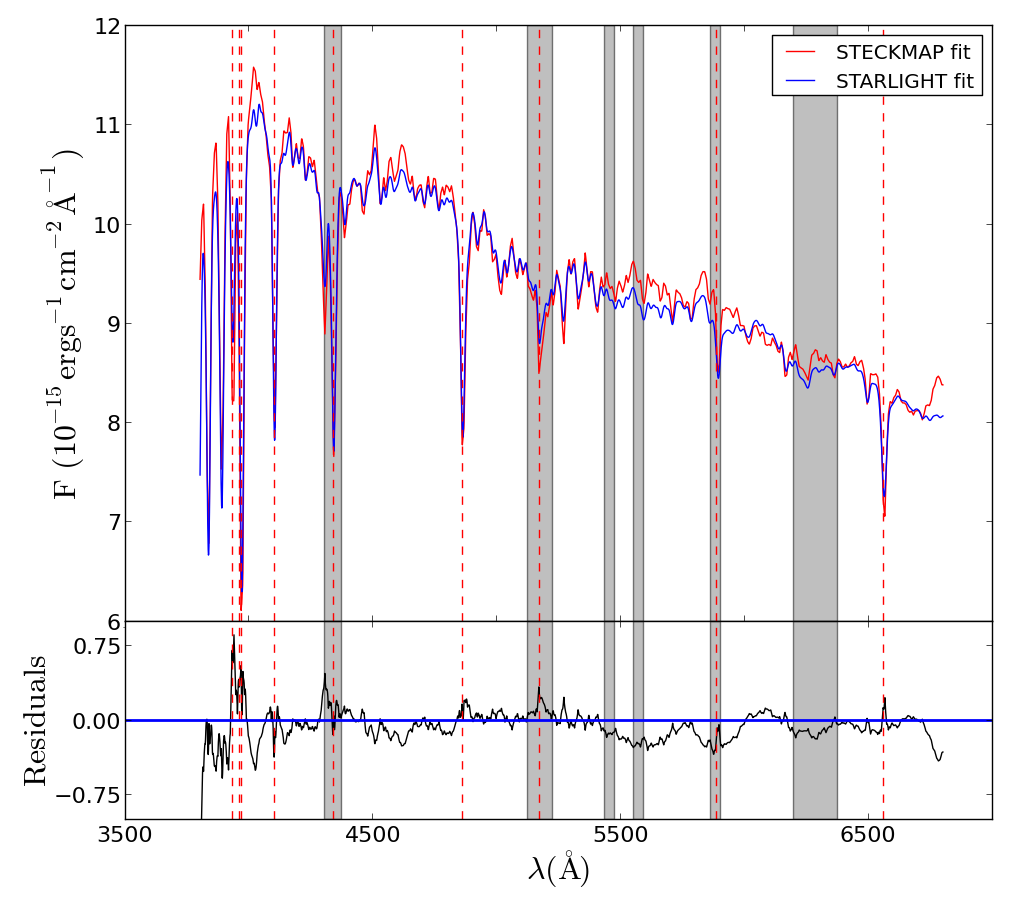}
\caption{Comparison between the {\tt STECKMAP} and {\tt STARLIGHT} (using spectral templates mimicking constant SFR) best models. Top panel: The {\tt STECKMAP} best model is plotted in red. The {\tt STARLIGHT} best model is plotted in blue. Bottom panel: Residuals of both best models ({\tt STECKMAP} - {\tt STARLIGHT}). The shaded regions of the spectrum are the masked regions (not considered in the fits). Vertical, dashed red lines are located at the wavelengths of the main stellar absorption features. The residuals are in units of 10$^{-15}$ erg$^{-1}$ cm$^{-2}$ \AA$^{-1}$.}
\label{STAR_vs_STECK}
\end{figure}

In Fig. \ref{LMC_fits} we analyze each spectral fit ({\tt STECKMAP}, {\tt ULySS} and {\tt STARLIGHT}) in order to assess where the observed discrepancies discussed above (SFH and age percentages) may come from. At first glance the three codes seem to properly fit the observed spectrum. The $rms$ of the residuals is 0.12 for the {\tt STECKMAP} fit, 0.21 for the {\tt STARLIGHT} fit and 0.18 for the {\tt ULySS} fit (both ``SSP'' and ``constant SFR'' tests). A careful inspection shows that {\tt STECKMAP} and {\tt ULySS} are able to better reproduce some individual spectral features than {\tt STARLIGHT} as a consequence of the polynomial fitting. Figure \ref{STAR_vs_STECK} shows a more detailed comparison between the {\tt STECKMAP} and {\tt STARLIGHT} best models (``constant SFR''), i.e. the spectrum corresponding to the recovered stellar content. We note some important differences in the shape of the continuum (wave-like features in the residual plot) as well as differences in some specific absorption features. Among the features with larger discrepancies we can highlight $H\epsilon$, $H\beta$, the \caii $ $ doublet (3933, 3969 \AA), the magnesium feature at 5175 \AA, and the sodium absorption line (5892 \AA). Every stellar feature is deeper in the {\tt STECKMAP} best model than in the case of the {\tt STARLIGHT} fit except for the sodium feature. These differences affect considerably the recovered stellar content.

Based on the work presented here we find that, although the techniques for the analysis of unresolved stellar populations give good results when compared to the CMD reconstruction, there is still room for improvement. For example, a wider range of spectral coverage might help to overcome the issues at young ages and to better constrain the recovered SFH. Few empirical stellar libraries are available in the ultraviolet (150 to 333\,nm) and near-infrared (up to 2 microns) and thus, theoretical SSPs limited to the optical range are used \citep[e.g.,][]{2009MNRAS.400..273R, 2012ApJ...755...87S, 2010MNRAS.406.2185M, 2013MNRAS.435.2861M}. In addition, we find that a successful way of overcoming the ill-conditioned problem of the reconstruction of the stellar content from an observed spectrum is by imposing smoothed solutions (which have more physical meaning), or by using spectral templates composed of complex SPs. We suggest that this could be a basis for improvement in the next generation stellar population synthesis codes from integrated spectra.

\section{Conclusions}
\label{conclusions}

In this paper we have studied the SFH and the AMR of the LMC bar using two distinct approaches. We compared the results from the reconstruction of an observed CMD reaching the oMSTO, obtained from WFPC2$@$HST data, with the results from the spectral fitting of an integrated spectrum taken at the 3.6\,m ESO telescope on La Silla, using EFOSC2. Due to the relatively high surface brightness of the LMC bar, and its close proximity, the chosen field is an unique one in which both an accurate, deep CMD and a high quality spectrum can be obtained. We have applied state-of-the-art models, stellar libraries, isochrones, and codes following each approach in order to derive the SFR\,(t) and AMR of this bar field. Different codes ({\tt STECKMAP}, {\tt ULySS}, and {\tt STARLIGHT}) have been used to recover the stellar content from the integrated spectrum and the results compared to the CMD analysis. The analysis of the integrated spectrum using each code has been performed in a consistent manner,  and in an effort to avoid any biases in the solutions, the CMD analysis was performed independently from the spectral analysis.

The best agreement between the integrated spectrum analysis and the CMD analysis was found using {\tt STECKMAP}, the only full spectrum fitting code that we tested with a penalization implemented. {\tt STECKMAP} produces SFR\,(t) and AMR in good agreement with those obtained from the CMD. All the spectral fitting codes used in this study are normally used with SSPs as spectral templates. {\tt ULySS} and {\tt STARLIGHT} do not use any penalization and, as a consequence, solutions dominated by episodic bursts are derived if SSPs are used as spectral templates. {\tt ULySS} is able to reproduce the overall shape of the SFR\,(t) but with 'bursts' of star formation. However, this code fails at reproducing the AMR, especially at young ages. {\tt STARLIGHT} is able to approximately reproduce the AMR (except at the youngest ages),  but has problems with the shape of SFR\,(t), especially at intermediate ages where no equivalent contribution in the CMD is found. We have been able to improve these results (both with {\tt ULySS}  and {\tt STARLIGHT}) by using a set of complex spectral templates constructed adopting a constant SFR\,(t) in bins of log-spaced ages instead of SSPs. This suggests that ``complex SPs'', rather than simple SP spectral templates should be preferred when analyzing the stellar content of composite stellar systems.

This is the first time that the results of these two different approaches for studying stellar populations in galaxies have been compared for an object with a complex SFH, and for which a CMD reaching the oMSTO could be obtained. Such studies are of crucial importance in order to test recent advances in the field, especially in the analysis of the integrated stellar populations. Important ongoing and upcoming projects such as CALIFA \citep[][]{2012A&A...538A...8S}, SAMI \citep[][]{2012MNRAS.421..872C} or MANGA will make use of these techniques to study the formation and evolution of galaxies. 

In future work we plan to expand our analysis to a sample of Local Group dwarf galaxies bracketing a range of properties. This will allow us to identify the impact that different factors (such as different SFHs, fractions of young vs. old stars, the existence of a blue horizontal branch or blue straggler stars) may have on the results. Not only will this comparison identify where full spectrum fitting techniques may fail and where they succeed, but it will also inform us on the spectral ranges and features which are more likely to improve these techniques. Understanding the limits of the reliability of SFHs obtained from integrated spectra provides an important basis for the understanding of galaxies at low and high redshift.

\begin{acknowledgements}
We thank the anonymous referee for very useful comments. We would like to thank Roberto Cid Fernandes and Brad Gibson for very helpful suggestions and discussions that have improved considerably this paper. It is a pleasure to thank the 3.6m telescope team in La Silla for their support during the observing run, in particular R. Athreya and E. Wenderoth. This research has been partly supported by the Spanish Ministry of Science and Innovation (MICINN) under grants AYA2011-24728 and Consolider-Ingenio CSD2010-00064; and by the Junta de Andaluc\'ia (FQM-108). We also acknowledge financial support from the research projects under grants AYA2010-21887-C04-03, AYA2010-21322-C03-03. TRL thanks the support of the Spanish Ministerio de Educaci\'on, Cultura y Deporte by means of the FPU fellowship. DA acknowledges the support of the Haute Provence Observatory as well as the support of the IAC and ESO for visits during which this study was finalized. MK is a postdoctoral fellow of the Fund for Scientific Research- Flanders, Belgium (FWO 65052 / 12E4115N LV). P.S-B acknowledges support from the Ram\'on y Cajal program, grant ATA2010-21322-C03-02 from the Spanish Ministry of Economy and Competitiveness (MINECO).
\end{acknowledgements}

%-------------------------------------------------------------------

\bibliographystyle{aa} % style aa.bst
\bibliography{bibliography} % your references Yourfile.bib

\appendix

\section{Analysis using a base of integrated cluster spectra (Bica method)}
\label{bica}

\begin{figure}
\centering
\includegraphics[scale=0.4]{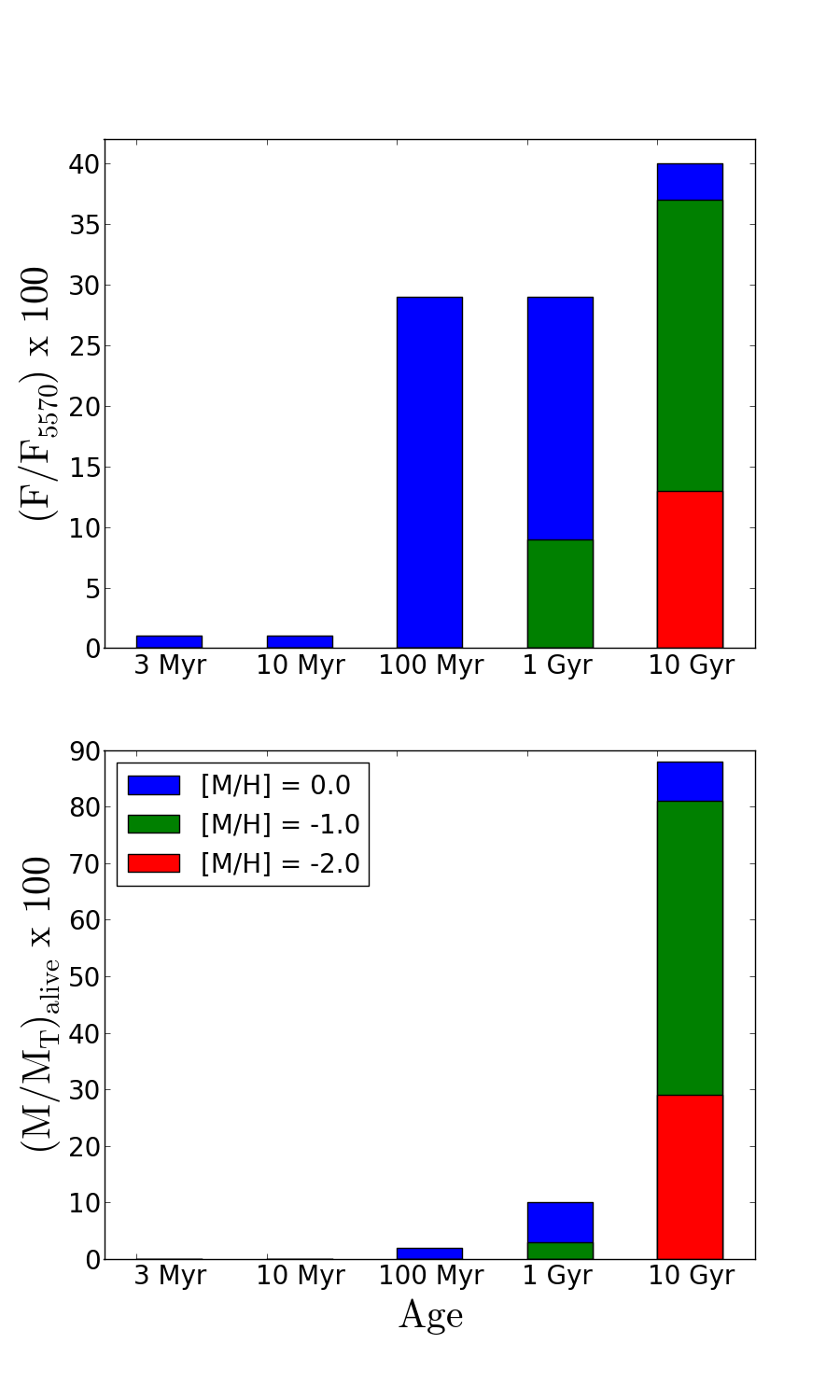} \\
\caption{Synthesis results using the Bica analysis. Top panel, flux fraction of the different base elements. Bottom panel, mass alive fraction of the different base elements.}
\label{Bica_plot}
\end{figure}

To historically link previous methods to study the stellar content in galaxies with modern SED and full spectrum fitting techniques, we analyze the data via a spectral population synthesis technique originally developed by \citet[][]{1988A&A...195...76B}, later updated by \citet[][]{1996MNRAS.278..965S}. This method aims at reproducing the observed equivalent widths (Ws) and the continuum ratios (Cs) using the integrated-light spectra of an ensemble of star clusters with different ages and metallicities  \citep[e.g.,][]{1986A&A...162...21B, 1986A&AS...66..171B, 1987A&AS...70..281B, 1988A&A...202....8B, 1994A&A...283..805B}. The Ws and Cs values from the cluster base are built in a grid parametrized by the age and the metallicity, and extrapolated in the case of high metallicities unreachable through observations. In the current analysis, we have used eight components to map the age-metallicity plane, with a constraint on their metallicities to be solar and subsolar. Indeed, in a low mass galaxy such as the LMC, we do not expect to find stellar components with metallicity above solar \citep[][]{1998MNRAS.299..535P}. The base elements used in the analysis are listed in table \ref{tab_bica_1}. The results are shown in Fig. \ref{Bica_plot}.

From the population analysis above it is clear that the 1 Gyr and 100 Myr components are very important in flux (see Fig. \ref{Bica_plot}, top panel). It is also important to compute how much they represent in terms of mass fractions. For such purposes we employed a flux-mass transformation method \citep[][]{1988A&A...202....8B}. This method uses different mass to V-light ratios ($M/L_V$) related to each age component. It also takes into account metallicity effects among old star clusters. We show in Fig. \ref{Bica_plot} (bottom panel) the mass distributions respectively for each component. We can see that the 1 Gyr and 100 Myr components are less than 10 $\%$ and 5 $\%$ in mass respectively.

We can interpret those results as star formation along three different age ranges (less than 0.5 Gyr, 0.5 -- 5 Gyr, and 5 -- 14 Gyr, see Table \ref{tab_bica_2}). If we consider that the mass at every age range has been formed uniformly and we compute a ``pseudo-SFR'' dividing the mass alive percentage by the age width for each population, that leads us to recent star formation of 4.4 Gyr$^{-1}$, followed by a drop (2.2 Gyr$^{-1}$), with a higher ``pseudo-SFR'' at older ages. These results are away from the STECKMAP and CMD results (see table 3), essentially because of a lack of time resolution at old ages (see table \ref{percentages}).

\begin{table}
\centering
\begin{tabular}{llllll}
\hline\hline
3 Myr &  10 Myr &  100 Myr &  1 Gyr &  10 Gyr &  [M/H] \\ \hline
   8  &  7      &  6       &  4     &  1      &   0.0  \\
      &         &          &  5     &  2      &  -1.0  \\
      &         &          &        &  3      &  -2.0  \\ \hline
\end{tabular}
\caption{Base elements in the age $\times$ metallicity plane for the \citet[][]{1988A&A...195...76B} analysis. E(B-V)$_i$ = 0.0. The different base elements are identified by numbers running from 1 to 8.}   
\label{tab_bica_1}
\end{table}

\begin{table}
\centering
\begin{tabular}{ll}
\hline\hline
Range &  $M_{alive}$  \\ 
  (Gyr)  &  $(\%)$    \\ \hline
5-14 &  88  \\
0.5-5   &  10  \\
< 0.5     &   2  \\ \hline
\end{tabular}
\caption{Results for the \citet[][]{1988A&A...195...76B} analysis. $M_{alive}$ represents the percentage in mass that is still present as stars.}
\label{tab_bica_2}
\end{table}

\section{Recovered Star Formation Histories for the {\tt STECKMAP} tests}
\label{plots_tests}

We include in this appendix the recovered SFH in the tests described in Sect.~\ref{cons_tests}. We show the stellar content for the 24 tests in a similar as in Fig.~\ref{SFH_steck_3}. As already outlined in the main body of the paper, we must highlight the consistency between different tests except in some extreme cases.

\begin{figure*}
\centering
\includegraphics[scale=0.31]{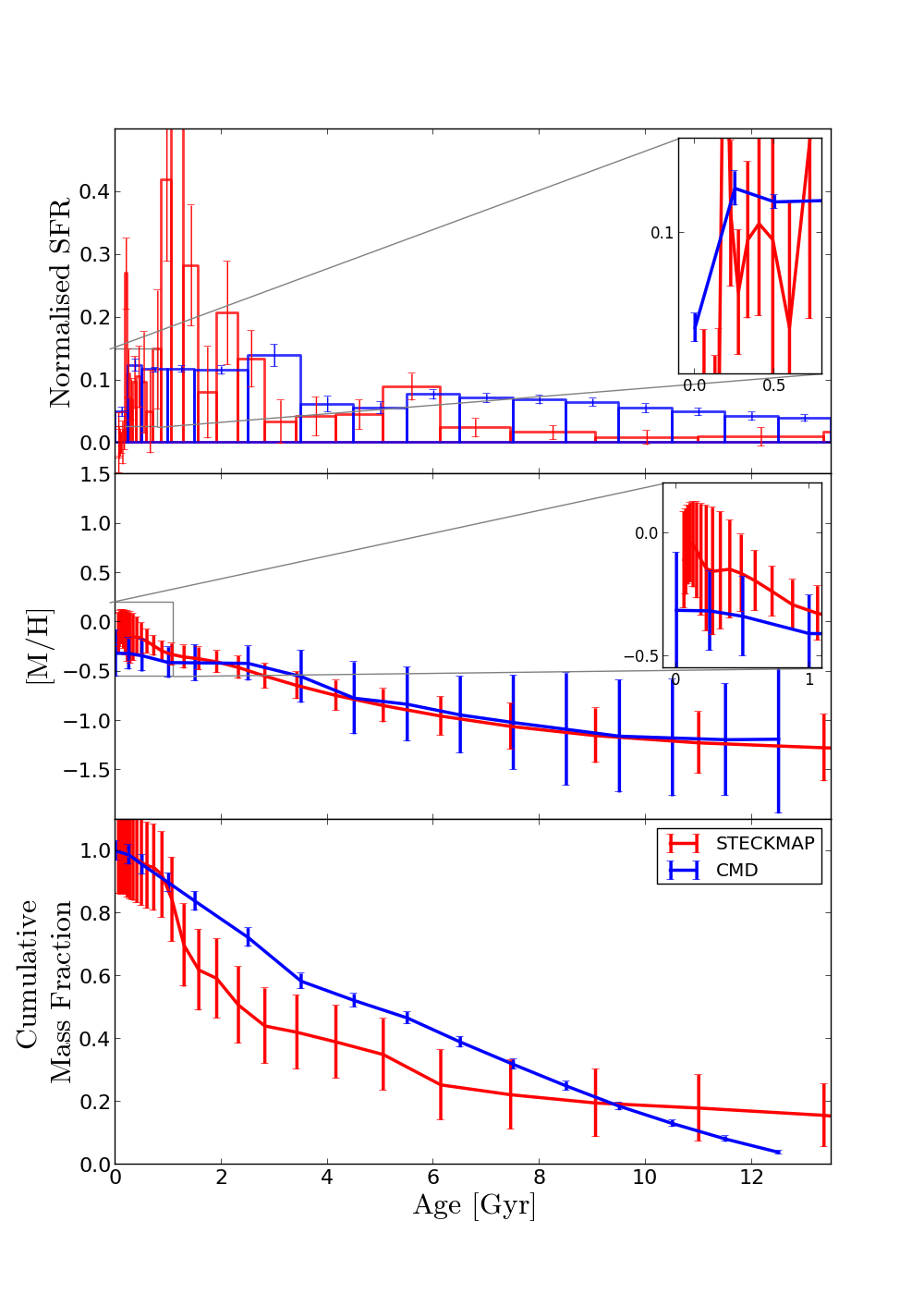} ~
\includegraphics[scale=0.31]{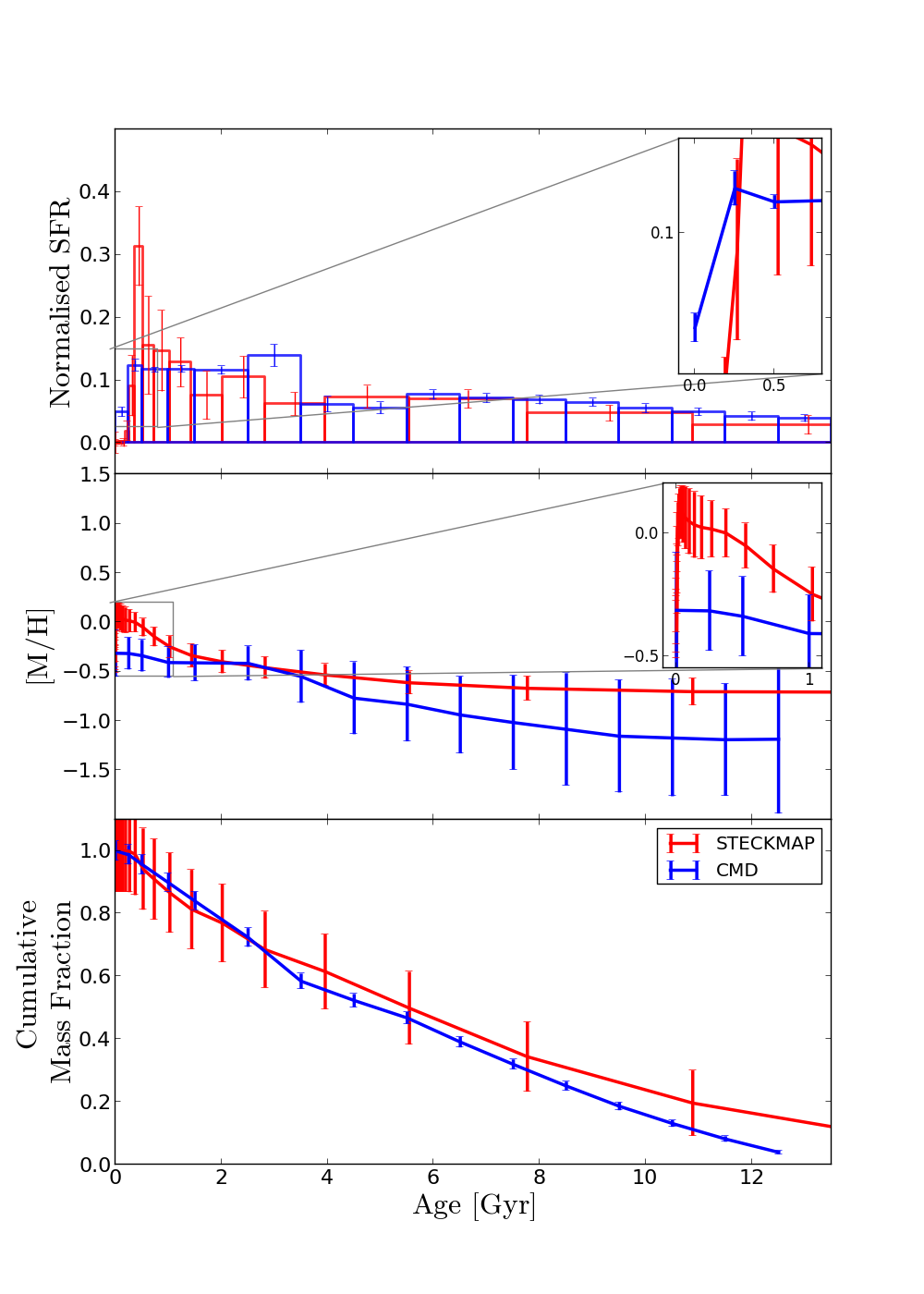} \\
\caption{Comparison between the LMC bar SFH from the CMD and the integrated spectrum using {\tt STECKMAP}. Left: Test 1; Right: Test 2. For further information see Fig.~\ref{SFH_steck_3}.}
\label{steck_tests_1}
\end{figure*}

\begin{figure*}
\centering
\includegraphics[scale=0.31]{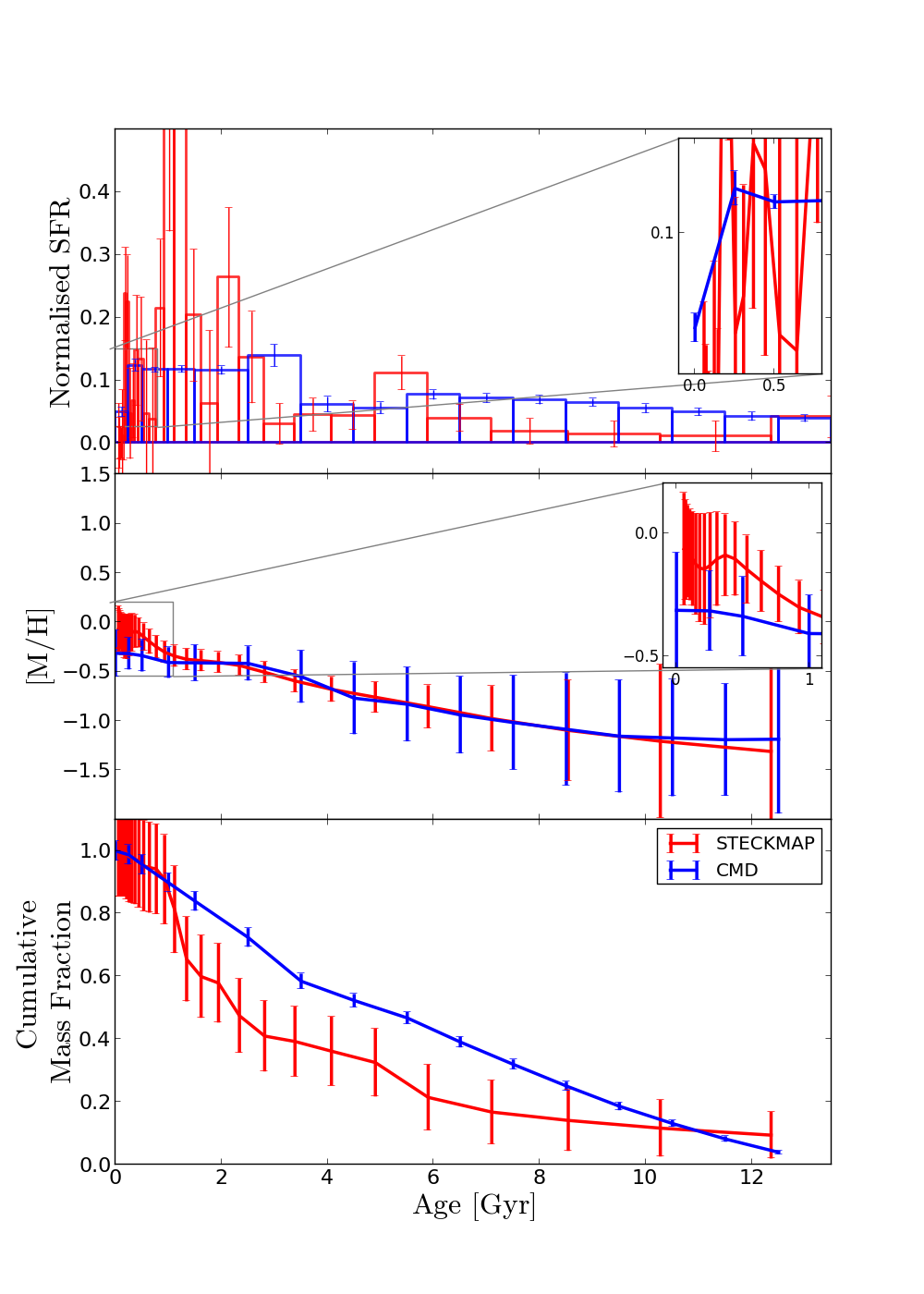} ~
\includegraphics[scale=0.31]{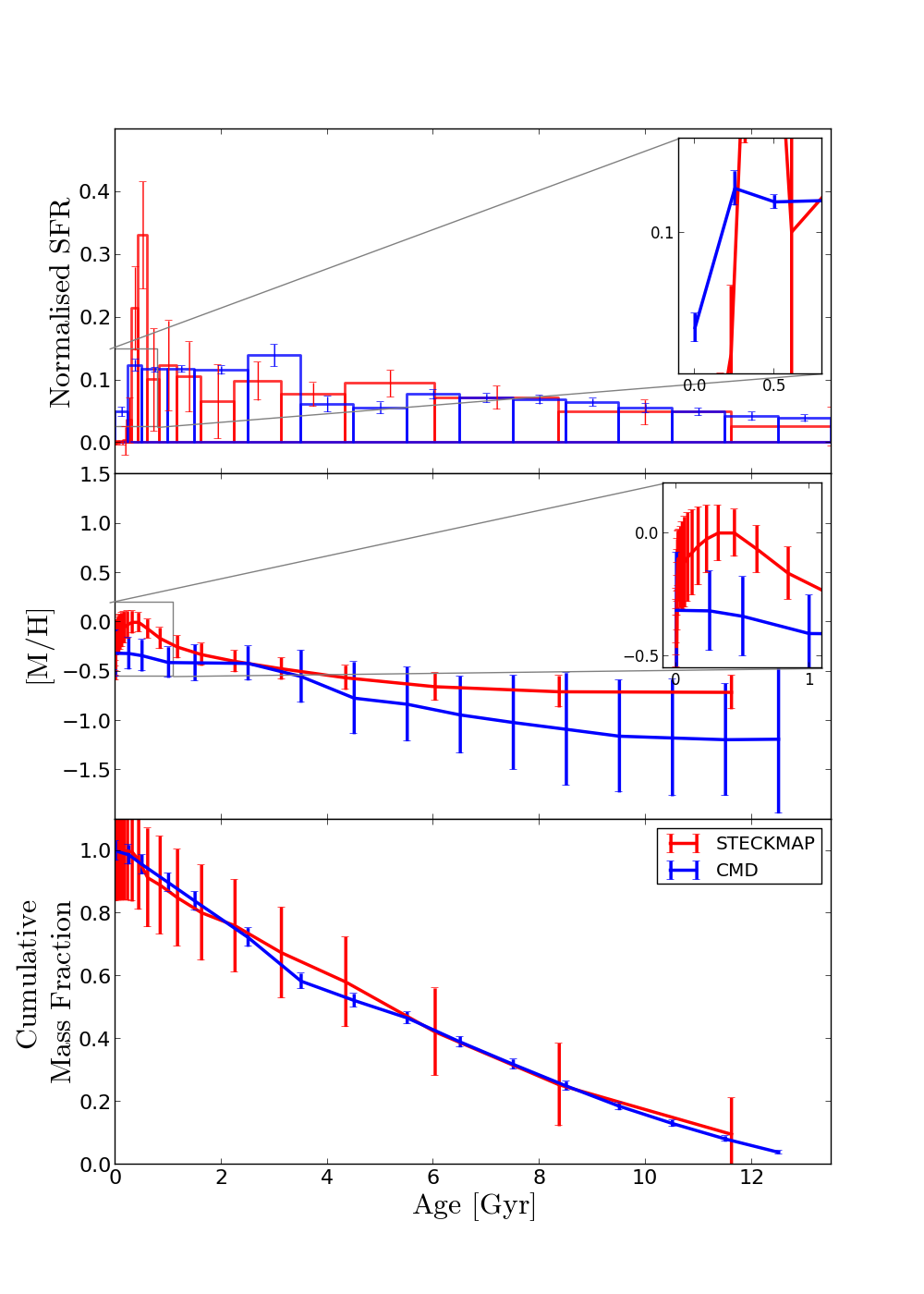} \\
\caption{Comparison between the LMC bar SFH from the CMD and the integrated spectrum using {\tt STECKMAP}. Left: Test 3; Right: Test 4. For further information see Fig.~\ref{SFH_steck_3}.}
\label{steck_tests_2}
\end{figure*}

\begin{figure*}
\centering
\includegraphics[scale=0.31]{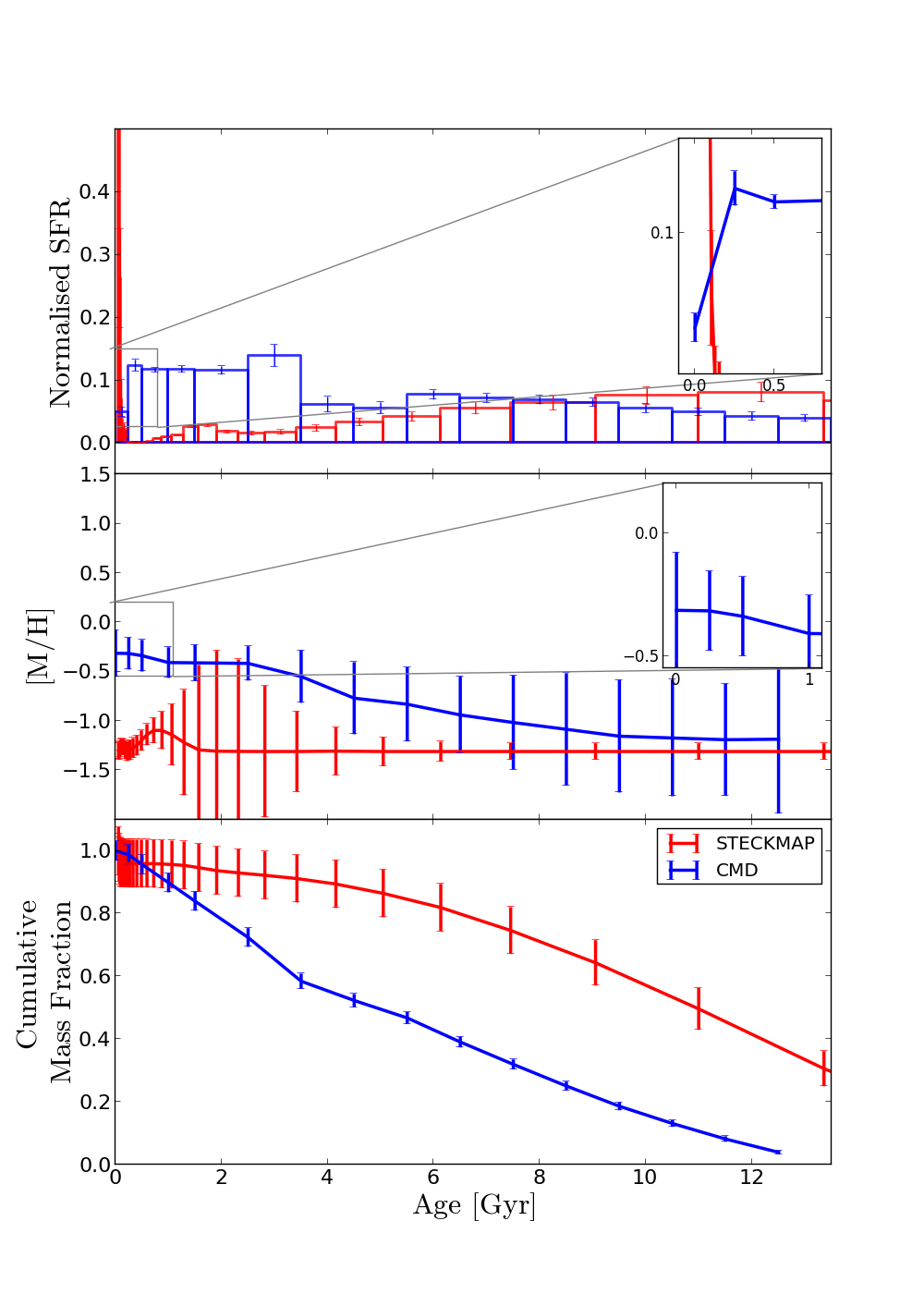} ~
\includegraphics[scale=0.31]{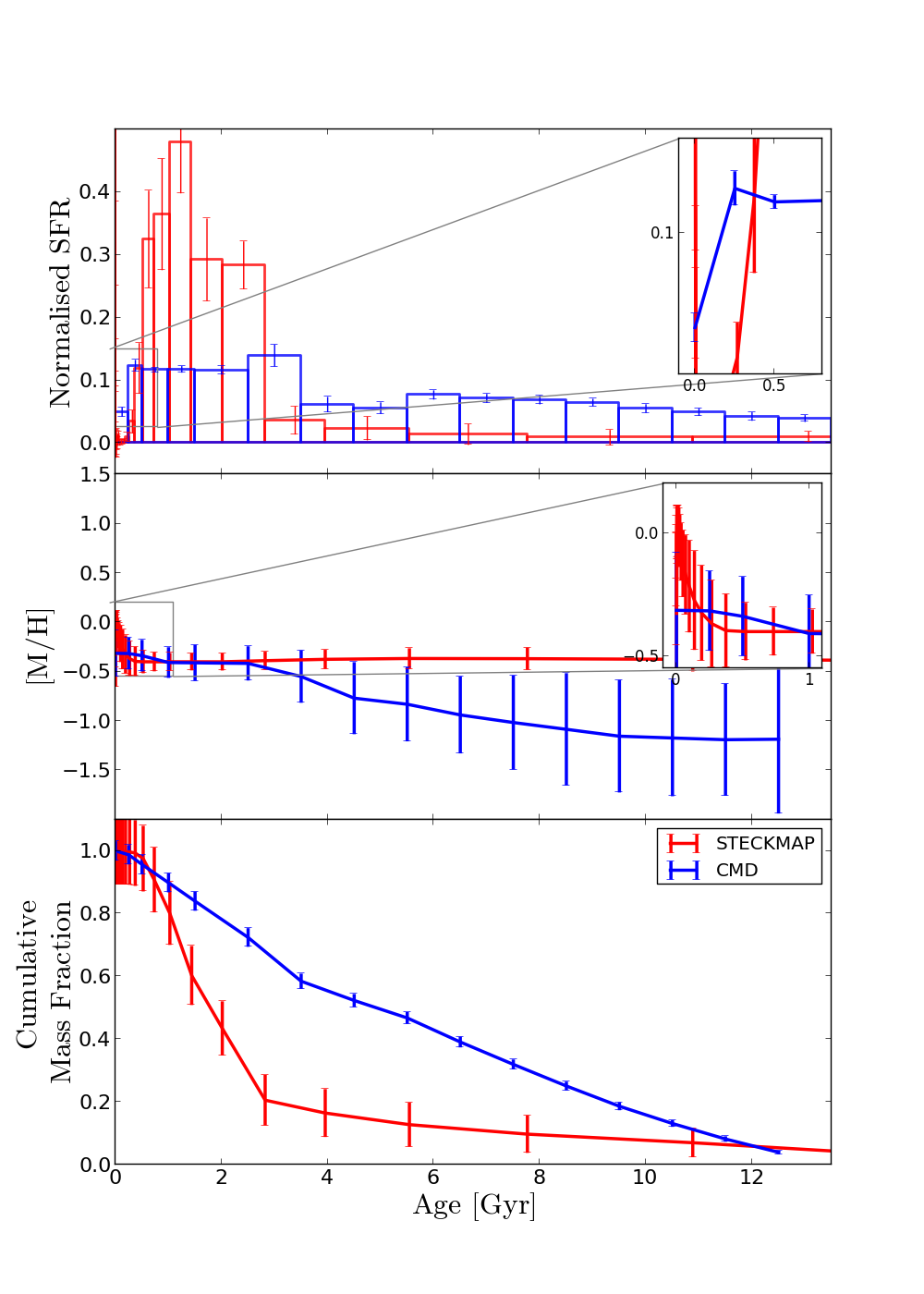} \\
\caption{Comparison between the LMC bar SFH from the CMD and the integrated spectrum using {\tt STECKMAP}. Left: Test 5; Right: Test 6. For further information see Fig.~\ref{SFH_steck_3}.}
\label{steck_tests_3}
\end{figure*}

\begin{figure*}
\centering
\includegraphics[scale=0.31]{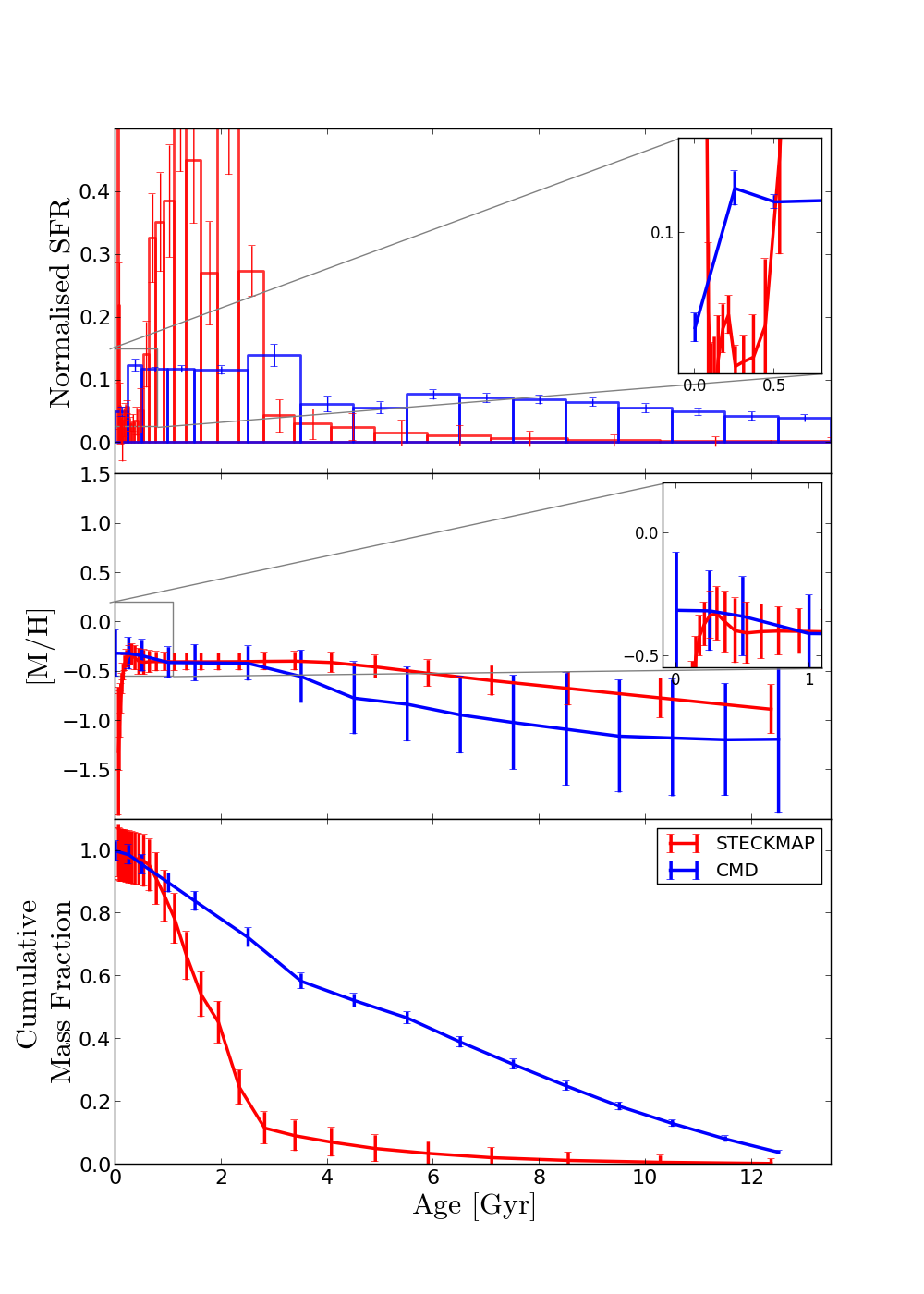} ~
\includegraphics[scale=0.31]{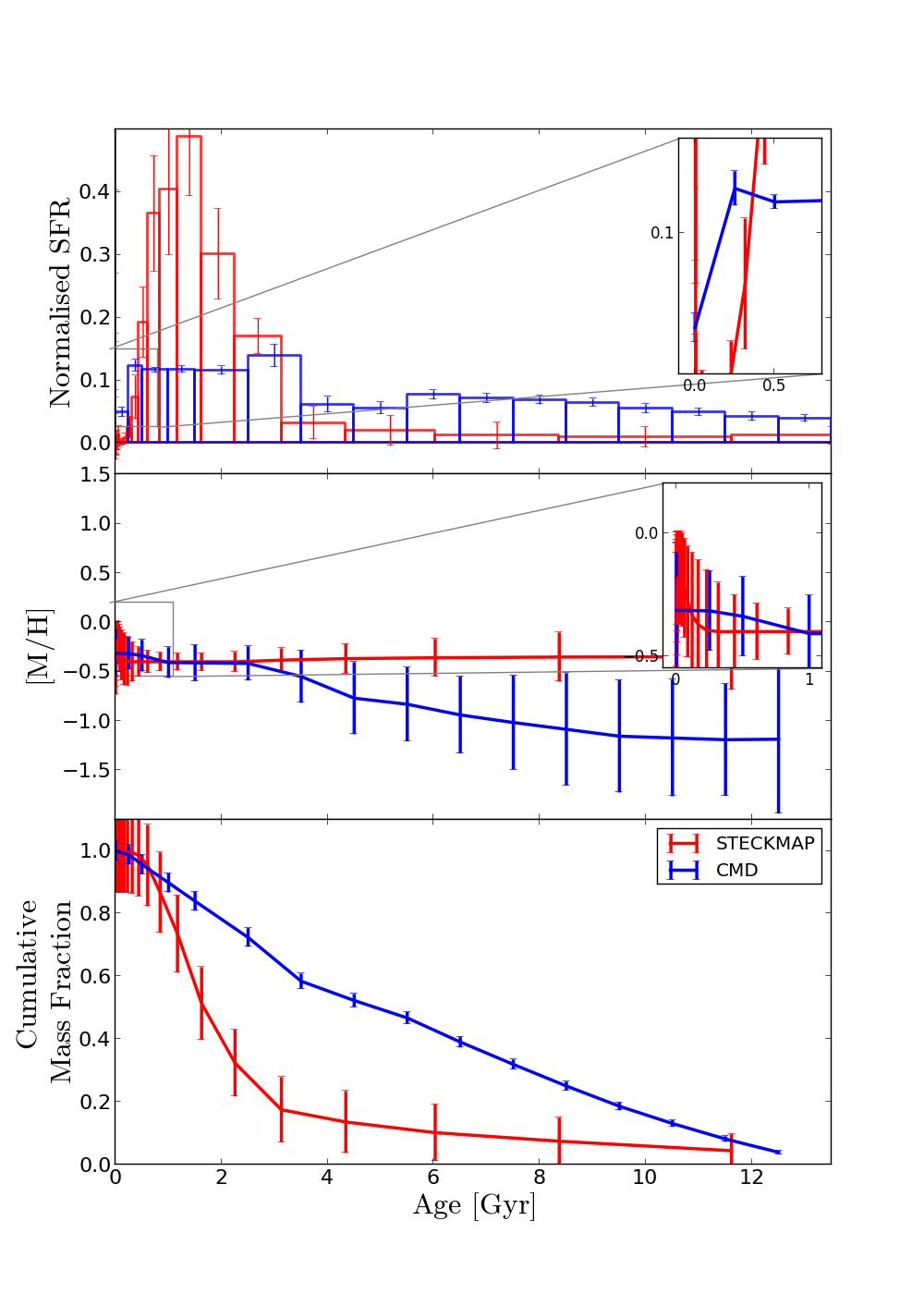} \\
\caption{Comparison between the LMC bar SFH from the CMD and the integrated spectrum using {\tt STECKMAP}. Left: Test 7; Right: Test 8. For further information see Fig.~\ref{SFH_steck_3}.}
\label{steck_tests_4}
\end{figure*}

\begin{figure*}
\centering
\includegraphics[scale=0.31]{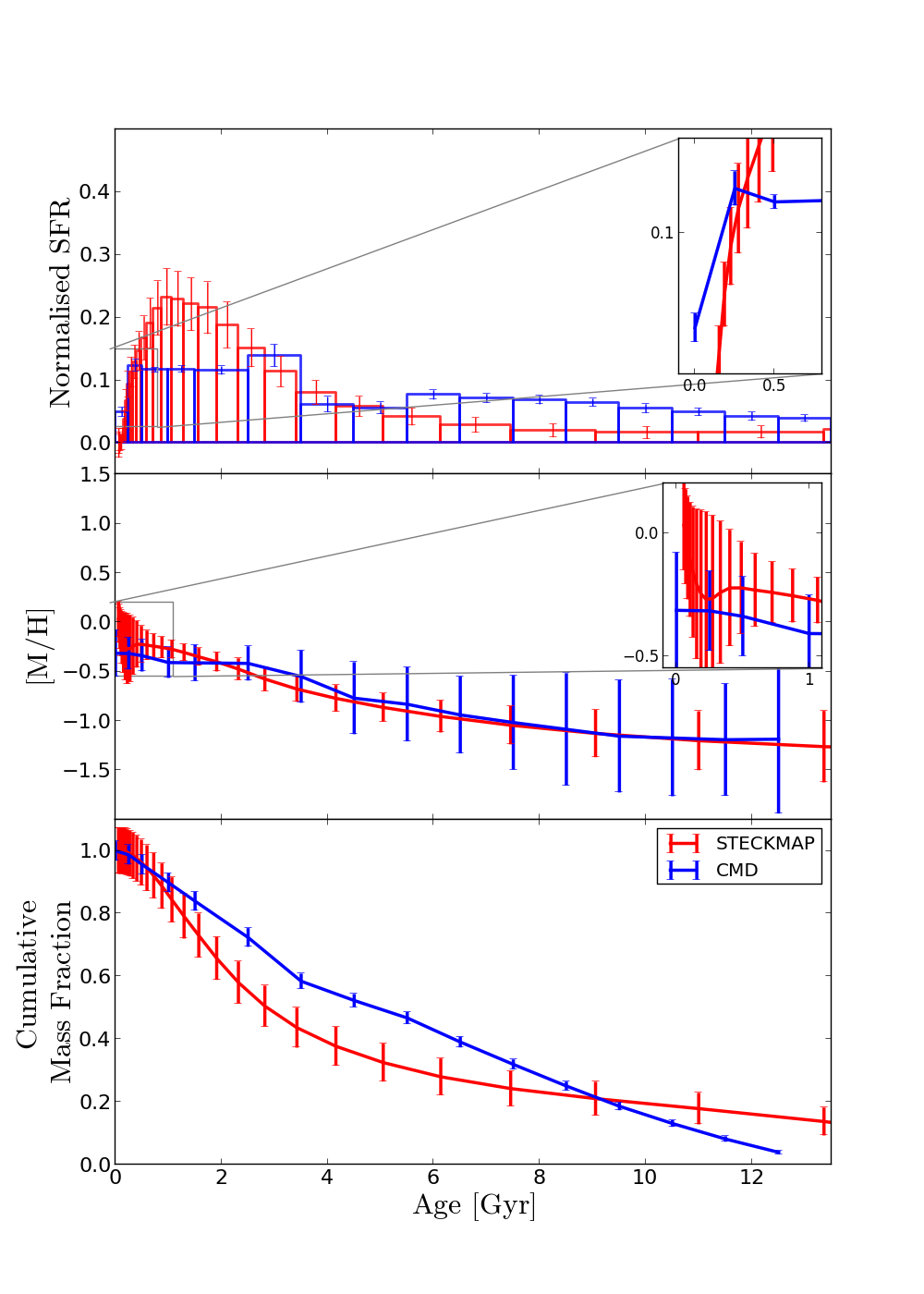} ~
\includegraphics[scale=0.31]{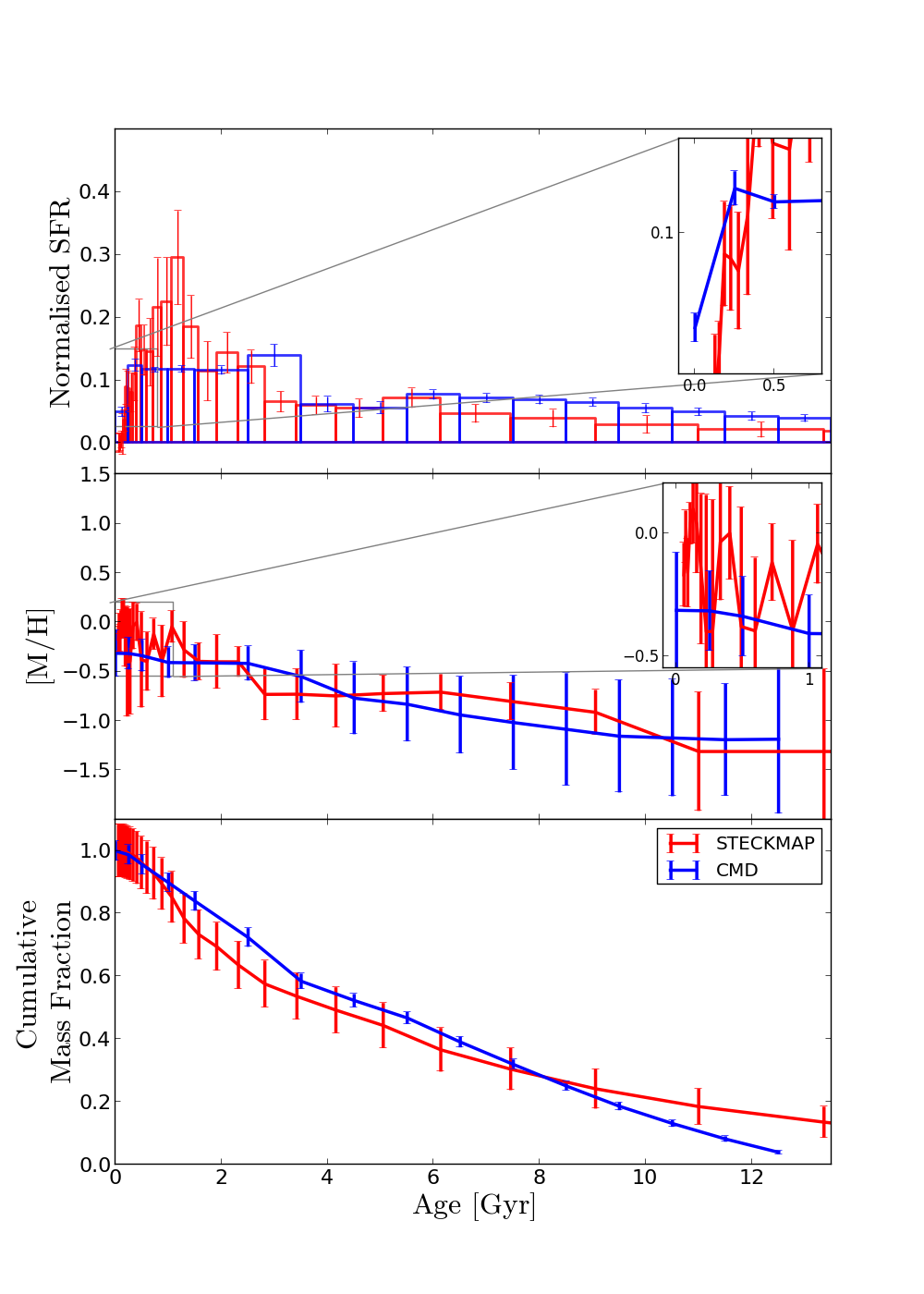} \\
\caption{Comparison between the LMC bar SFH from the CMD and the integrated spectrum using {\tt STECKMAP}. Left: Test 9; Right: Test 10. For further information see Fig.~\ref{SFH_steck_3}.}
\label{steck_tests_5}
\end{figure*}

\begin{figure*}
\centering
\includegraphics[scale=0.31]{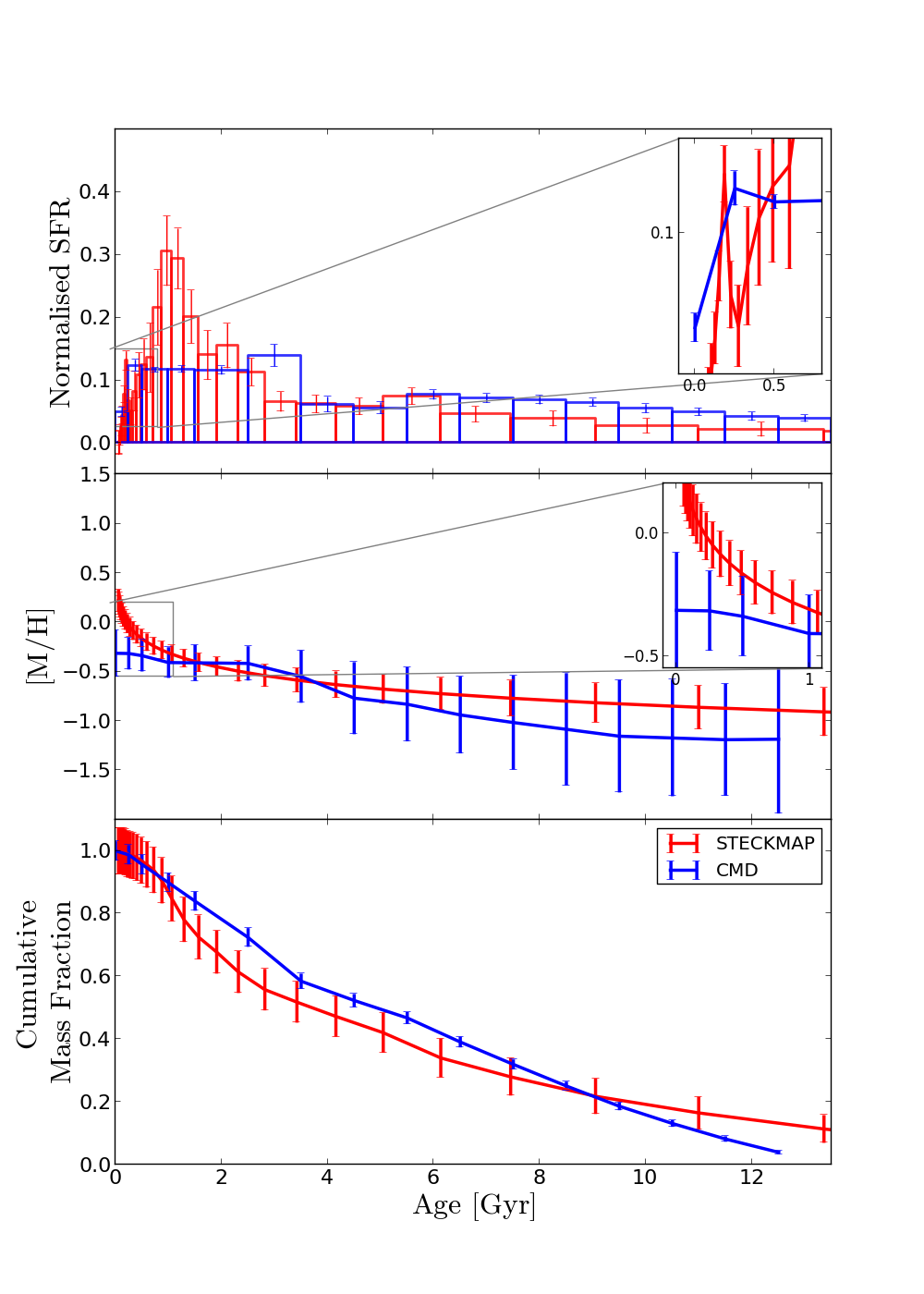} ~
\includegraphics[scale=0.31]{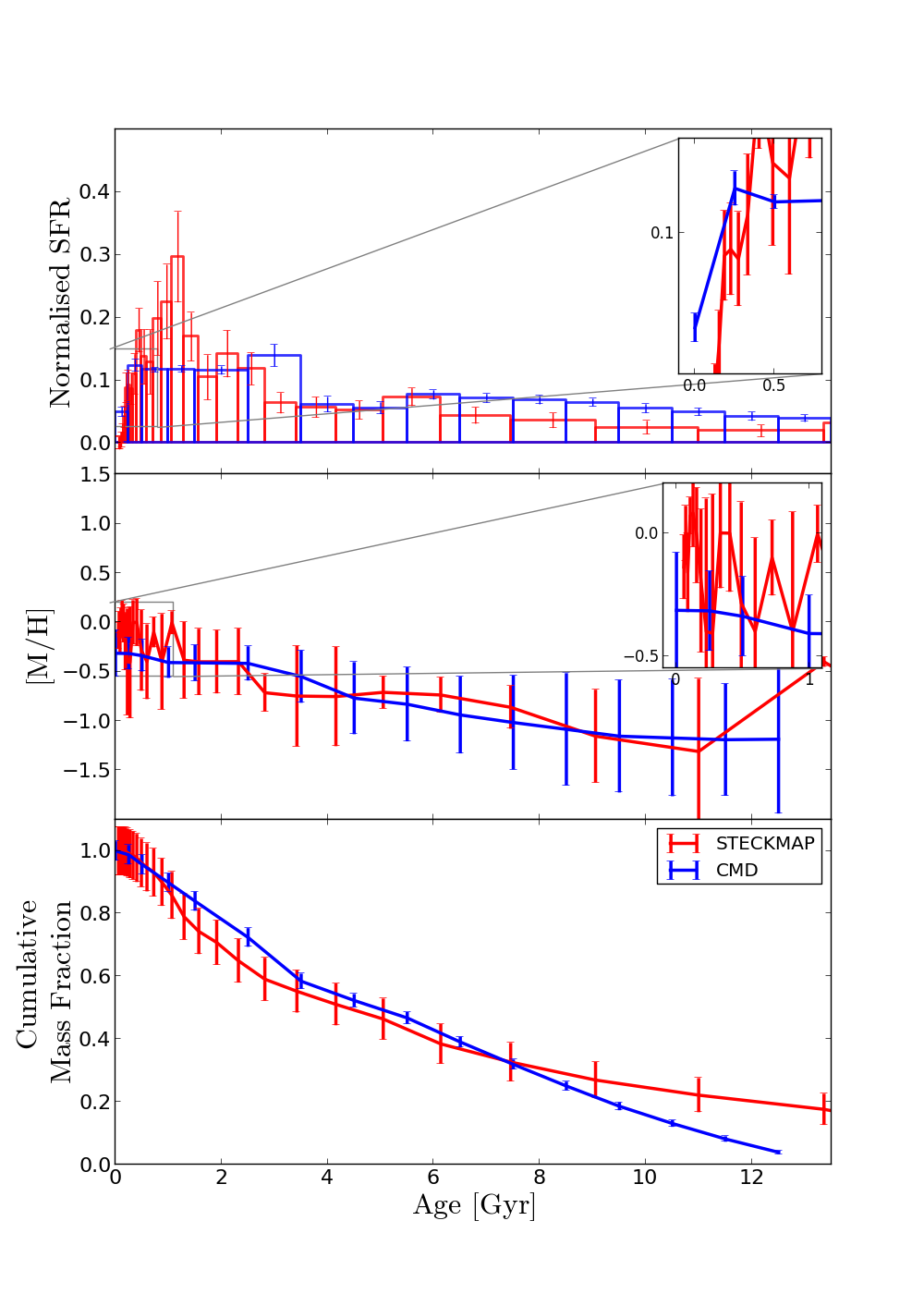} \\
\caption{Comparison between the LMC bar SFH from the CMD and the integrated spectrum using {\tt STECKMAP}. Left: Test 11; Right: Test 12. For further information see Fig.~\ref{SFH_steck_3}.}
\label{steck_tests_6}
\end{figure*}

\begin{figure*}
\centering
\includegraphics[scale=0.31]{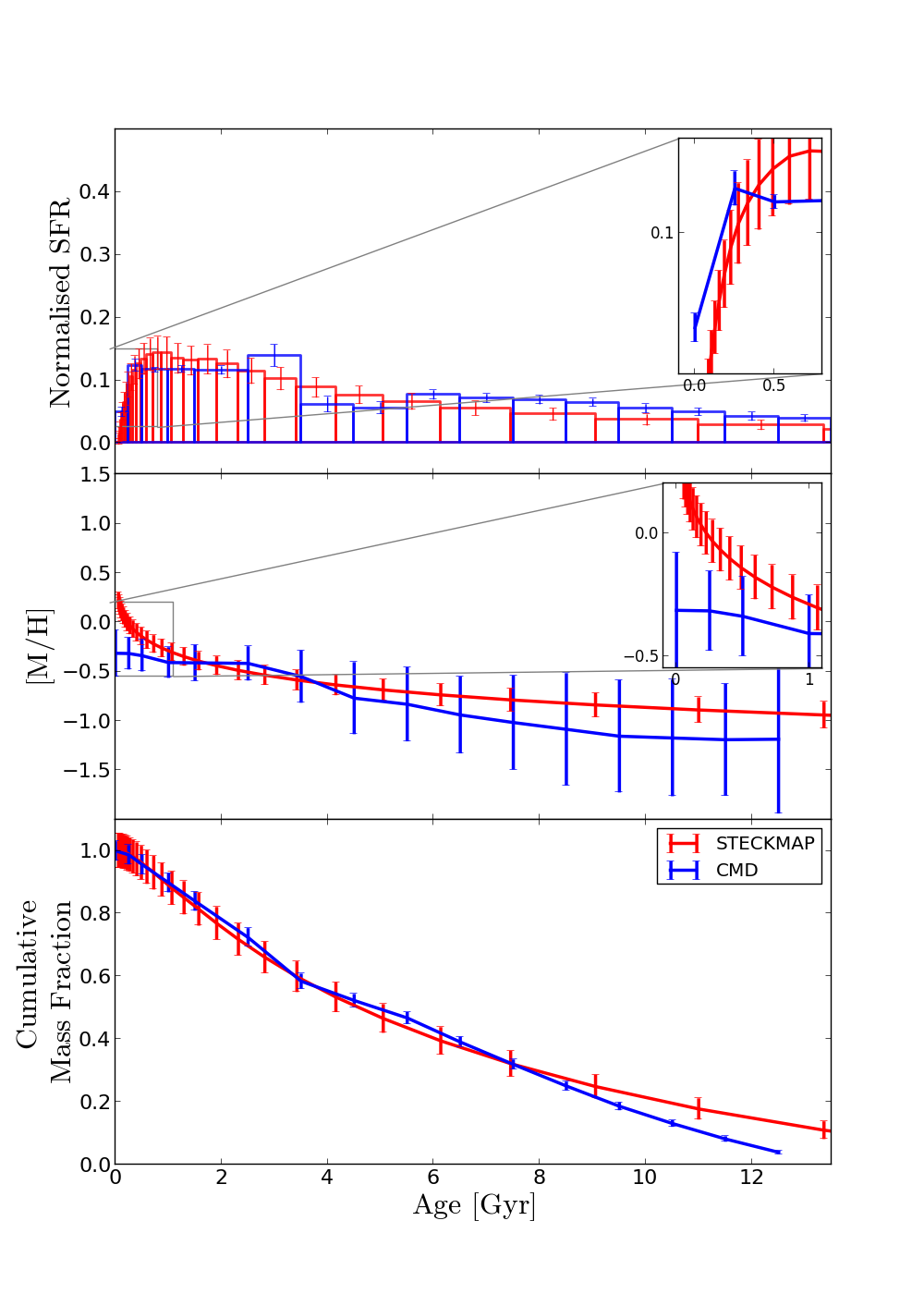} ~
\includegraphics[scale=0.31]{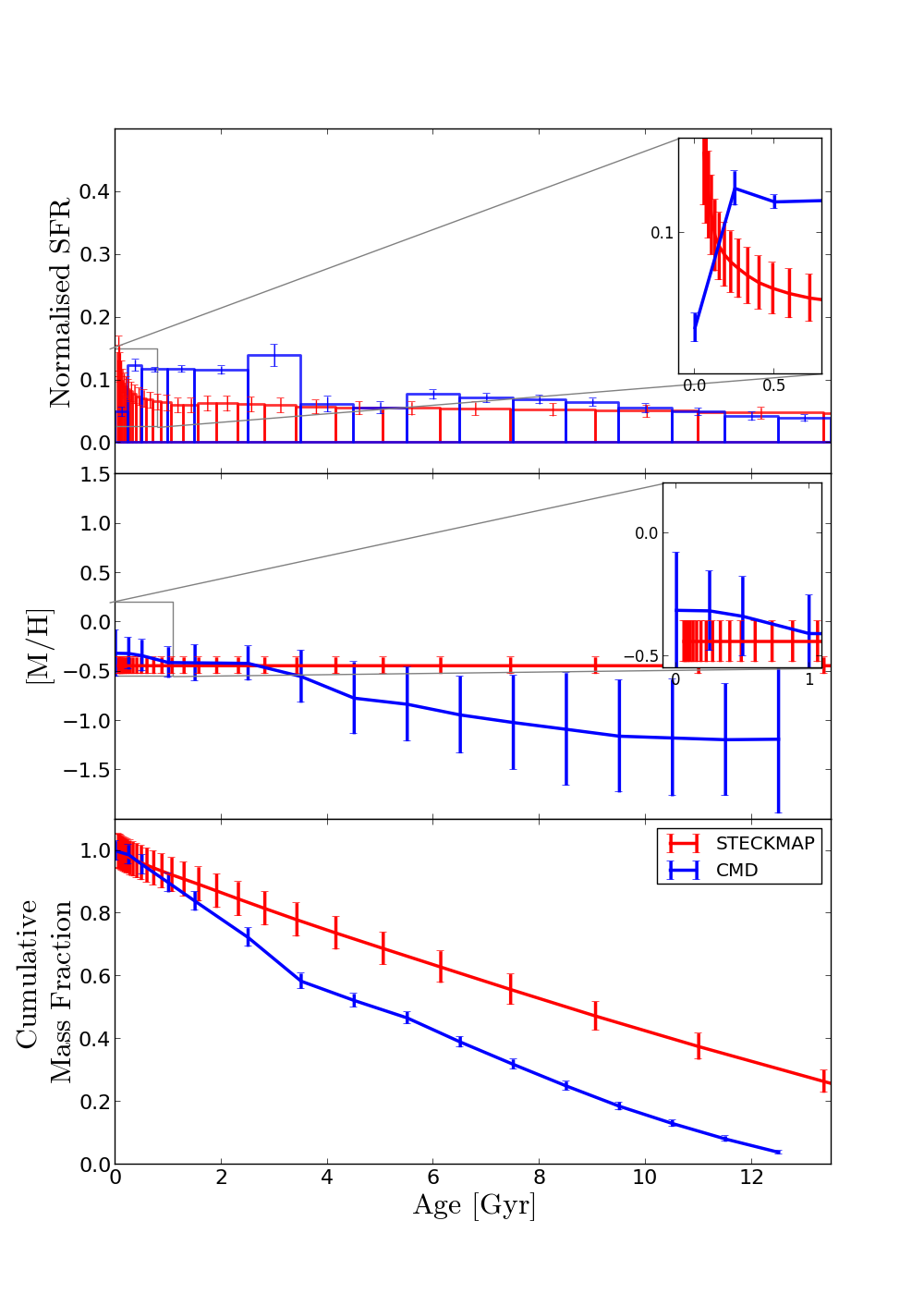} \\
\caption{Comparison between the LMC bar SFH from the CMD and the integrated spectrum using {\tt STECKMAP}. Left: Test 13; Right: Test 14. For further information see Fig.~\ref{SFH_steck_3}.}
\label{steck_tests_7}
\end{figure*}

\begin{figure*}
\centering
\includegraphics[scale=0.31]{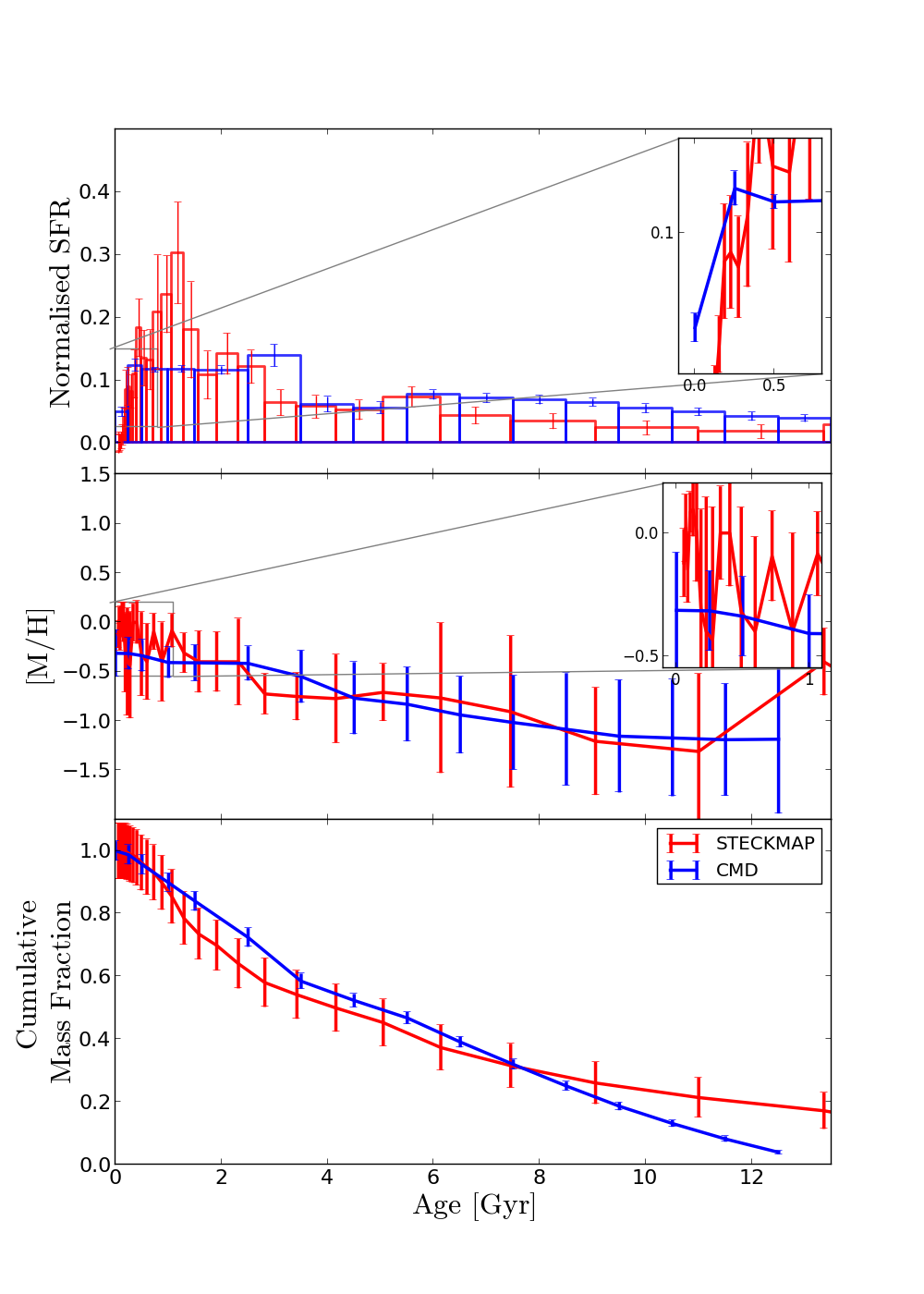} ~
\includegraphics[scale=0.31]{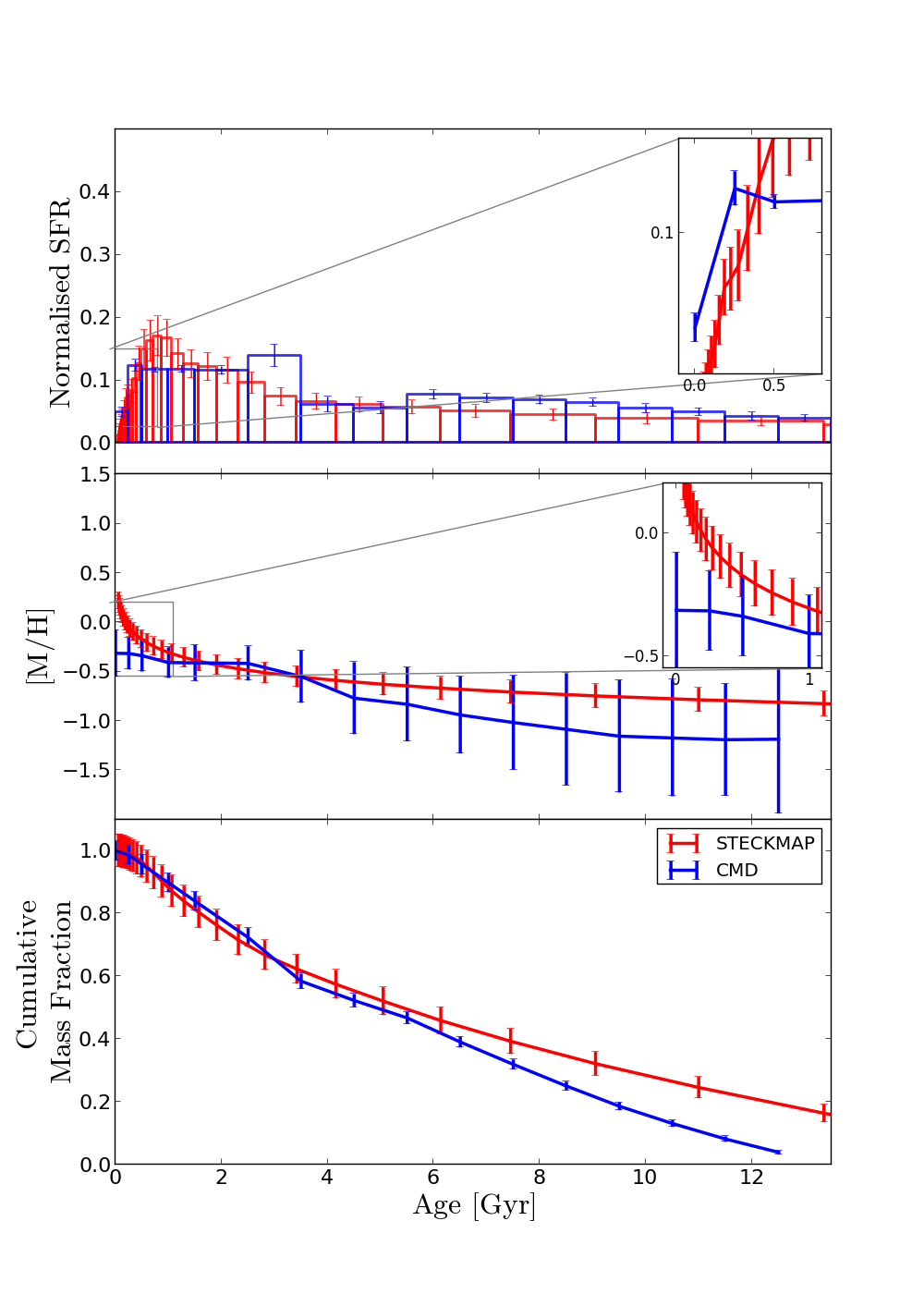} \\
\caption{Comparison between the LMC bar SFH from the CMD and the integrated spectrum using {\tt STECKMAP}. Left: Test 15; Right: Test 16. For further information see Fig.~\ref{SFH_steck_3}.}
\label{steck_tests_8}
\end{figure*}

\begin{figure*}
\centering
\includegraphics[scale=0.31]{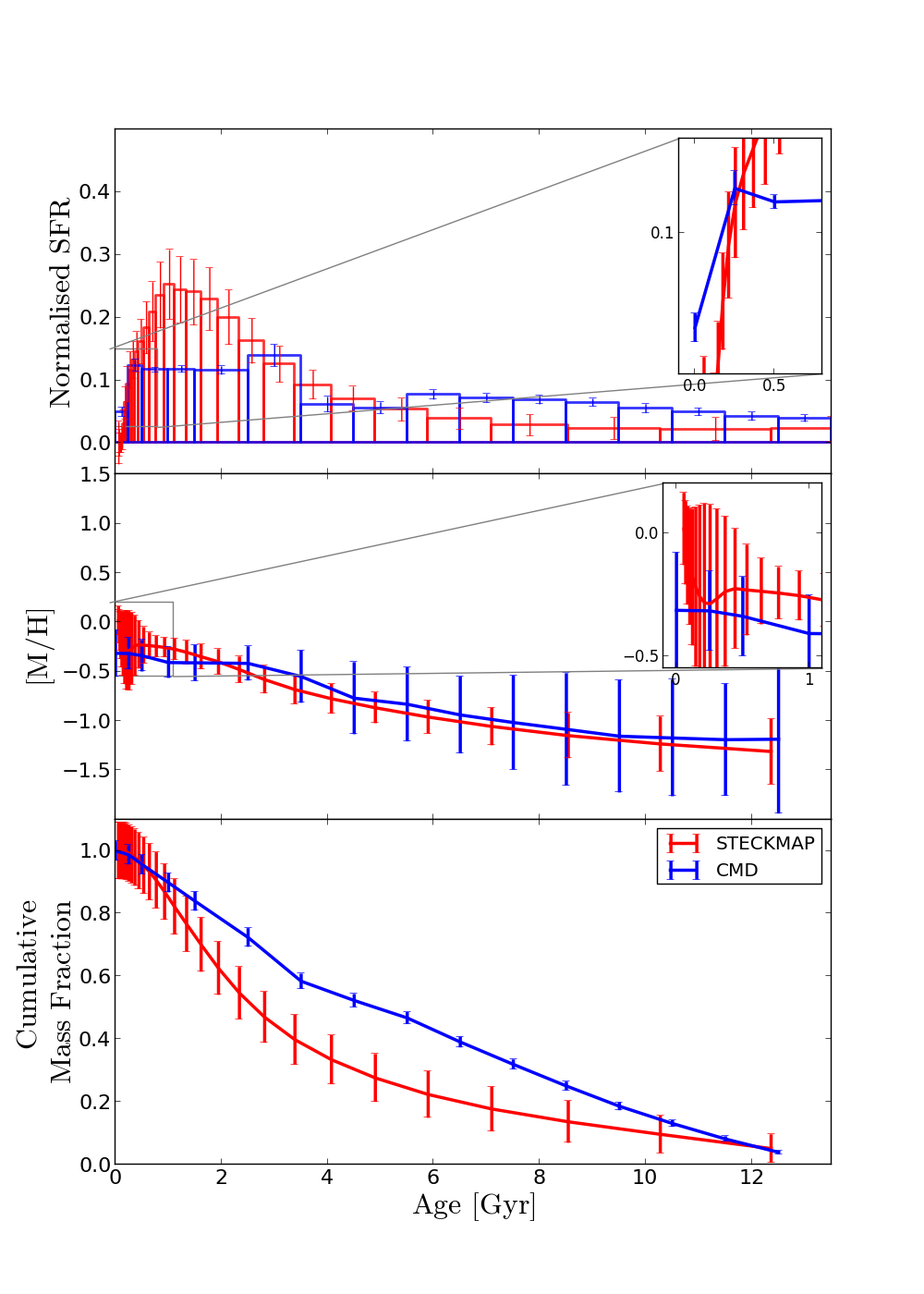} ~
\includegraphics[scale=0.31]{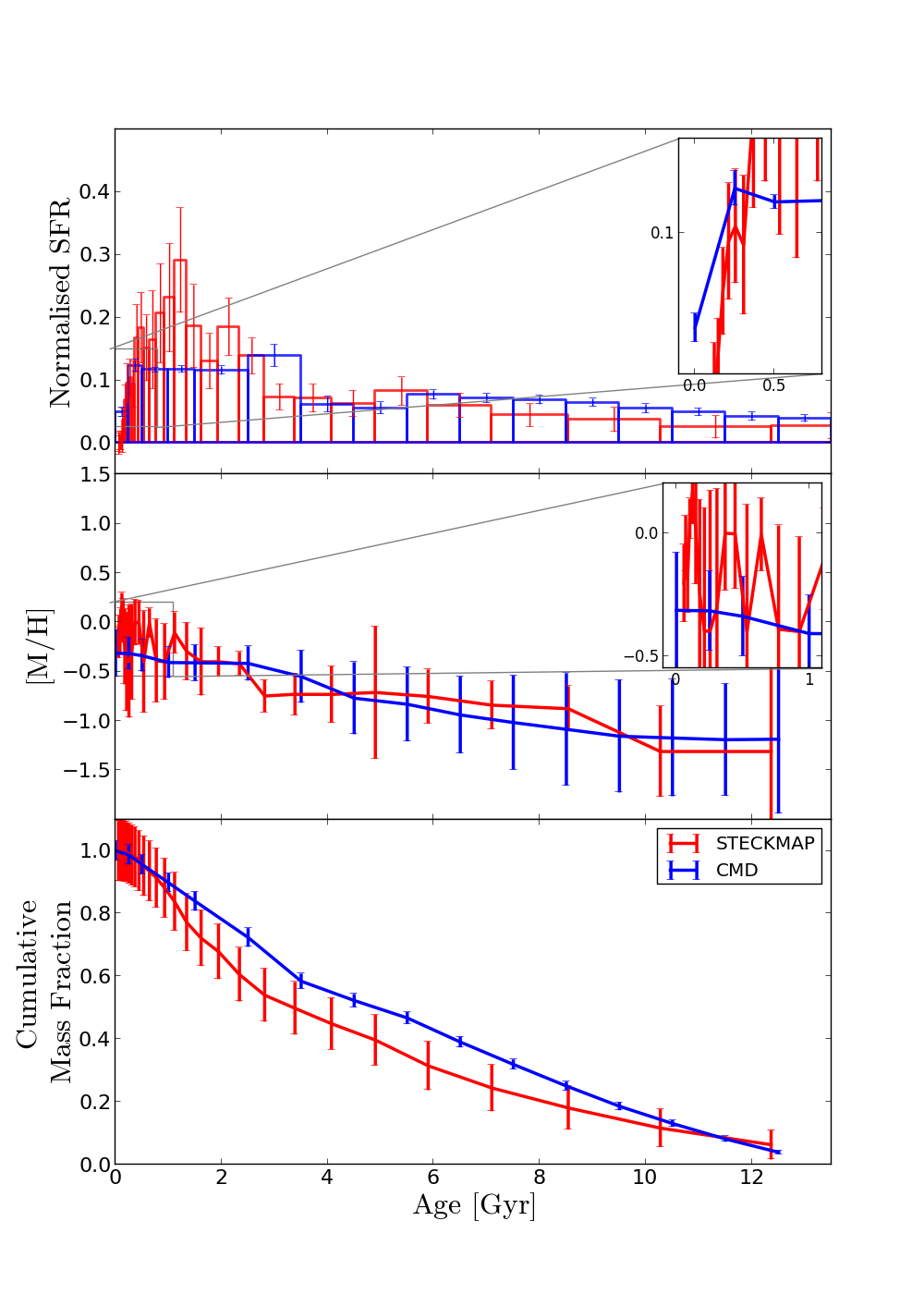} \\
\caption{Comparison between the LMC bar SFH from the CMD and the integrated spectrum using {\tt STECKMAP}. Left: Test 17; Right: Test 18. For further information see Fig.~\ref{SFH_steck_3}.}
\label{steck_tests_9}
\end{figure*}

\begin{figure*}
\centering
\includegraphics[scale=0.31]{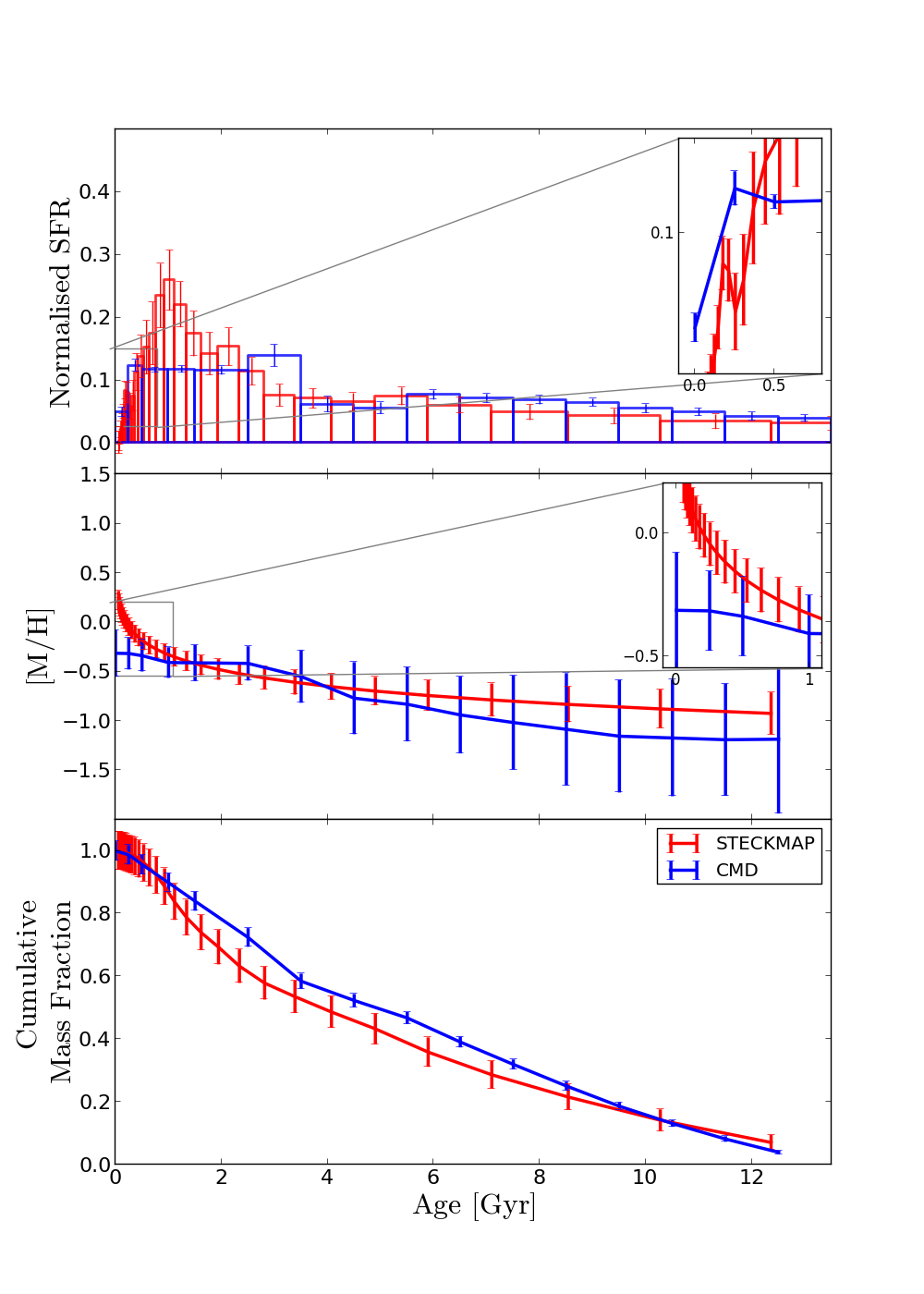} ~
\includegraphics[scale=0.31]{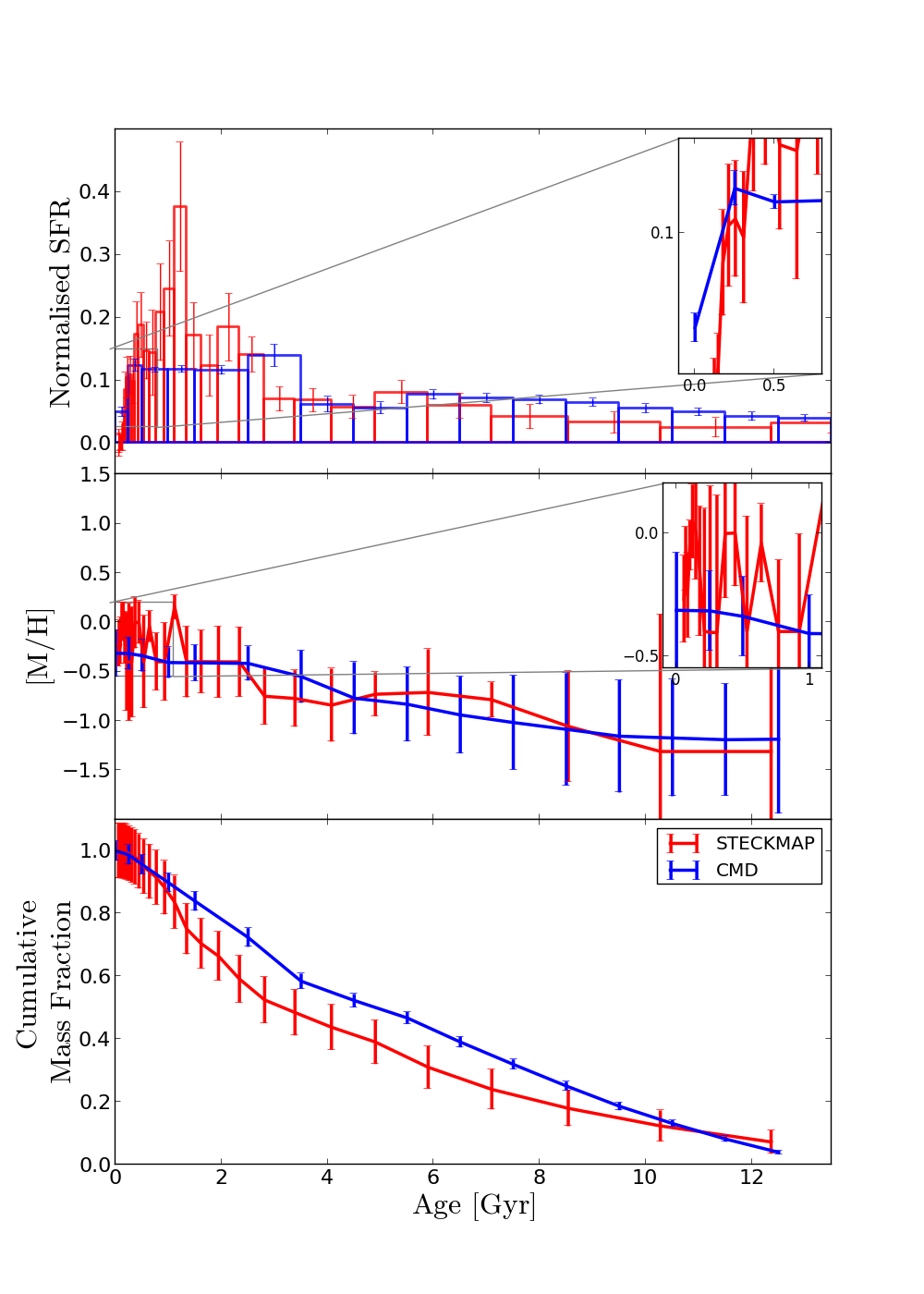} \\
\caption{Comparison between the LMC bar SFH from the CMD and the integrated spectrum using {\tt STECKMAP}. Left: Test 19; Right: Test 20. For further information see Fig.~\ref{SFH_steck_3}.}
\label{steck_tests_10}
\end{figure*}

\begin{figure*}
\centering
\includegraphics[scale=0.31]{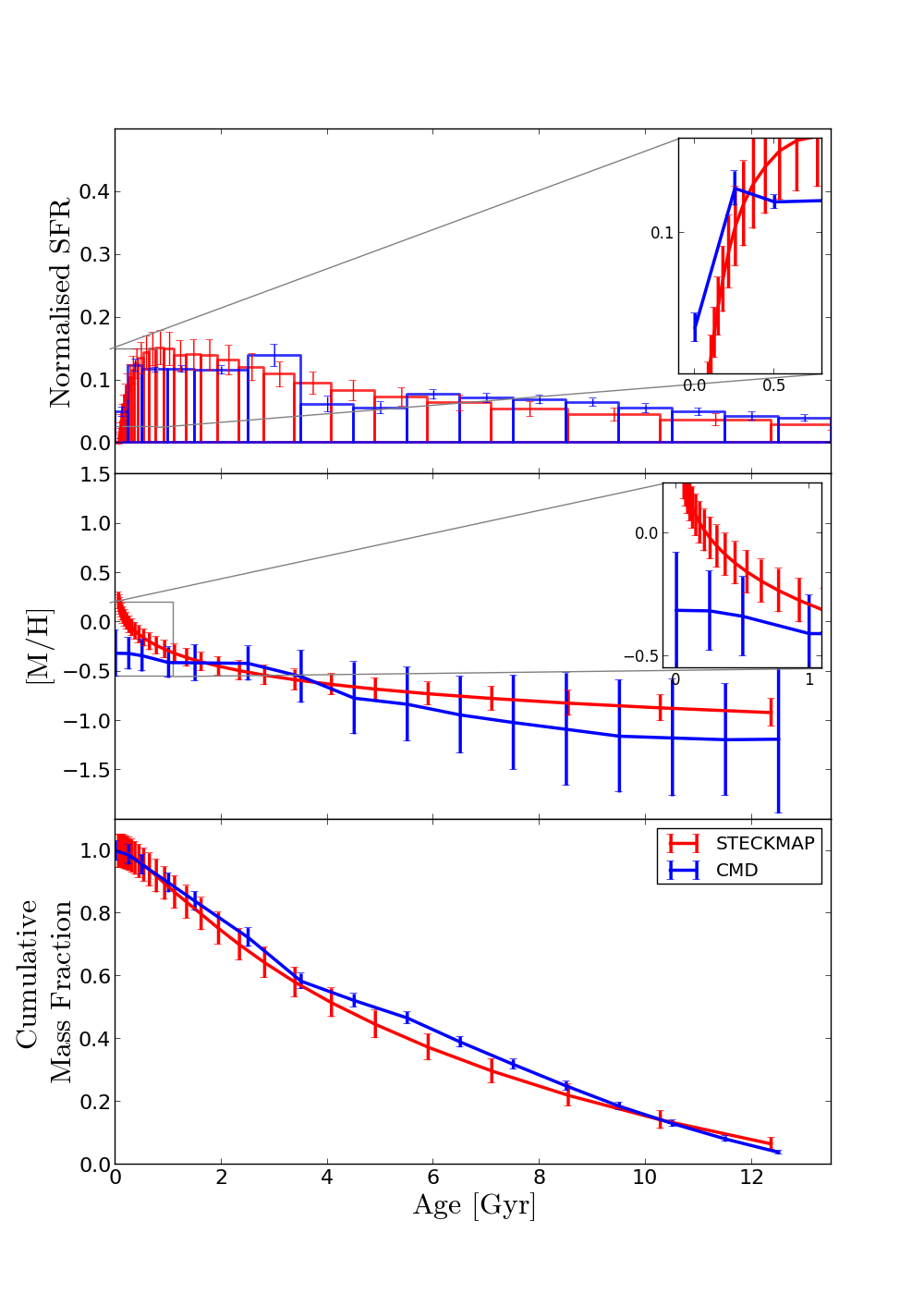} ~
\includegraphics[scale=0.31]{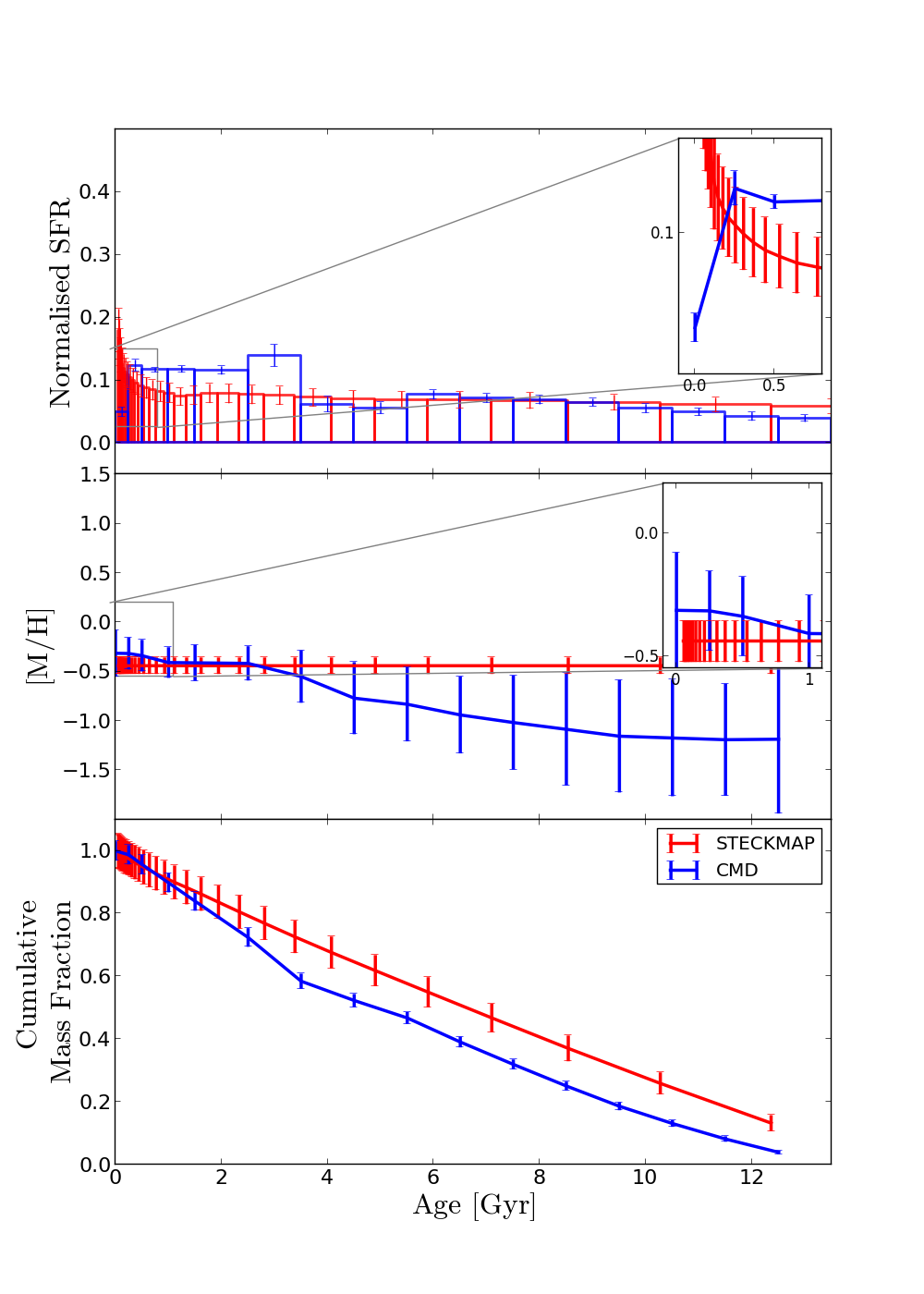} \\
\caption{Comparison between the LMC bar SFH from the CMD and the integrated spectrum using {\tt STECKMAP}. Left: Test 21; Right: Test 22. For further information see Fig.~\ref{SFH_steck_3}.}
\label{steck_tests_11}
\end{figure*}

\begin{figure*}
\centering
\includegraphics[scale=0.31]{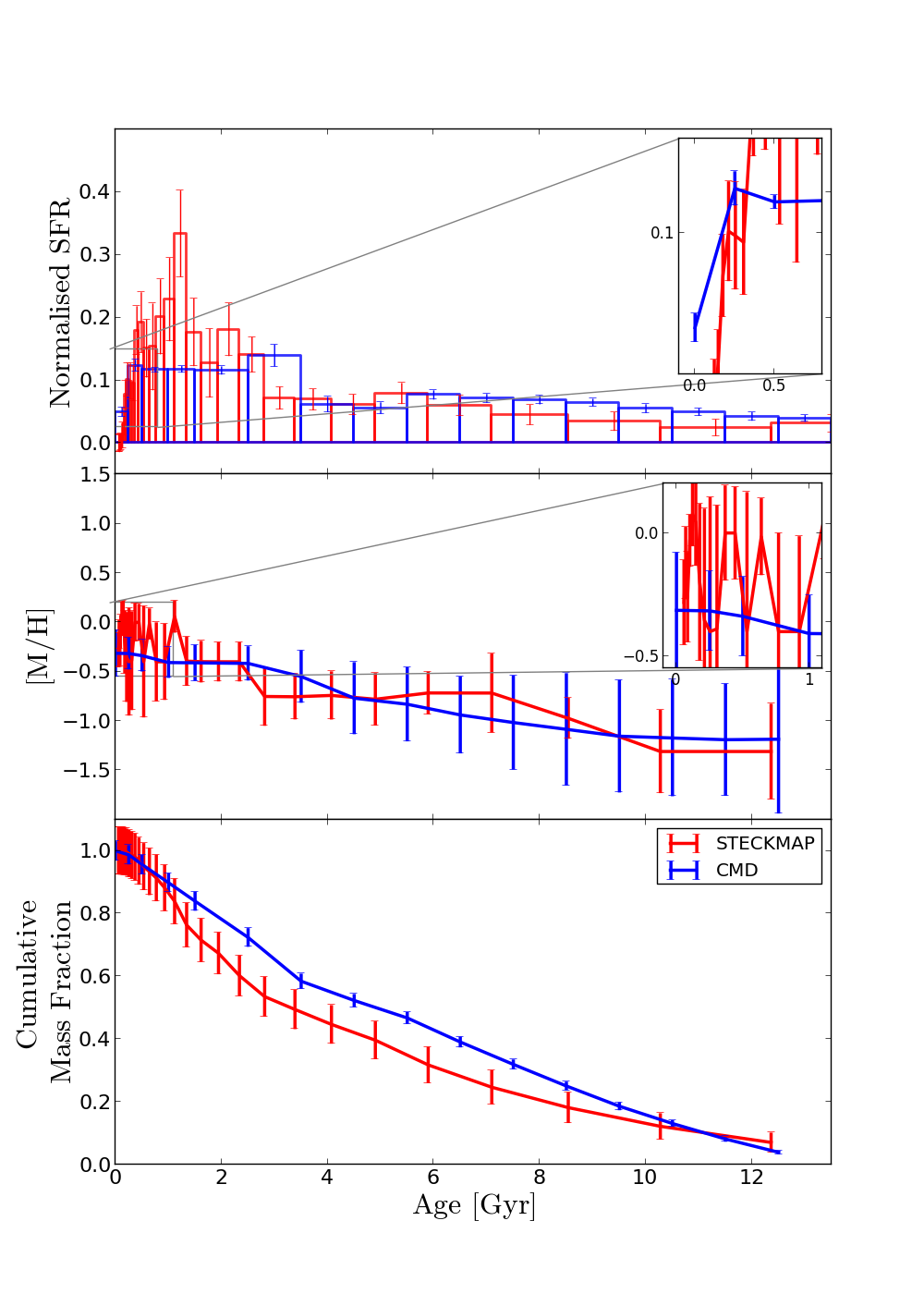} ~
\includegraphics[scale=0.31]{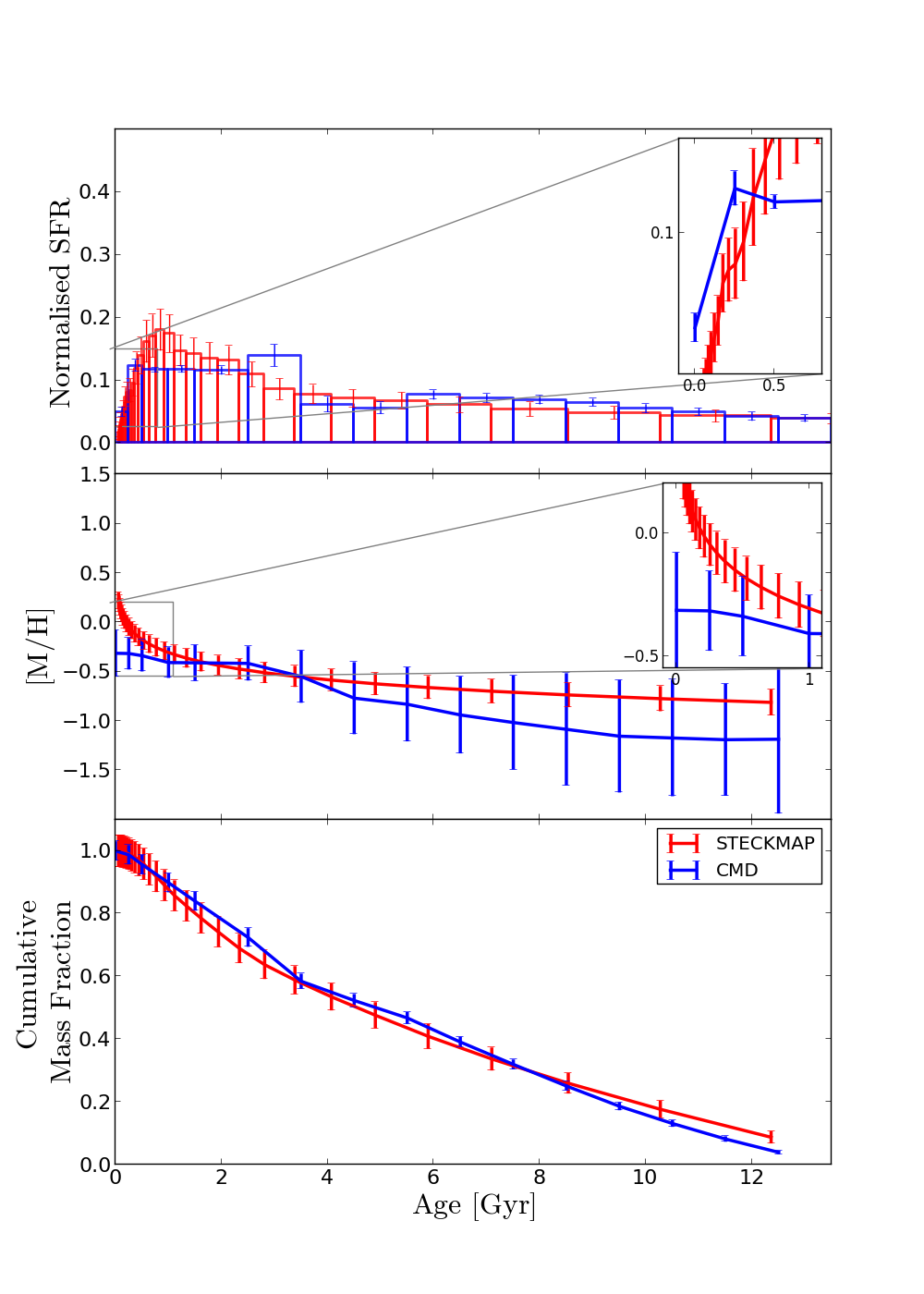} \\
\caption{Comparison between the LMC bar SFH from the CMD and the integrated spectrum using {\tt STECKMAP}. Left: Test 23; Right: Test 24. For further information see Fig.~\ref{SFH_steck_3}.}
\label{steck_tests_12}
\end{figure*}

\end{document}